\documentclass[opre, nonblindrev]{informs3_noMNSC_removethis_}
\usepackage[T1]{fontenc}
\usepackage[latin9]{inputenc}
\usepackage{geometry}
\usepackage{array}
\usepackage{units}
\usepackage{amsmath}
\usepackage{amssymb}
\usepackage{graphicx}
\usepackage[authoryear]{natbib}
\usepackage[unicode=true,pdfusetitle,
 bookmarks=true,bookmarksnumbered=true,bookmarksopen=false,
 breaklinks=false,pdfborder={0 0 0},pdfborderstyle={},backref=false,colorlinks=false]
 {hyperref}

\makeatletter

\providecommand{\tabularnewline}{\\}

\DoubleSpacedXI

\usepackage{natbib}
 \bibpunct[, ]{(}{)}{,}{a}{}{,}%
 \def\bibfont{\small}%
\TheoremsNumberedThrough     
\ECRepeatTheorems

\EquationsNumberedThrough    

\usepackage{chngcntr}

\makeatother

\begin{document}
\RUNTITLE{Big data Bayesian risk assessment} 
\RUNAUTHOR{Shang, Dunson, Song}
\TITLE{Exploiting Big Data in Logistics Risk Assessment via Bayesian Nonparametrics}

\ARTICLEAUTHORS{
	\AUTHOR{Yan Shang}
	\AFF{Fuqua School of Business, Duke University;
		\EMAIL{yan.shang@duke.edu}}
	\AUTHOR{David Dunson}
	\AFF{Department of Statistical Science, Duke University;
		\EMAIL{dunson@duke.edu}}
	\AUTHOR{Jing-Sheng Song}
	\AFF{Fuqua School of Business, Duke University;
		\EMAIL{jingsheng.song@duke.edu}}}

\ABSTRACT{In cargo logistics, a key performance measure is transport risk, defined as the deviation of the actual arrival time from the planned arrival time. Neither earliness nor tardiness is desirable for customer and freight forwarders. In this paper, we investigate ways to assess and forecast transport risks using a half-year of air cargo data, provided by a leading forwarder on 1336 routes served by 20 airlines. Interestingly, our preliminary data analysis shows a strong multimodal feature in the transport risks, driven by unobserved events, such as cargo missing flights. To accommodate this feature, we introduce a Bayesian nonparametric model -- the probit stick-breaking process (PSBP) mixture model -- for flexible estimation of the conditional (i.e., state-dependent) density function of transport risk. We demonstrate that using alternative methods can lead to misleading inferences. Our model provides a tool for the forwarder to offer customized price and service quotes. It can also generate baseline airline performance to enable fair supplier evaluation. Furthermore, the method allows us to separate recurrent risks from disruption risks. This is important, because hedging strategies for these two kinds of risks are often drastically different.}

\KEYWORDS{Bayesian statistics, big data, disruptions and risks, empirical, international air cargo logistics, nonparametric, probit stick-breaking mixture model}

\HISTORY{First submitted on December 17, 2014; revised on February 23, 2016}

\maketitle

\section{Introduction}

Global trade has grown considerably in recent decades; many companies
now have overseas facilities and supply chain partners. International
cargo logistics management thus plays an increasingly important role
in the global economy. Air transport delivers goods, that are time-sensitive,
expensive, perishable or used in just-in-time supply networks, at
competitive prices to customers worldwide. Indeed, air cargo transports
goods worth in excess of \$6.4 trillion annually. This is approximately
35\% of world trade by value \citep{iata_cargo_2014}. This industry,
including express traffic, is forecast by Boeing to grow at an average
4.7\% annual rate in the next two decades to reach a total of more
than twice the number of revenue tonne-kilometers (RTK) logged in
2013. However, attention paid to this industry is surprisingly little:
air cargo industry \textquoteleft .. has remained the poor cousin
to the more glamorous passenger side of the business (passenger air
transport industry)\textquoteright{} \citep{morrell_moving_2011}. 

The consequences of this neglect are significant as the service level
of cargo transport has become firms\textquoteright{} big concern.
In cargo logistics, a key (service) performance measure is\emph{ transport
risk} (or delivery reliability), defined as the deviation of the actual
arrival time from the planned arrival time, 
\[
\mbox{transport risk}=\mbox{actual arrival time}-\mbox{planned arrival time}.
\]
Neither earliness nor tardiness is desirable for customer and freight
forwarders. While tardiness causes delay in production and product/service
delivery to all downstream customers, earliness incurs additional
storage and handling costs. Extreme risks, such as more than 48 hour
delays or more than 24 hours earliness, is defined as \emph{(transport)
disruption} \emph{risks}, because they severely impact the operations
of the customers and the freight forwarders. To distinguish disruption
risks from the routine deviations within a day, we refer to the latter
as \emph{recurrent risks}. According to a 2011 PRTM survey, 69\% of
companies named improving delivery performance as their top supply
chain management strategy. In a 2010 report of Infosys, \textquotedblleft carrier
delays and non-performance on delivery\textquotedblright{} is ranked
as the leading risk in the logistics industry. Furthermore, in a 2014
survey conducted by the International Air Transport Association (IATA)
to major freight forwarders and their customers, low reliability is
perceived as the second most important factor (next to transportation
cost). 

In this paper, we study the transport risks of international air cargo
based on a half-year of air cargo data between 2012 and 2013, provided
by a leading forwarder on 1336 routes served by 20 airlines. Using
a Bayesian nonparametric (BNP) model \textendash{} the Probit stick-breaking
(PSBP) mixture model \textemdash{} we obtain accurate estimates of
transport risk distributions and disruption risk probabilities. Our
model provides a tool for the forwarder to offer customized price
and service quotes. It can also generate baseline airline performance
to enable fair supplier evaluation. 

We make several contributions to the Operations Management (OM) and
Transportation literature as outlined below. 

\subsection{Empirical Air Cargo Transport Risk Distribution}

Our work appears to be the first empirical study of global air logistics
in the supply chain literature. One interesting phenomenon observed
from the data is that the distribution of transport risk, conditional
on predictors (i.e., independent variables including airline, route,
shipping time, cargo weight etc), is a \textit{multimodal distribution},
as shown in Figure 
\begin{figure}[tbph]
\begin{centering}
\begin{minipage}[t]{0.49\columnwidth}%
\includegraphics[width=1\columnwidth]{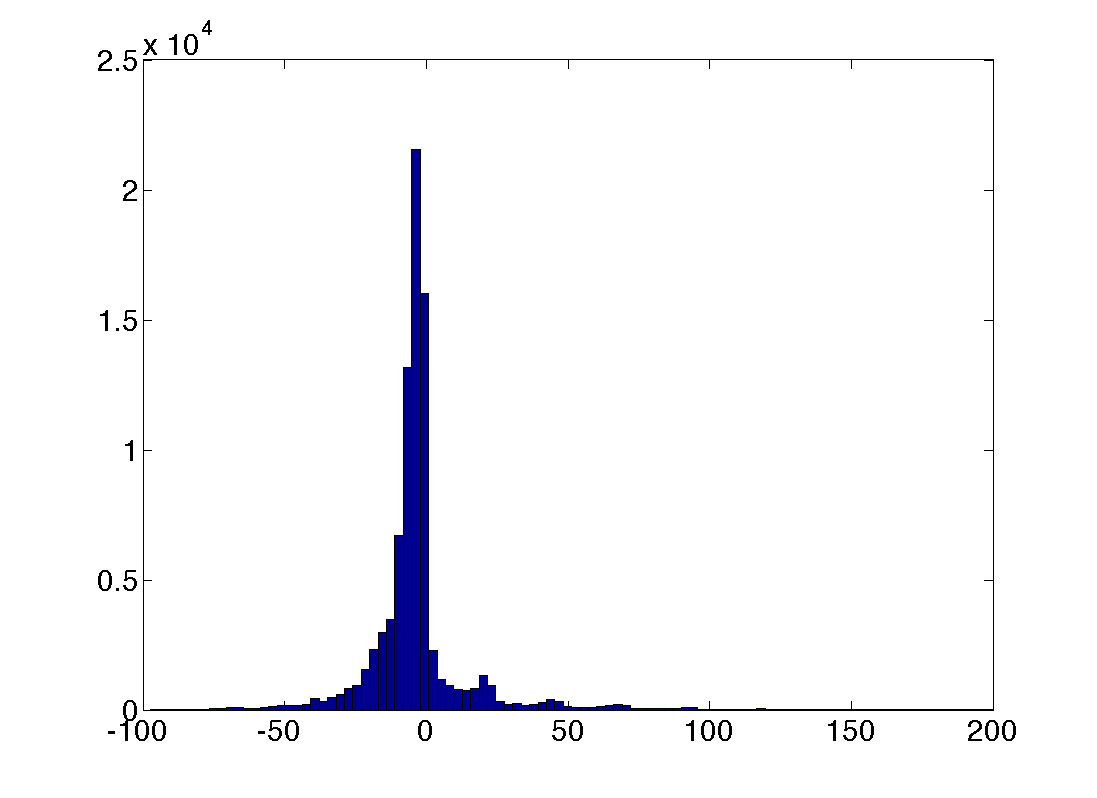}%
\end{minipage}\hfill{}%
\begin{minipage}[t]{0.49\columnwidth}%
\includegraphics[width=1\columnwidth]{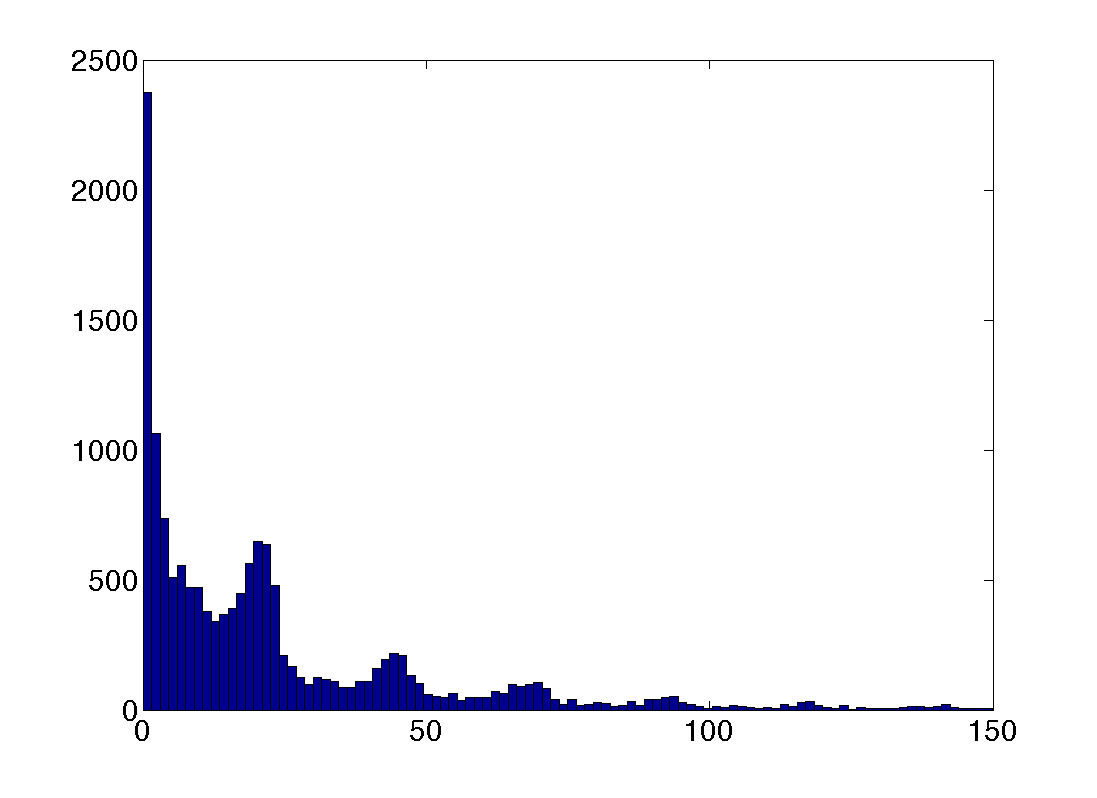}%
\end{minipage}
\par\end{centering}
\centering{}\caption{Left \textemdash{} Histogram of transport risk in hours; Right \textemdash{}
Histogram of positive transport risk in hours }
\label{fig: arrival deviation}
\end{figure}
\ref{fig: arrival deviation}. The left side of Figure \ref{fig: arrival deviation}
is the empirical distribution of transport risks of all shipments
observed in the data (almost 90 thousand shipments), which, clearly,
is a non-symmetric, long-tail distribution with several bumps at the
distribution's positive part. To better observe the bumps, we only
plot the data that falls in the range $\left(0,\;150\right)$ on the
right side of Figure \ref{fig: arrival deviation}. Here, we can see
clearly that big bumps concentrate around days (at 24 hours, 48 hours,
and 72 hours, etc.) and small bumps concentrate between days. These
systematic peaks are largely due to the fact that a cargo that failed
to be loaded onto its scheduled flight was loaded onto a flight on
the same route later. The scheduled gap between flights, which depends
heavily on the route, for example,\textbf{ }is usually around 24 hours
for international flights and several hours for domestic flights.
The time gaps between scheduled flights thus transfer to the gaps
between different peaks in the conditional distribution of transport
risk to form a multimodal distribution; see ${\cal x}$3 for more
detail. 

Previous empirical studies primarily focus on domestic passenger flight
arrival or departure delays; see \citet{deshpande_impact_2012} for
a review. Most of this literature assumes delays follow unimodal distributions,
adopting linear models; e.g, \citet{shumsky_dynamic_1995} and \citet{mueller_analysis_2002}.
However, delay distributions exhibit clear multimodality, making linear
models unsuitable for air cargo transport risk assessment and prediction.
The previous focus on linear models may have been due to the use of
data from the US Department of Transportation (DOT) collected at the
level of each flight. Our data are instead collected at the level
of each cargo trip, including information on a trip from the beginning
to end, usually consisting of several connecting flights. Data on
the full trip allow us to explore new transport uncertainties not
considered before. Specifically, we include information on delays
due to missed flights within transportation risk. The clear multimodality
in the full trip delay distributions motivates the new modeling approaches
proposed in this paper. However, the modeling methods we develop are
not restricted to cargo transport risk but can also be applied to
other transport risks (e.g., passenger air transport). For an air
passenger, the transport risk is determined by when the passenger
arrives at the destination. Passenger arrival time can be different
from the arrival time of the planned last flight since the passenger
might miss the final flight due to a delay of his/her previous flight
or some errands at the connecting airport. From this perspective,
the passenger transport risk problem is similar to the cargo transport
risk problem that we study in this paper. 

To our best knowledge, the closest work to ours is \citet{tu_estimating_2008}.
The authors studied the departure push-back delay of flights using
a model consisting of three parts accounting for seasonality, daily
trends and a residual error modeled by a mixture distribution with
Gaussian kernels and mixing weights fixed. The authors used the model
to fit one year of data on flights originating from Denver International
Airport and operated by United Airlines. However, the simplicity of
their model makes it inapplicable to our study with much more volatile
transport risks (as explain above) and also a much bigger data set
containing cargo shipments to more than 200 airports operated by more
than 20 airlines in 95 countries. To accommodate this complexity and
better fit our data, we use a Bayesian nonparametric model. In extensive
model comparisons with alternative models (see Appendix $\mathsection$B.5
for more details), including a flexible mixture model generalizing
\citet{tu_estimating_2008}, our model shows superior performance. 

\subsection{BNP Model and Conditional Distribution Function}

Our second contribution is methodological. To accommodate the multimodal
feature in the empirical transport risk distribution, we introduce
a state-of-the-art Bayesian statistics tool \textendash{} the BNP
mixture model. To the best of our knowledge, no prior work has used
related techniques in empirical OM, which so far predominantly applies
frequentist statistics, such as ordinary least square estimation or
maximum likelihood estimation, see, e.g., \citet{deshpande_impact_2012},
\citet{li_are_2014} and the references therein. 

Bayesian statistics has experienced rapid development in the past
two decades accelerated by ever-increasing computational power. Among
these tools, BNP mixture models have become popular in the last several
years, with applications in fields as diverse as finance, econometrics,
genetics, and medicine (refer to \citet{rodriguez_nonparametric_2011}
for references therein). A nonparametric mixture model can be expressed
as follows: in the case where we are interested in estimating a single
distribution from an independent and identically distributed ($i.i.d$)
sample $y_{1},\cdots,y_{n}$, observations arise from a convolution
\[
y_{j}\sim\int k\left(\cdot\mid\boldsymbol{\psi}\right)G\left(\mbox{d}\boldsymbol{\psi}\right)
\]
where $k\left(\cdot\mid\boldsymbol{\psi}\right)$ is a given parametric
kernel indexed by $\boldsymbol{\psi}$ (we use bold symbol to indicate
vector), and $G$ is a mixing distribution assigned a discrete form
\[
G\left(\boldsymbol{\psi}\right)=\sum_{l=1}^{L}\omega_{l}\delta_{\boldsymbol{\psi}_{l}},\;\mbox{where }\sum_{l=1}^{L}\omega_{l}=1\mbox{ and }\omega_{l}\ge0,\;\forall l=1,\cdots,L
\]
and $L$ can be finite or infinite. For example, assuming that $G$
follows a Dirichlet process prior leads to the well known Dirichlet
process mixture model (\citealp{escobar_bayesian_1995}). 

For our application, we adopt a specific BNP model \textemdash{} the
PSBP mixture model, which was formally developed in \citet{rodriguez_nonparametric_2011}.
This method is known for its flexibility, generality, and importantly,
computational tractability. In addition, PSBP leads to consistent
estimation of any conditional density under weak regularity conditions
as shown in \citet{pati_posterior_2013}. \citet{rodriguez_bayesian_2009}
used this technique to create a nonparametric factor model to study
genetic factors predictive of DNA damage and repair. \citet{chung_nonparametric_2009}
applied this tool to develop a nonparametric variable selection framework.
Our model is designed to capture the transport risk distribution characteristics
in all ranges, covering both recurrent and disruption risks. 

Particularly, we focus on modeling the conditional distribution of
transport risks, within the PSBP framework. Modeling the conditional
distribution allows us to investigate the relationship between transport
risks and potential predictors, including airline, route, shipping
time, cargo weight etc, based on which we can further explore ways
to improve transport reliability. We will explain this in more details
in ${\cal x}$3.1. 

To demonstrate the value of PSBP, we compare our transportation risk
estimation with that obtained from a naive linear model (see Equation
\eqref{eq: OLS} in Appendix $\mathsection$B.5.1 for details). We
show that the two methods deliver dramatically different results.
For instance, the naive linear model fails to capture the critical
roles airlines play in transport service levels, and more importantly,
underestimates disruption risks, which can result in insufficient
risk management strategies. We further compare our model with two
generalized and advanced alternative models: generalized additive
models (GAM) (see Equation \eqref{eq: GAM} in Appendix $\mathsection$B.5.3
for details) and flexible mixture models (see Equation \eqref{eq: Flexmix}
in Appendix $\mathsection$B.5.3 for details). Overall, our PSBP model
shows a strong in-sample and out-of-sample predictive power but is
relatively heavy in computation time. For the detailed model comparison,
please refer to ${\cal x}$3.5 and Appendix $\mathsection$B.5.

\subsection{Data-Driven Risk Assessment Tool}

Our method suggests a powerful and general tool to help supply chain
risk assessment, a topic that has not received the attention it deserves.
In particular, while supply chain risk management is gaining increasing
attention from both practitioners and academics, a recent McKinsey
\& Co. Global Survey of Business Executives shows that \textquotedblleft nearly
one-quarter of firms say their company doesn't have formal risk assessment.\textquotedblright{}
On the other hand, as articulated in \citet{van_mieghem_risk_2011},
managing risk through operations contains 4 steps: 1. identification
of hazards; 2. risk assessment; 3. tactical risk decisions; 4. implement
strategic risk mitigation or hedging. These four steps must be executed
and updated recurrently. Among the four steps, step 1 is more experience
and context based, which typically involves information from anecdotal
records or long experience with the specific business processes. Step
4 is more action-based, requiring detailed organizational design and
information systems to carry out the hedging strategies developed
in step 3. These two steps may not need quantitative methods. Steps
2 and 3, on the other hand, require rigorous analysis and quantification,
and therefore call for analytical research. While most of the supply
chain risk management literature focuses on the third step, which
involves developing strategies for reducing the probabilities of negative
events and/or their consequences should they occur, this paper focuses
on step 2 \textendash{} risk assessment. 

Risk assessment involves estimation of two components: (a) risk likelihood,
i.e., \textquotedblleft the probability that an adverse event or hazard
will occur\textquotedblright{} and (b) risk impact, i.e., \textquotedblleft the
consequences of the adverse event\textquotedblright{} \citep{van_mieghem_risk_2011}.
The long-term expected risk is the integration of these two parts.
\citet{kleindorfer_accident_2003} assess risk impact (part (b)) of
catastrophic chemical accidents using data collected by the Environmental
Protection Agency. \citet{kleindorfer_managing_2005} presented a
conceptual framework for risk assessment and risk mitigation for supply
chains facing disruptions. Different from these studies, our work
focuses on using statistical methods to accurately estimate the risk
likelihood (part (a)), which calls for more advanced scientific computation
and analysis tools. Correctly identifying hazards and assessing risk
has important implications for the effectiveness of alternative management
policies \citep{cohen_operations_2007}. Our study shows that a careful
risk assessment is critical to developing tailored services for customers
(i.e., shippers) of different types and selecting service suppliers. 

The transport risk studied in this paper resembles the random yield/capacity
risks in manufacturing studied by many authors; see, e.g., \citet{federgruen_optimal_2009},
\citet{wang_mitigating_2010}. Also, transportation disruption risk
is an important type or component of random supply disruption risks
considered by \citet{song_inventory_1996,tomlin_value_2006}, etc.
While most of these authors focus on risk mitigation strategies assuming
a particular risk distribution, such as a Bernoulli distribution for
disruption risks, the Bayesian PSBP mixture model introduced here
can be used to generate empirical random yield distributions and disruption
probabilities, when data are available. 

The reminder of the paper is organized as follows: in ${\cal x}$2,
we give a brief introduction of the air cargo logistics industry and
its challenges, the data we used for this study and the research questions
we ask. In ${\cal x}$3, we describe approaches for model selection,
we introduce the PSBP mixture model and the algorithm for posterior
computation, and we compare our model with other models based on goodness
of fit and predictive performance. In ${\cal x}$4 we explain the
results. In ${\cal x}$5 we propose several applications of our model
to design more efficient operational strategies. In ${\cal x}$6,
we conclude the paper and discuss future directions. Appendix A contains
data cleaning steps, and the tables and figures illustrating the data.
Appendix B contains certain algorithm details, model implementation
steps, model checking and comparison results. Appendix C contains
estimation results and selected figures. 

\section{Industry Background, Data Source, and Research Questions}

Though a crucial part of global operations, the air cargo industry
is less known to the public because it operates behind the scenes.
For this reason, in order to understand our model and analysis, it
is necessity to provide a brief background of the industry, which
also explains the initial motivation for the industry to develop a
standardized \textit{\small{}Cargo 2000} process. Our data is \textit{\small{}Cargo
2000} standardized. 

\subsection{Service Chain Structure}

First, we examine the shipping process. Typically, an air cargo transport
involves four parties: \textit{shippers} (e.g., manufacturers), \textit{freight
forwarders} (\textit{forwarders} in short), \textit{carriers }\textit{\emph{(i.e.,
airlines)}} and \textit{consignees }\textit{\emph{(e.g., downstream
manufacturers or distributors)}}; see Figure 
\begin{figure}[tbph]
\centering{}\includegraphics[scale=0.5]{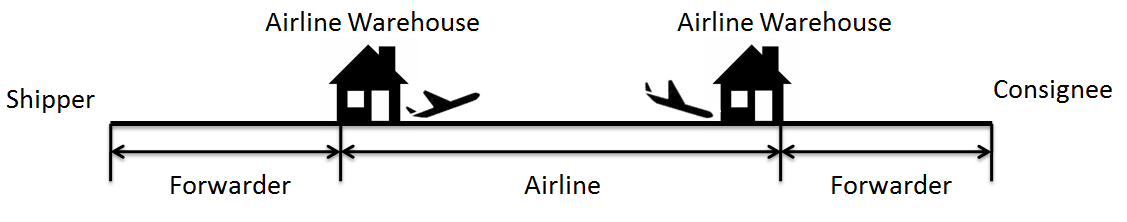} \caption{Cargo flows from the shipper to the forwarder; then from the forwarder
to the airline; then from the airline to the same forwarder. In the
end, the forwarder delivers the cargo to the consignee}
\label{fig: cargo flow} 
\end{figure}
 \ref{fig: cargo flow} for an illustration. These four parties form
a chain structure, usually called the air transport supply chain.
A shipper initiates a transaction by providing the forwarder company
with ``(1) origin/destination; (2) collection/delivery date; (3)
shipment details (cargo pieces, weight and volume); (4) shipper/consignee
information; (5) product/shipping service required''\citep{iata_c2k_2014}.
Following their route map, the forwarder picks up cargoes from the
shipper at the required time, consolidating cargoes sharing the same
route if possible, and then sends cargoes to the selected airline
at an origin airport. The airline takes charge of cargoes until arriving
at the destination airport. An airline might use a direct flight or
2 \textendash{} 3 connecting flights based on the route map. The forwarder
accepts cargoes at the destination, and delivers them to consignees. 

To simplify terms, we refer to both the shipper and the consignee
as the ``customers''. Customers use forwarders in 90\% of air cargo
shipments. A forwarder is a service provider for its customers, while
it in turn uses airlines as service providers. Upon receiving a shipping
request, a forwarder sends a booking request to several airlines,
choosing the most economic one that satisfies the agreed upon timetable.
Large forwarders typically reserve a certain percentage (e.g., 30\%)
of the total space on most airlines, including passenger and cargo
airlines. 

\subsection{\textit{Cargo 2000} (C2K) Standards}

To compete against integrators (service providers who arrange door-to-door
transportation by combining mode(s) of transportation, such as DHL,
UPS etc.), \textit{Cargo 2000} (C2K) was founded by a group of leading
airlines and freight forwarder companies, ``IATA Interest Group'',
in 1997. This initiative was designed to enable industry-wide participants
to ``provide reliable and timely delivery shipments through the entire
air transport supply chain'' (C2K Master Operating Plan (MOP) Executive
Summary\citep{iata_c2k_2014}). Specifically, they developed a system
of shipment planning and performance monitoring for air cargo which
allows proactive and realtime event processing and control. Currently
C2K is composed of more than 80 major airlines, forwarders, ground-handling
agents, etc (see Figure \ref{fig: c2k members} in Appendix for the
current members of C2K). C2K Quality Management System is implemented
with two different scopes: Airport-to-Airport (A2A) and Door-to-Door
(D2D). In this paper, we focus on the A2A level shipments due to data
constraints. 

The following describes how C2K is used to create a shipping plan,
and how airlines and forwarders monitor, control, intervene and repair
each shipment in real-time.

\subsubsection{Plan }

After a carrier has confirmed requested capacity on planned flights,
it creates an A2A route map (RMP) and shares it with the forwarder.
A RMP describes the path the shipment follows, including flight information,
milestones and the latest-by time for the fulfillment of milestones
along the transport chain. See Table A.1 and Figure \ref{fig: milestone explanation}
in Appendix ${\cal x}$A.2 for an illustration. If a customer agrees
on the plan, the RMP is set alive. Otherwise, modifications will be
made until agreement is achieved. Essentially, each route map is a
combination of a station profile and milestones. Station profiles,
which contain information on the duration for completion of each process
step, are kept by forwarders and carriers. The milestones are defined
by the C2K MOP. 

\subsubsection{Monitor, Control, Intervene and Repair}

After a route map is issued, the shipping process is monitored against
this map. The completion of every milestone triggers updates on both
the airline's and forwarder's IT systems. Any deviation from the plan
triggers an alarm, which allows for corrections to be taken by the
responsible party in order to bring the shipment back on schedule.
If necessary, a new RMP is made for the remaining transport steps.
Meanwhile, an exception record is entered into the system recording
the necessary information such as time, location, and reasons. See
Table \ref{table: except} in Appendix ${\cal x}$A.2 for an illustration. 

\subsubsection{Report}

At the end of the shipment process, a report, including whether or
not the delivery promise was kept and which party was accountable
for the failure, is generated. This allows the customers to directly
compare the performance of their C2K enabled forwarders, carriers
and logistics providers. 

\subsection{Forwarder\textquoteright s Frustration and Our Objectives}

Even with current systems, the service level remains unsatisfying.
As a result, forwarders risk loosing customers even though forwarders
have no direct control of A2A, which is the most uncertain part of
shipping. Questions for the \textit{forwarder} to solve include: (1)
how to predict transport risks so as to prepare for risks and inform
customers in advance and (2) how to improve transport reliability
in each route by selecting the best supplier? We aim to help address
these questions. Suppose a customer comes to the forwarder with a
fixed route (origin-destination), time of shipping, weight and volume
of cargo. We aim to provide the forwarder with a distribution of transport
risk conditional on demand variables (route, month, cargo weight/volume)
and decision variables (airline, number of flight legs, planned duration,
initial deviation time) with $95\%$ uncertainty interval. See Table
\begin{table}[tbph]
\centering{}{\small{}\caption{Potential predictors}
\label{table: predictors}}%
\begin{tabular}{l|>{\raggedright}p{11cm}}
\hline 
\textbf{\textit{\small{}demand variables}} & \tabularnewline
\hline 
{\small{}route ($r$)} & {\small{}an origin-destination airport pair combination (captures
all the fixed effects on a particular route).}\tabularnewline
{\small{}month ($m$)} & {\small{}month when the shipping is finished}\tabularnewline
{\small{}cargo weight ($wgt$)} & {\small{}total weight of the cargo (kilograms)}\tabularnewline
{\small{}cargo number-of-pieces ($pcs$)} & {\small{}total number of pieces of the cargo (unit load)}\tabularnewline
\hline 
\textbf{\textit{\small{}decision variables}} & \tabularnewline
\hline 
{\small{}airline ($a$)} & {\small{}the airline transported the cargo}\tabularnewline
{\small{}number of legs ($leg$)} & {\small{}number of connecting flights taken to arrival at destination}\tabularnewline
{\small{}planned duration ($dur$)} & {\small{}total time (days) planned to take to finish the transport}\tabularnewline
{\small{}initial deviation ($dev_{start}$)} & {\small{}deviation (days) between actual and planned check-in time
at airline origin warehouse }\tabularnewline
\hline 
\end{tabular}
\end{table}
 \ref{table: predictors} for descriptions of these variables. Specifically,
demand variables are determined by the shipper's demand requirement
which can not be changed. On the other hand, the decision variable
can be chosen by the shipper at the time of purchasing shipping services.
Based on this information, an optimal route can be chosen to match
the customer's cost/utility function, providing different options
to different customers. Please refer to $\mathsection$5.1 for application
illustration. Next, we elaborate how the above mentioned demand and
decision variables affect the transport risk. 

\subsubsection{Effect of Demand Variables}

\textit{1. Route:} service level differs dramatically across routes
depending on (a) supply-demand of air transport service and (b) congestion
level and infrastructure at visited airports. We use a route-level
effect to absorb all these factors.

\textit{2. Month:} demand (e.g., holiday shipping) and weather (e.g.,
winter snow) both have a seasonal trend, which results in different
perceived air cargo transport service levels in different months.
We used the month, in which each shipment completes, as the predictor.
Since shipments only take 1.7 days to complete on average, essentially
identical results would be achieved using the month of transport start.

\textit{3. Cargo weight and volume:} each flight has a maximum weight
and volume (cargo volume is approximated by cargo pieces in this paper).
Larger cargoes may be more likely to fail to be loaded onto the scheduled
flight due to (1) airlines overselling capacities and (2) changes
of currently available capacity, such as more luggage from passengers.
However, larger cargoes are usually more valuable, thus may have higher
transport priority. 

\subsubsection{Effect of Decision Variables }

\textit{1. Airline:} transportation delays are expected to vary substantially
across airlines due to a wide variety of factors, and hence we added
(1) the interaction of airline and route and (2) the interaction of
airline and number of legs into the model. 

\textit{2. Number of legs:} number of legs increases the probability
for a cargo to miss connecting flights, so it is a strong predictor
of transport risk.

\textit{3. Planned duration:} even conditional on route, airline and
number of legs, planned duration differs greatly. This reflects cushions
added to the shipping time. 

\textit{4. Initial deviation:} if the cargo is sent to the airline
earlier than scheduled, it can be loaded onto an earlier flight and
otherwise the cargo might miss it's planned flight. Before the trip
starts, the forwarder can use 0 as the default value to make transport
risk prediction. As soon as the forwarder has sent the cargos to the
airline, a new prediction can be made with the new time information. 

\subsubsection{Other Potential Predictors}

There are other factors, such as price and weather, that may also
affect the risk distribution, but are not available in our data. Our
model indirectly captures these effects through allowing the distribution
of risk to vary flexibly with the demand and decision variables mentioned
above. Different definitions of demand/decision variables can be adopted,
such as to replace our ``route'' with ``path'' (with connecting
airports information). This can potentially improve the predictive
accuracy. However since our data is sparse and our major focus is
to present fundamental modeling details, we choose to retain our current
settings while these specifics can be easily modified in our model. 

\subsection{Data and Summary Statistics}

Our data contain a leading freight forwarder company's C2K standard
air freight shipments from October 2012 to April 2013. The data contain
real-time milestone updates, similar to the data shown in Table A.1
in Appendix $\mathsection$A.2, and route maps for each shipment.
The last route map before the shipment is used to measure risk. After
cleaning (see Appendix ${\cal x}$A.1 for cleaning steps), the data
include 86,149 shipments on 1336 routes operated by 20 airlines. Freights
are shipped from 58 countries to 95 countries. In Appendix ${\cal x}$A.3
are summary statistics. In sum, we observe that: (i) European airlines,
such as Lufthansa and KLM, play a significant role in the data; (ii)
More than 50\% of shipments are transported on routes served by more
than 1 airline. For example, around 30\% of shipments are on routes
served both by direct flights and 2-leg service; (iii) There are more
than 50\% of shipments transported on routes where services of different
legs are available. This confirms the need for a careful assessment
of the impact of different choices, which can lead to a higher utility
if service levels vary significantly. 

\section{Model }

In this section, we explain the model in details. ${\cal x}$3.1 provides
motivation for estimating the conditional distribution of transport
risk and advantages of using the PSBP mixture model. The model can
be decomposed into two parts: mixture weight and mixture kernel, detailed
in ${\cal x}$3.2 and ${\cal x}$3.3, respectively. The Bayesian posterior
sampling algorithm to estimate unknown parameters in weights and kernels
is presented in ${\cal x}$3.4. We also discuss model selection and
comparison in ${\cal x}3.5$. Detailed supplementary materials can
be found in Appendix ${\cal x}$B.

\subsection{Conditional Risk Distribution}

\begin{figure}[tbph]
\begin{centering}
\begin{minipage}[t]{0.49\columnwidth}%
\includegraphics[width=1\columnwidth]{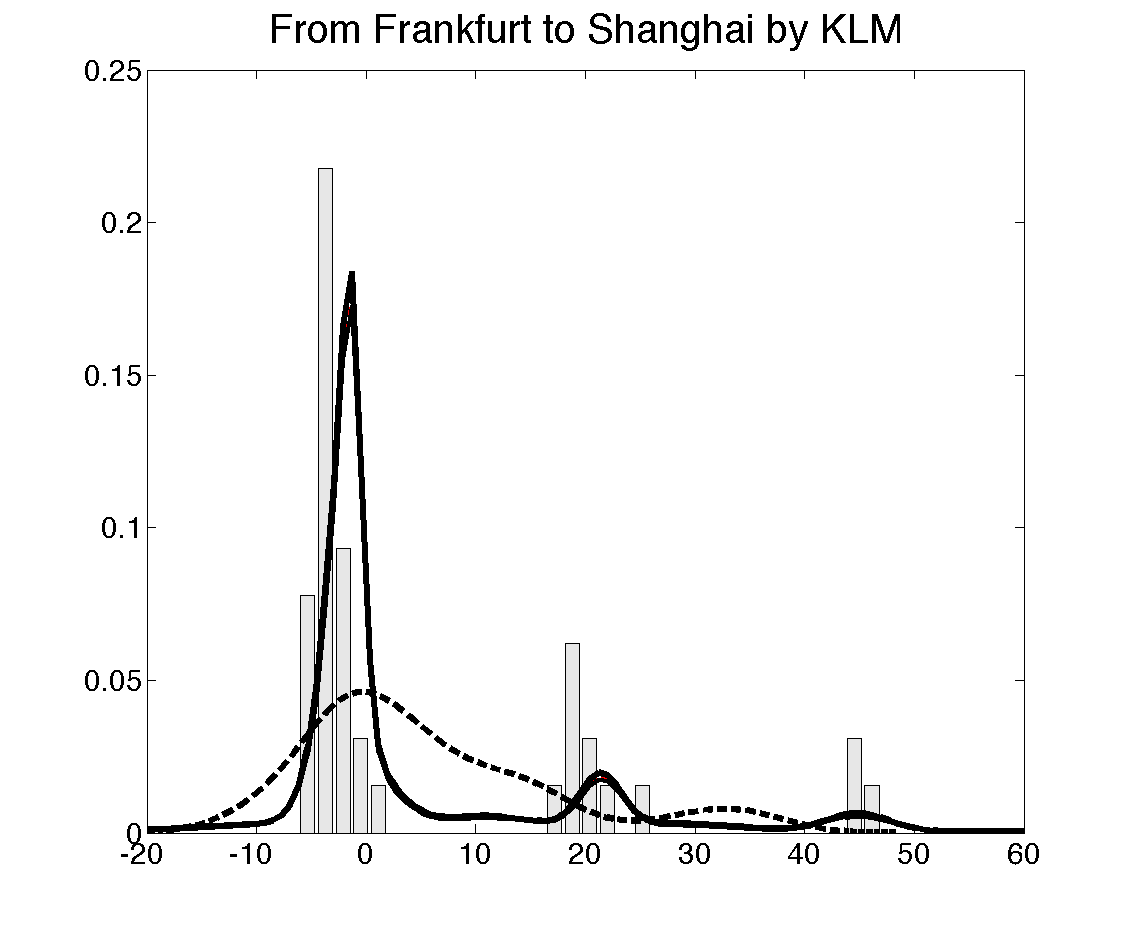}%
\end{minipage}\hfill{}%
\begin{minipage}[t]{0.49\columnwidth}%
\includegraphics[width=1\columnwidth]{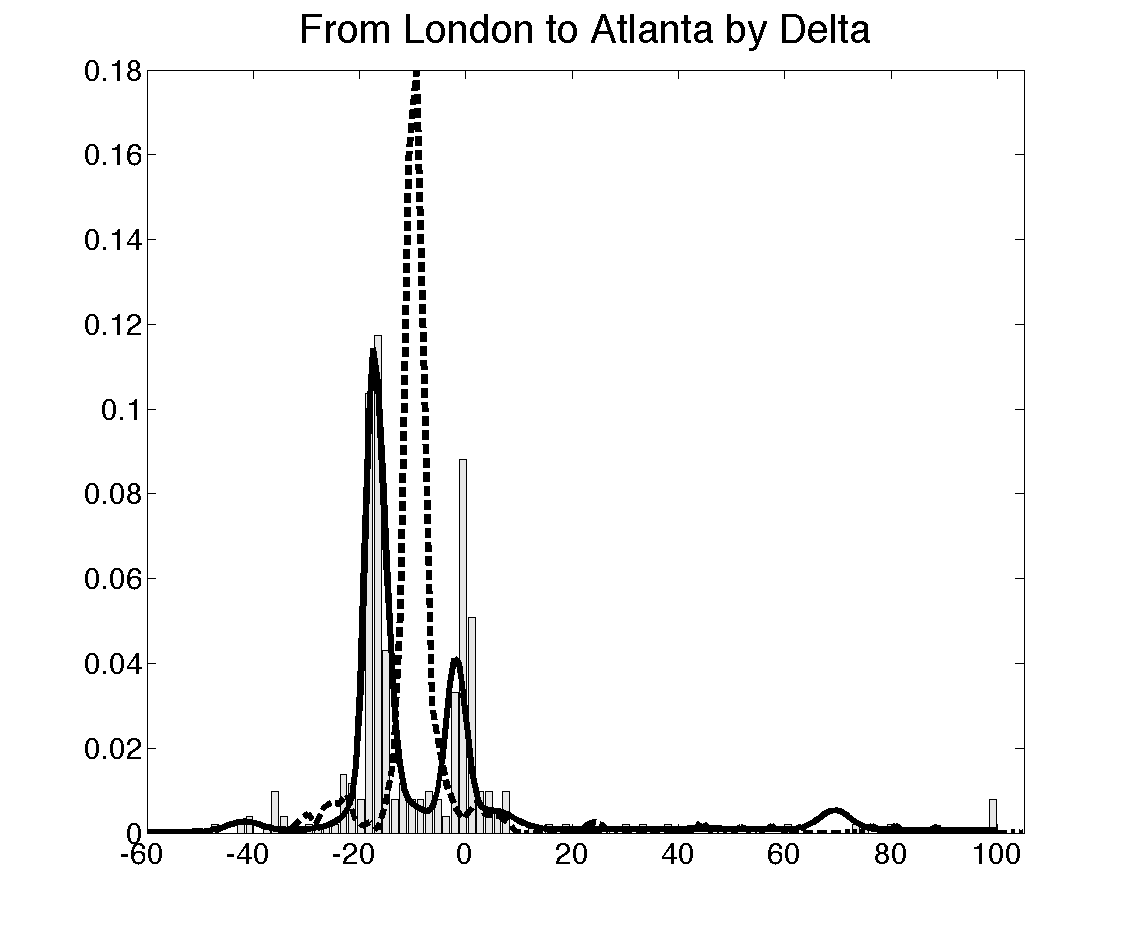}%
\end{minipage}
\par\end{centering}
\caption{True data (histogram), PSBP predictive (solid curve) and naive linear
model predictive (dashed curve) mean conditional response density
$\hat{f}(y\mid{\bf x})$. Left: route = Frankfurt to Shanghai, airline
= KLM; Right: route = London to Atlanta, airline = Delta. For the
exact method to calculate predictive conditional response density,
please refer to ${\cal x}$5.2. }
\label{fig: sample routes}
\end{figure}
The multimodal feature is not only present at the aggregate data level,
see Figure \ref{fig: arrival deviation}, but also at the granular
level, such as each route or route-airline level. The histograms in
Figure \ref{fig: sample routes} show the empirical distributions
on two sample routes served by two airlines. In order to make accurate
predictions and inferences based on such data, the first step is to
choose a model flexible enough to fit the data well. Usual choices
of models for multimodal data rely on mixtures, e.g., mixtures of
Normal kernels, which are known to provide an accurate approximation
to any unknown density.

We cannot rely on simple mixture models, as we are investigating the
distribution of transport risks conditional on demand and decision
variables, including both categorical and continuous predictors. This
leads to a problem of\textbf{\textit{ conditional distribution estimation}}.
One stream of literature on flexible conditional distribution estimation
uses frequentist methods. \citet{fan_estimation_1996} proposed a
double-kernel local linear approach, and related frequentist methods
have been considered by \citet{hall_methods_1999} and \citet{hyndman_nonparametric_2002}
among others. The other popular choice is a BNP mixture model. \citet{muller_bayesian_1996}
proposed a Bayesian approach to nonlinear regression, in which the
authors modeled the joint distribution of dependent variable and independent
variables using a Dirichlet process mixture of Normals (\citealt{lo_class_1984,escobar_bayesian_1995}).
This type of approach induces a model for the conditional distribution
of the response through a joint model for the response and predictors.
Although such joint models are provably flexible, in practice they
can have clear disadvantages relative to models that directly target
the conditional response distribution without needing to model the
high-dimensional nuisance parameter corresponding to the joint density
of the predictors. Such disadvantages include treating the independent
variables as random, while they are often designed variables (e.g.,
it seems unnatural to consider route or airline as random), and relatively
poor practical performance in estimating the conditional distribution.

We instead focus on direct modeling of the unknown conditional distribution
of transport risk $y$ given predictors ${\bf x}=\left(x_{1},\cdots,x_{p}\right)^{'}\in{\cal X}$
(${\cal X}$ is the sample space for the predictors ${\bf x}$) without
specifying a model for the marginal of ${\bf x}$. In our context,
predictors ${\bf x}$ = \{airline $\left(a\right)$, route $\left(r\right)$,
month $\left(m\right)$, number of legs $\left(leg\right)$, initial
deviation $\left(dev_{start}\right)$, planned duration $\left(dur\right)$,
cargo weight $\left(wgt\right)$, cargo number of pieces $\left(pcs\right)$\}
(as specified in Table \ref{table: predictors}). In particular, we
assume the transport risk $y$ arises from a convolution

\begin{equation}
y\mid{\bf x}\sim\int k\left(y\mid\boldsymbol{\psi}\right)G_{{\bf x}}\left(\mbox{d}\boldsymbol{\psi}\right)\label{eq: BNP Mixture}
\end{equation}
where $k\left(\cdot\mid\boldsymbol{\psi}\right)$ is a given parametric
kernel indexed by parameters $\boldsymbol{\psi}$ (e.g., Normal kernel
$k\left(\cdot\mid\boldsymbol{\psi}\right)$ is indexed by $\boldsymbol{\psi}=$(mean,
standard deviation)), and the mixing distribution $G_{{\bf x}}$ is
allowed to vary flexibly with predictors ${\bf x}\in{\cal X}$. The
typical form in the BNP literature (refer to \citet{rodriguez_nonparametric_2011}
for references) lets 
\begin{equation}
G_{{\bf x}}=\sum_{l=1}^{L}\omega_{l}\left({\bf x}\right)\delta_{\boldsymbol{\psi}_{l}\left({\bf x}\right)},\mbox{ where }\sum_{l=1}^{L}\omega_{l}\left({\bf x}\right)=1\mbox{ and }\omega_{l}\left(x\right)\ge0\label{eq: BNP}
\end{equation}
where the atoms $\left\{ \boldsymbol{\psi}_{l}\left({\bf x}\right):{\bf x}\in{\cal X}\right\} _{l=1}^{L}$
are $i.i.d$ sample paths from a stochastic process over ${\cal X}$,
and $\left\{ \omega_{l}\left({\bf x}\right),{\bf x}\in{\cal X}\right\} $
are predictor-dependent probability weights that sum to one for all
$x$. The above form is too general to be useful and it is necessary
to make some simplifications for practical implementation. One common
possibility is to introduce predictor dependence only in the $G_{{\bf x}}$
atoms, $\psi_{l}\left(x\right)$, while keeping weights, $\omega_{l}\left({\bf x}\right)=\omega_{l}$,
fixed. However, this approach tends to have relatively poor performance
in our experience, including the air cargo transport risk data, we
have also shown this in model comparison with Flexmix model (see Appendix
$\mathsection$B.5.3 for details). 

In our case, the peak locations of the dependent variable, transport
risk, are almost constant (i.e., daily peaks for international shipments,
and some additional few-hourly peaks for domestic shipments besides
the daily peaks). However, the heights of the peaks change greatly
along with ${\bf x}$ (e.g., route, airline, demand variables). The
height of each peak represents (roughly) the probability for the observation
to fall into the kernel centered around that peak. For example, if
conditional on certain ${\bf x}_{1}$, the peak around 24 hours is
relatively high, then a shipment, conditional on ${\bf x}_{1}$, has
a large probability of being delayed for 24 hours. On the other hand,
if conditional on certain ${\bf x}_{2},$ there is only one significant
peak around 0 hours, then a shipment, conditional on ${\bf x}_{2}$,
probably arrives close to the planned arrival time. So, in our context,
to find out how the height of each peak depends on ${\bf x}$ is of
central interest. 

Inducing dependence structure in the weights can be difficult and
lead to complex and inefficient computational algorithms, limiting
the applicability of the models. To overcome these difficulties, we
adopt the PSBP mixture model, which has the advantages of computational
tractability and consistency under weak regularity conditions. 

\subsection{Bayesian Probit Stick-breaking Process}

Recalling the general form of the mixing measure in Equation (\ref{eq: BNP}),
\textbf{\textit{stick-breaking}} weights are defined as $\omega_{l}=u_{l}\prod_{p<l}\left(1-u_{p}\right)$,
where the stick-breaking ratios are independently distributed $u_{l}\sim H_{l}$
for $l<L$ and $u_{L}=1$ for the case of finite $L$. In the baseline
case in which there is no predictor, \textbf{\textit{Probit stick-breaking
}}weights are constructed as
\[
u_{l}=\Phi\left(\gamma_{l}\right),\;\gamma_{l}\sim\mathsf{N}\left(\mu,\phi\right)
\]
where $\Phi\left(\cdot\right)$ denotes the cumulative distribution
function (cdf) for the standard normal distribution. $\mu$ is the
mean and $\phi$ is the precision (the reciprocal of the variance)
of a normal distribution such that for $x\sim\mathsf{N}\left(\mu,\phi\right)$,
the probability density function (pdf) is $f(x)=\sqrt{\frac{\phi}{2\pi}}\exp\left\{ -\frac{\phi}{2}\left(x-\mu\right)^{2}\right\} $.
For a finite $L$, the construction of the weights ensures that $\sum_{l=1}^{L}\omega_{l}=1$.
When $L=\infty$, $\sum_{l=1}^{\infty}\omega_{l}=1$ almost surely
(\citealt{rodriguez_nonparametric_2011}). 

The use of Probit transformation to define the weights builds a mapping
between a real number $\gamma_{l}$ from $-\infty$ to $+\infty$
into $u_{l}\in(0,1)$. Thus, the transformation allows researchers
to restate the model using normally distributed latent variables $\gamma_{l}$,
facilitating computation via data augmentation Gibbs sampling algorithms
presented in $\mathsection$3.4. This transformation also makes model
extensions to include additional structure (e.g,. predictors) straightforward.
Additionally, the Probit transformation simplifies prior elicitation
as presented at the end of ${\cal x}3.2$. 

In order to make $\omega_{l}\left({\bf x}\right)$ predictor-dependent,
we further express the latent variables $\gamma_{l}$ as a linear
function of ${\bf x}$, $\left\{ \gamma_{l}\left({\bf x}\right),{\bf x}\in{\cal X}\right\} $
(In this paper we use superscript as an index rather than the exponent
of the parameter):
\begin{eqnarray}
\omega_{l}\left({\bf x}\right) & = & \Phi\left(\gamma_{l}\left({\bf x}\right)\right)\prod_{p<l}\left(1-\Phi\left(\gamma_{p}\left({\bf x}\right)\right)\right)\label{eq: stick-breaking}\\
\gamma_{l}({\bf x}) & = & \theta_{l}^{1}+\theta_{a}^{2}+\theta_{r}^{3}+\theta_{\left(a,r\right)}^{4}+\theta_{m}^{5}+\theta_{leg}^{6}+\theta_{\left(a,leg\right)}^{7}+f_{1}\left(dev_{start}\mid\boldsymbol{\theta}^{8}\right)\nonumber \\
 &  & +f_{2}\left(dur\mid\boldsymbol{\theta}^{9}\right)+f_{3}\left(\mbox{log}\left(wgt\right)\mid\boldsymbol{\theta}^{10}\right)+f_{4}\left(\mbox{log}\left(pcs\right)\mid\boldsymbol{\theta}^{11}\right)\label{eq: full model}
\end{eqnarray}
where $\{\theta_{l}^{1}\}$ controls the baseline probability of latent
class $l$ ($l=1,\cdots,L$), $\{\theta_{a}^{2}\}$ controls the baseline
heterogeneity of airline $a$ ($a=1,\cdots,20$), $\{\theta_{r}^{3}\}$
controls the heterogeneity of route $r$ ($r=1,\cdots1336$), $\left\{ \theta_{\left(a,r\right)}^{4}\right\} $
represents the dependence of weights on possible interactions between
airlines and routes, and the meanings of $\left\{ \theta_{m}^{5}\right\} $,
$\left\{ \theta_{leg}^{6}\right\} $,$\left\{ \theta_{\left(a,leg\right)}^{7}\right\} $
are similar. In addition, $f_{1}$, $f_{2}$, $f_{3}$ and $f_{4}$
are spline functions expressed as a linear combination of B-splines
of degree 4 (see Appendix $\mathsection$B.4 for details of the smooth
splines used), where the knots of $dev_{start}$ are {[}-3, -2, -1,
0, 1, 2, 3{]}, the knots of $dur$ are {[}1, 2, 4, 6, 8, 10{]}, the
knots of $log(weight)$ are {[}2, 4, 6, 8{]} and the knots of $log(pcs)$
are {[}1, 3, 5{]}. Here we use the logarithm form of cargo weight
$\left(wgt\right)$ and number of pieces $\left(pcs\right)$ as the
predictors, since the original distributions are highly skewed. To
ensure identification of the parameters, we let $\theta_{1}^{2}=\theta_{1}^{3}=\theta_{\left(1,r\right)}^{4}=\theta_{\left(a,1\right)}^{4}=\theta_{1}^{5}=\theta_{(1,leg)}^{6}=\theta_{(a,1)}^{7}=0$
for all $a$, $r$ and interactions in sample space ${\cal X}$. 

\subsection{Posterior Computation}

In Bayesian statistics, the posterior distribution is typically not
available analytically, involving an intractable normalizing constant.
For this reason, posterior calculations usually rely on either large
sample approximations, which may have questionable accuracy in our
transportation risk applications, or Markov chain Monte Carlo (MCMC)
sampling. The basic idea in MCMC sampling is to construct a Markov
chain having stationary distribution corresponding to the joint posterior
distribution of the model parameters, with this done in a manner that
avoids ever having to calculate the intractable constant. In order
for the Markov chain to have the appropriate behavior, the Markov
transition kernel needs to be carefully chosen, with usual choices
corresponding to either Metropolis-Hastings (MH) or Gibbs sampling.
MH can involve a lot of tuning in models with many parameters, while
Gibbs avoids tuning by sampling sequentially from the conditional
posterior distributions of subsets of parameters given current values
of the other parameters. Gibbs sampling relies on a property known
as conditional conjugacy. Focusing on a subset of the model parameters
and conditioning on the other parameters, the prior probability distribution
is conditionally conjugate if the conditional posterior distribution
takes the same form as the prior. The specific choices of our model
form and prior distributions (to be described below) are motivated
by retaining conditional conjugacy.

In order to obtain conditional conjugacy for blocks of parameters,
we follow a common strategy known as data augmentation. The basic
idea in data augmentation is that one may obtain conditional conjugacy,
and hence a simple Gibbs sampling algorithm, by introducing latent
variables in a careful manner. The MCMC algorithm is then run for
both the latent variables and the model parameters; although this
increases the number of unknowns to sample, it can lead to greater
efficiency by allowing model parameters to be sampled in blocks directly
from full conditional posterior distributions. Similar augmentation
strategies are routinely used in frequentist statistics; e.g., to
fix mixture models with the EM algorithm. Here, we follow the augmentation
strategy of \citep{rodriguez_bayesian_2009}. First we focus on case
when $L<\infty$. For each observation $y_{j}\mid{\bf x}$, (corresponding
to replicate $j$ conditional on ${\bf x}$, $j=1,\cdots,n\left({\bf x}\right)$
if there are $n({\bf x})$ replicates, otherwise $j$ is dropped if
there are no replicates, i.e., $n({\bf x})=1$), we introduce a latent
indicator variable $s_{j}\left({\bf x}\right)$ such that $s_{j}\left({\bf x}\right)=l$
if and only if observation $y_{j}\mid{\bf x}$ is sampled from mixture
component $l$ ($l=1,2,\cdots,L$). The use of these latent variables
is standard in mixture models. With the help of latent indicators
$s_{j}({\bf x})$, Gibbs sampling of more than 2000 model parameters
can be classified into four categories, as presented in Appendix $\mathsection$B.1.1
$\sim$ $\mathsection$B.1.5.

\subsection{Distributional Choices}

To complete a specification of our model, we require a specific choice
for the kernel in the kernel mixture, as well as prior probability
distributions for each of the model parameters. These choices are
described below.

\subsubsection{Normal Kernel}

A mixture of a moderate number of Normals is known to produce an accurate
approximation of any smooth density. Also motivated by computational
tractability of the Normal distribution (i.e., conditional conjugacy),
we specify the parametric kernel, $k\left(\cdot\mid\boldsymbol{\psi}\right)$,
of the PSBP mixture model as a Normal distribution, $\mathsf{N}\left(\mu,\phi\right)$,
where $\boldsymbol{\psi}=\left(\mu,\phi\right)$. Recalling that our
mixture model takes the form in Equation (\ref{eq: BNP Mixture}),
we replace the kernel in the above equation with Normal and use the
PSBP specified prior $G_{{\bf x}}$. Then the conditional distribution
of $y$ can be expressed in the simple form 
\begin{equation}
y\mid{\bf x}=\sum_{l=1}^{L}\omega_{l}\left({\bf x}\right)\mathsf{N}\left(y\mid\mu_{l},\phi_{l}\right)\label{eq: likelihood}
\end{equation}
The prior of atoms $\left\{ \left(\mu_{l},\phi_{l}\right),l=1,2,\cdots,L\right\} $
is $\mathsf{N}\mathsf{G}(\zeta_{\mu},\xi_{\mu},a_{\phi},b_{\phi})$,
a conditionally-conjugate Normal-Gamma prior such that
\begin{eqnarray*}
\mu_{l}\sim\mathsf{N}\left(\zeta_{\mu},\xi_{\mu}\phi_{l}\right), &  & \phi_{l}\sim\mathsf{G}\left(a_{\phi},b_{\phi}\right).
\end{eqnarray*}
where $l=1,2,\cdots,L$. The specification of prior $\zeta_{\mu},\xi_{\mu},a_{\phi}$
and $b_{\phi}$ is discussed in Appendix ${\cal x}$B.2. 

\subsubsection{Prior for Parameters in Weight }

We choose Normal priors for parameters $\Theta=$\{$\left\{ \theta_{l}^{1}\right\} $,
$\left\{ \theta_{a}^{2}\right\} $, $\left\{ \theta_{r}^{3}\right\} $,
$\left\{ \theta_{\left(a,r\right)}^{4}\right\} $, $\left\{ \theta_{m}^{5}\right\} $,
$\left\{ \theta_{leg}^{6}\right\} $, $\left\{ \theta_{\left(a,leg\right)}^{7}\right\} $,
$\boldsymbol{\theta}^{8}$, $\boldsymbol{\theta}^{9}$, $\boldsymbol{\theta}^{10}$,
$\boldsymbol{\theta}^{11}$\} 
\[
\theta_{j}^{i}\sim\mathsf{N}\left(\nu^{i},\epsilon^{i}\right),\mbox{ for }i=8,\cdots,11\mbox{ and }j=1,\cdots,n(i)
\]
where $n(i)$ is the number of B-spline basis used for predictor $i$.
For the coefficients of 7 categorical independent variables $\boldsymbol{\theta}^{1},\cdots,\boldsymbol{\theta}^{7}$
(i.e., $\boldsymbol{\theta}^{1}=\left\{ \theta_{l}^{1}\right\} $
etc), we build a hierarchy, which enables information borrowing among
parameters in one category
\begin{eqnarray*}
\theta_{l}^{1}\sim\mathsf{N}\left(\Phi^{-1}\left(\frac{1}{L-l+1}\right),\epsilon^{1}\right), & \theta_{a}^{2}\sim\mathsf{N}\left(0,\epsilon^{2}\right),\cdots & \theta_{\left(a,leg\right)}^{7}\sim\mathsf{N}\left(0,\epsilon^{7}\right).
\end{eqnarray*}
where $\epsilon^{i}\sim\mathsf{G}\left(c_{i},d_{i}\right)$ for $i=1,2,\cdots,7$.
Here $\mathsf{G}(a,b)$ is a Gamma distribution such that for $x\sim\mathsf{G}(a,b)$
the pdf is $f(x)=\frac{b^{a}}{\Gamma\left(a\right)}x^{a-1}e^{-bx}$.
We use the specially designed prior of $\theta_{l}^{1}$ to enforce
the same prior baseline probability of each cluster $l=1,2,\cdots,L$.
The specification of $\left\{ \left(c_{i},d_{i}\right),\mbox{ for }i=1,2,\cdots,7\right\} $
and $\left\{ \left(\nu^{i},\epsilon^{i}\right),i=8,9,\cdots,12\right\} $
is discussed in Appendix ${\cal x}$B.2. 

\subsection{Model Fitting Assessment}

In order to select a specific set of predictors to include in our
PSBP mixture model, we rely on comparing different possibilities using
cross validation. In particular, we select a model having the best
out-of-sample predictive performance. Details are provided in Appendix
${\cal x}$B.4, and the selected model is shown in Equation (\ref{eq: used model}).
This model is used for later model comparison and application illustration.
\begin{eqnarray}
\gamma_{l}({\bf x}) & = & \theta_{l}^{1}+\theta_{a}^{2}+\theta_{r}^{3}+\theta_{\left(a,r\right)}^{4}+\theta_{m}^{5}+\theta_{leg}^{6}+f_{1}\left(dev_{start}\mid\boldsymbol{\theta}^{8}\right)+f_{2}\left(dur\mid\boldsymbol{\theta}^{9}\right)+f_{3}\left(\mbox{log}\left(wgt\right)\mid\boldsymbol{\theta}^{10}\right)\label{eq: used model}
\end{eqnarray}
It is important to assess how well this model fits the data, while
verifying that it does not overfit. In general, Bayesian methods are
protected against overfitting due to the implicit penalty on model
complexity that appears in the posterior distribution but not in maximum
likelihood estimation. In mixture models, overfitting can occur by
using too many mixture components. However, the PSBP mixture model
and related Bayesian nonparametric models automatically favor placing
all but a negligible amount of the probability weight on a few components.
This is consistent with our observation of superior out-of-sample
predictive performance relative to flexible frequentist regression
models.

We follow a common Bayesian strategy of goodness of fit checking by
using posterior predictive plots. In particular, the basic idea is
to generate new data from the posterior predictive distribution under
our PSBP mixture model and see how the observed data relate to these
model generated data. If the model fits poorly, a systematic deviation
will show up. We observe an excellent goodness of fit based on these
assessments. On the contrary, a naive linear model, shown in Equation
\eqref{eq: OLS} in Appendix $\mathsection$B.5, shows extremely poor
fit. We additionally compare our model with frequentist generalized
additive models (GAMs) and flexible mixture models using both in-sample
and out-of-sample prediction residuals. Overall, our model presents
a superior performance. For interested readers, please refer to Appendix
${\cal x}$B.5 for more details.

\section{Results}

Table \ref{table: posterior parameters} in Appendix ${\cal x}$C.1
shows the posterior mean and 95\% probability interval of (selected)
model parameters. There are several things to note from the table:
\begin{enumerate}
\item The 50 kernel means, $\mu_{1},\mu_{2},\cdots,\mu_{50}$, range from
-70.0 to 77.5 (hours), indicating the model predicted deviation concentrates
within -3 to 3 days, consistent with the data. The 50 kernel standard
deviations, $1/\sqrt{\phi_{1}},1/\sqrt{\phi_{2}},\cdots,1/\sqrt{\phi_{50}}$,
range from 0.62 to 84.4, meaning the Normal kernels can be very narrow
or flat, allowing for flexible estimation. 
\item Level parameters, $\theta_{1}^{1},\theta_{2}^{1},\cdots,\theta_{49}^{2}$,
vary from -10.9 to 6.74, and the wide range suggests strong variation
in risk. For example, if an airline-route pair has $\gamma_{l}\left({\bf x}\right)-\theta_{l}^{1}$
close to zero, then for certain $l$ with $\theta_{l}^{1}$ smaller
than -5, the weight $\propto$ $\Phi\left(\gamma_{l}\left({\bf x}\right)\right)$
$\approx$ $\Phi\left(-5\right)$ $\approx$ 0, thus eliminating the
inclusion of this component. By similar arguments, $\theta_{l}^{1}$
can also help determine for which $\gamma_{l}\left({\bf x}\right)-\theta_{l}^{1}$
component $l$ plays major role. 
\item The posterior distributions of all the coefficients are substantially
more concentrated than their prior distributions, suggesting that
the data provide substantial information to update the priors; in
addition, the 95\% probability intervals are narrow. 
\item The posterior estimation of airline coefficient (we disguise the names
of airlines for confidentiality reasons. The airline index used here
is randomly assigned), $\theta_{a}^{2}$, shows great heterogeneity,
and the large standard deviation, $1/\sqrt{\epsilon^{2}}$, which
measures the variations among airlines, confirms this from one other
aspect. Closer inspection reveals that except A1, whose coefficient
is fixed at zero for identification, 18 of the remaining 19 airlines'
95\% probability intervals don't include 0. Furthermore, many of them
are far from zero, implying large impact on transport risk. However,
based on the linear model (see Equation \eqref{eq: OLS} in Appendix
$\mathsection$B.5), only 2 of the 19 airlines are significantly different
from 0 at 5\% confidence level. This huge difference underlies the
principle of the two estimation methods. The naive linear model focuses
in estimating the effects of independent variables on distribution
\textbf{\textit{mean}}, and its results indicate airlines don't necessarily
affect the mean of transport risk much. However, PSBP's results show
that airlines are playing an important role on selecting and weighting
possible kernels, which affects the tail shape, number of peaks, probability
of extreme observation etc. These results and comparison once again
show that the linear model, which cannot detect the airlines' (and
some other predictors' including routes' etc) impact on transport
risk in this case, would lose considerable valuable information. 
\item Since the number of routes and their interactions with airline are
large, 1336 and 587 respectively, we don't include their posterior
summaries in Table \ref{table: posterior parameters}. However, posterior
summaries of hyper-parameters standard deviation, $1/\sqrt{\epsilon^{3}}$,
illustrate the large heterogeneity between routes. More importantly,
the large standard deviation, $1/\sqrt{\epsilon^{4}}$, represents
possibly huge differences in terms of the distribution of transport
risks on the same route while by different airlines. This suggests
that a careful selection of carriers can result in dramatically different
shipping experiences.
\end{enumerate}

\section{Applications}

Estimates of predictive conditional probability density functions
(Cpdfs) are key to generating data-driven operation strategies. In
this section, we provide several examples of how posterior Cpdfs can
aid decision making. We note that there are other applications of
our transport risk models. 

\subsection{Service Comparison for One Shipment}

The most straightforward use of PSBP posterior estimation is to provide
predictive Cpdf of transport risk to shippers based on their predetermined
demand variables and selectable decision variables (see Table \ref{table: predictors}).
This not only helps the shipper to find a preferable service but also
helps the forwarder to set a price quote. Assume a customer comes
with predetermined demand requirement $d=\left\{ r,m,wgt\right\} $
and is choosing from services $s=\left(a,leg,dur\right)\in S\left(d\right)$,
where $S\left(d\right)$ is the set of services available given demand
factors, $d$. Here, even though the initial deviation, $dev_{start}$,
is one of the decision variables, we set it to 0 because this variable
is unknown and not selectable before shipping starts. Let $f\left(risk\mid d,s\right)$
be the predictive distribution of transport risk conditional on $d$
and a chosen $s$, and $l_{i}\left(risk\right)$ be customer $i$'s
loss function. The optimal conditional choice of $s$, which minimizes
expected transport loss, is defined as
\begin{eqnarray}
\left(s\mid d\right)_{i}^{*} & \triangleq & \mbox{argmin}_{s\in S\left(c\right)}Loss_{i}\left(s\mid d\right)\nonumber \\
Loss_{i}\left(s\mid d\right) & = & \int l_{i}\left(risk\right)f\left(risk\mid d,s\right)\mbox{d}dev\label{eq: optimal choice}
\end{eqnarray}
where $Loss_{i}\left(s\mid d\right)$ is customer $i$'s expected
loss of choosing $s$ given $d$. Estimating each customer's unknown
loss function $l_{i}\left(dev\right)$ is another interesting study
of practical value, but is outside the scope of this paper. Here we
use several generic loss functions to illustrate how to use predicted
$f\left(risk\mid d,s\right)$ to aid service selection.

In Figure 
\begin{figure}[tbph]
\begin{centering}
\begin{minipage}[t]{0.33\textwidth}%
\includegraphics[width=1\textwidth]{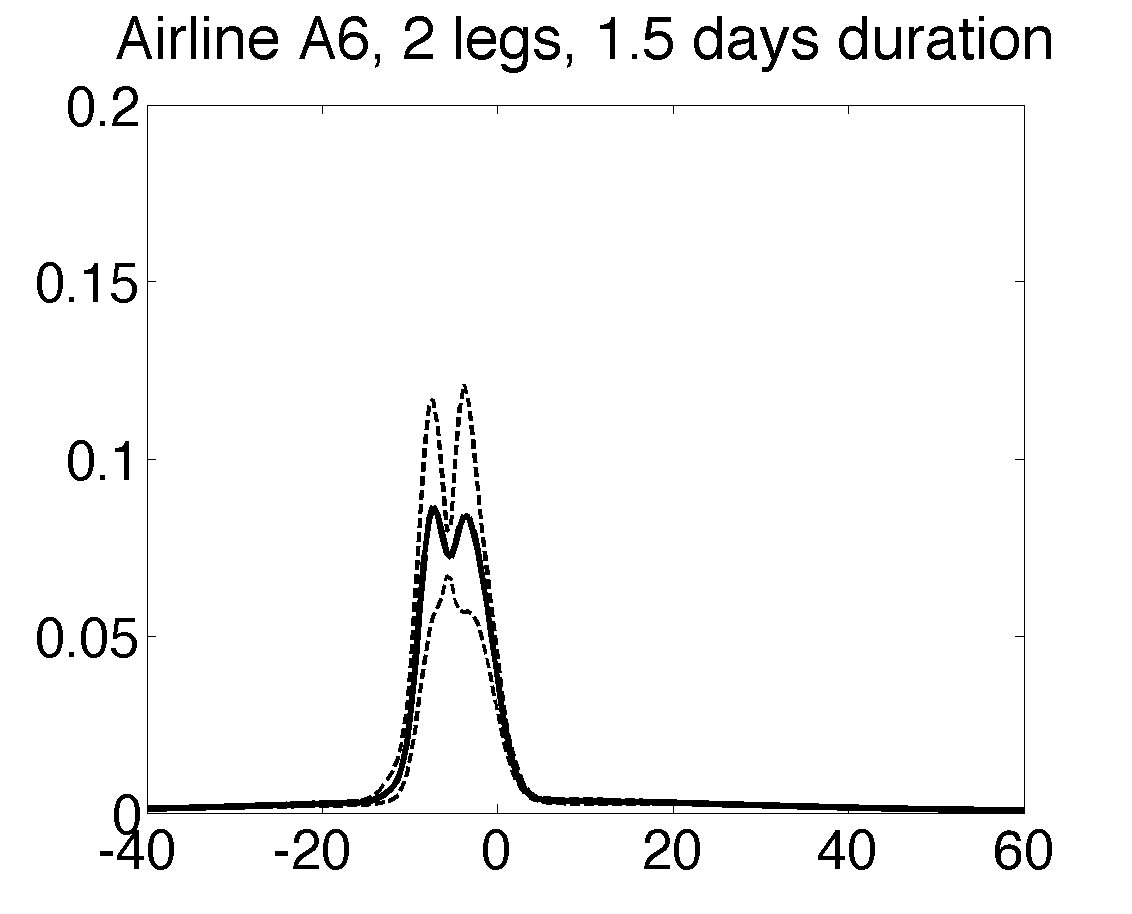}%
\end{minipage}%
\begin{minipage}[t]{0.33\textwidth}%
\includegraphics[width=1\textwidth]{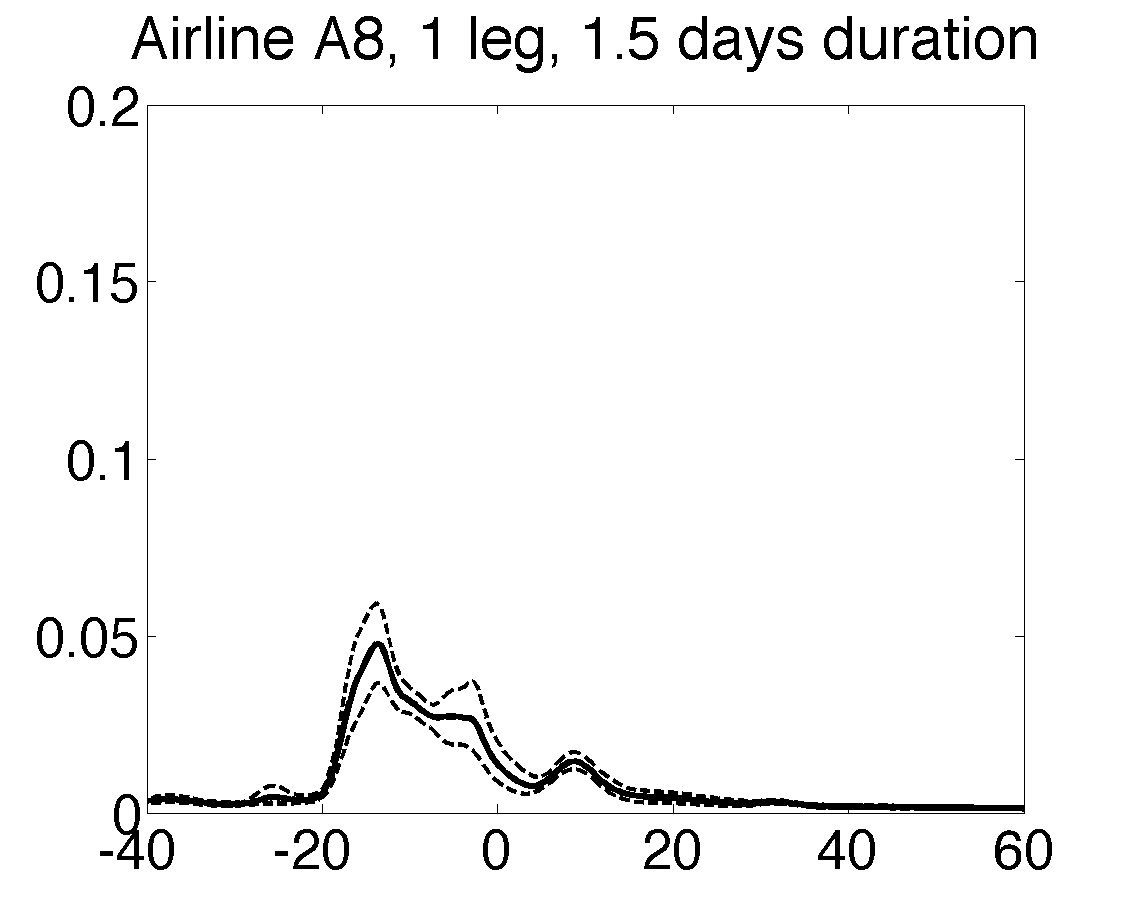}%
\end{minipage}%
\begin{minipage}[t]{0.33\textwidth}%
\includegraphics[width=1\textwidth]{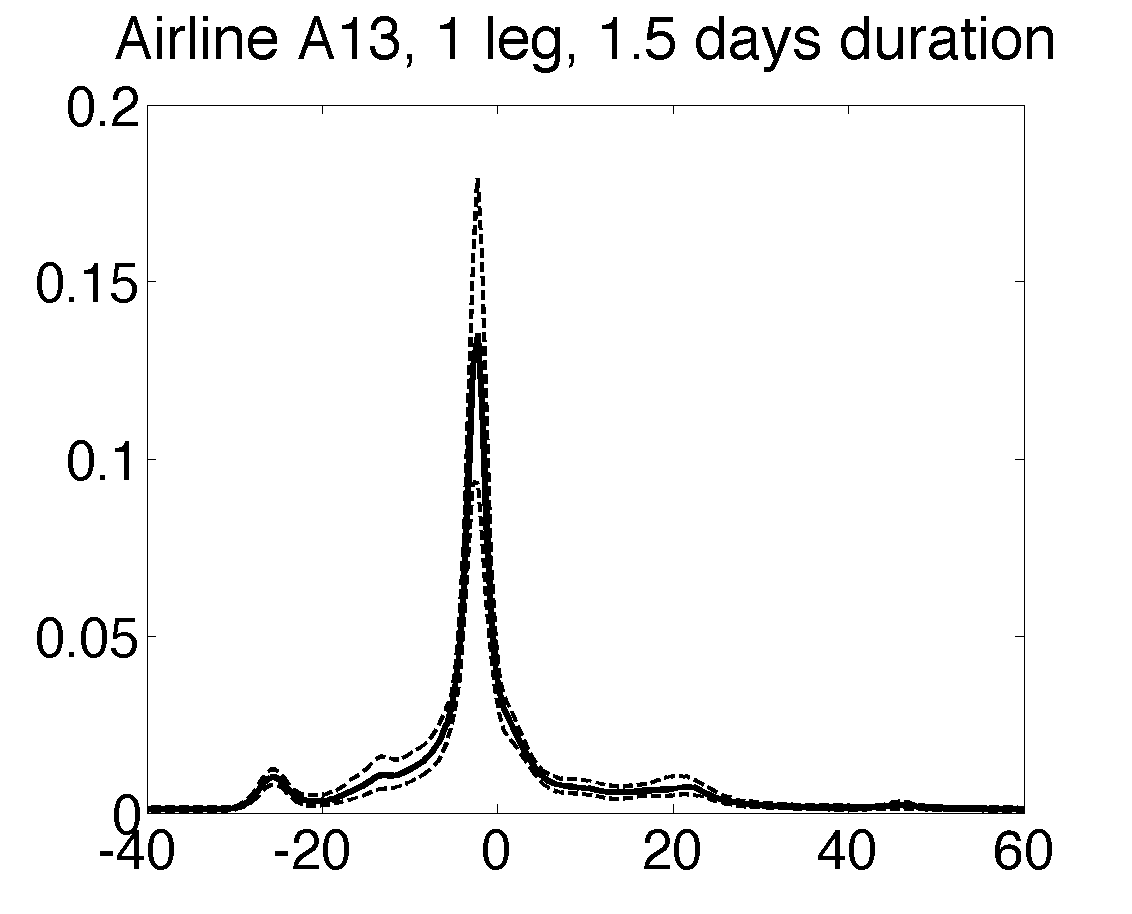}%
\end{minipage}
\par\end{centering}
\begin{centering}
\begin{minipage}[t]{0.33\textwidth}%
\includegraphics[width=1\textwidth]{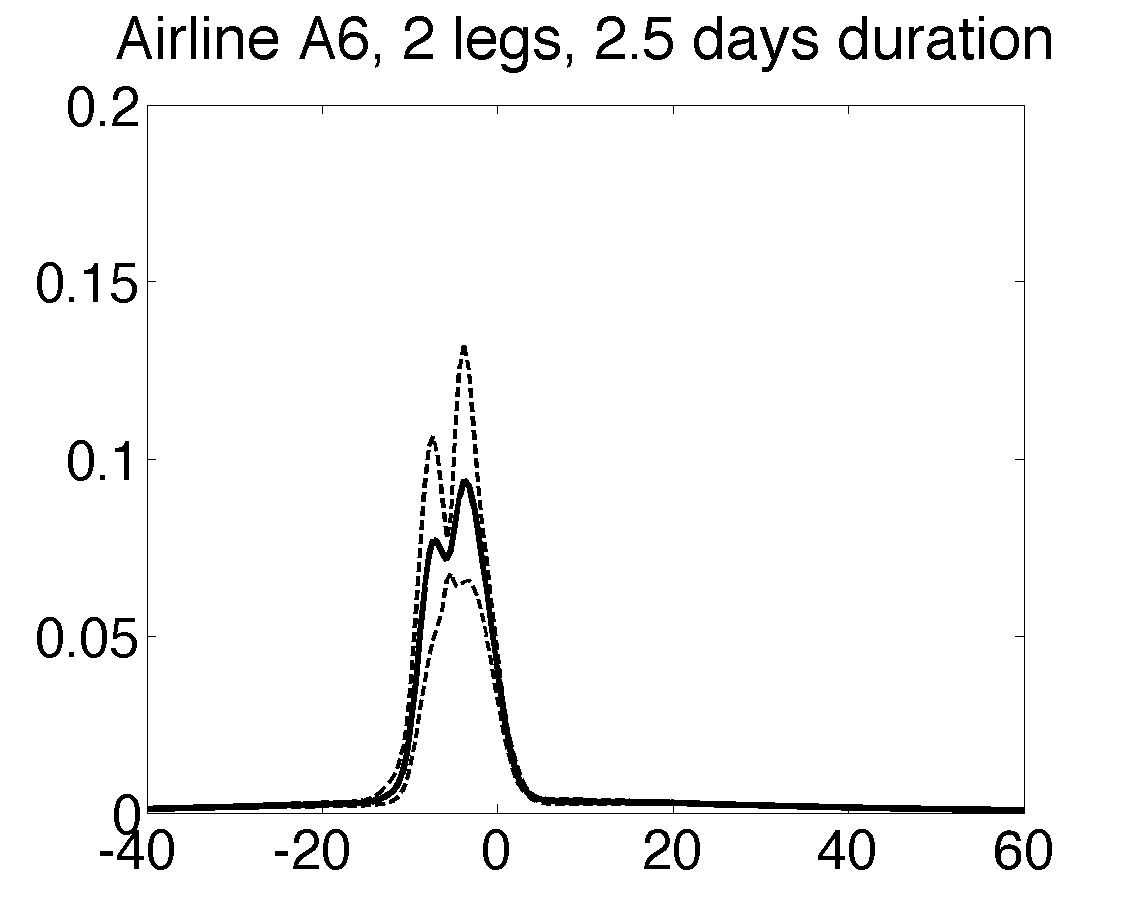}%
\end{minipage}%
\begin{minipage}[t]{0.33\textwidth}%
\includegraphics[width=1\textwidth]{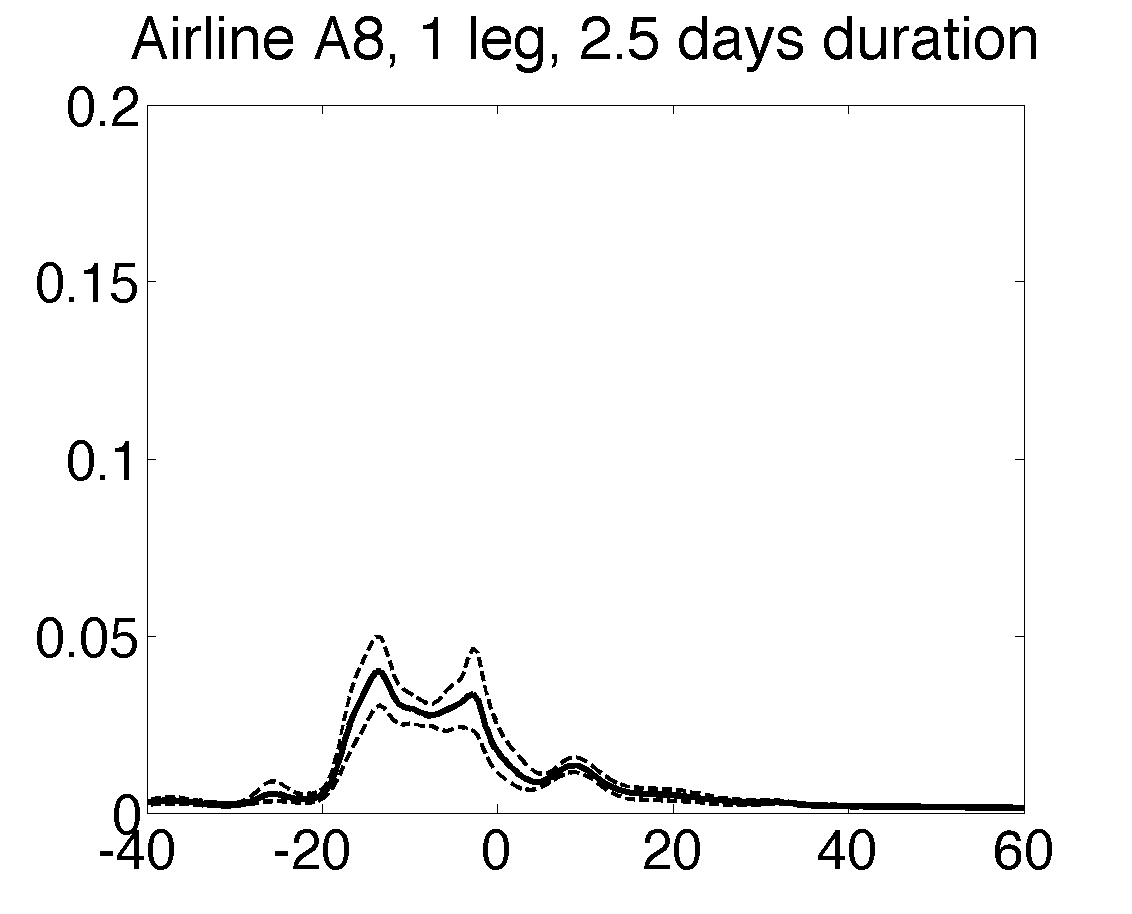}%
\end{minipage}%
\begin{minipage}[t]{0.33\textwidth}%
\includegraphics[width=1\textwidth]{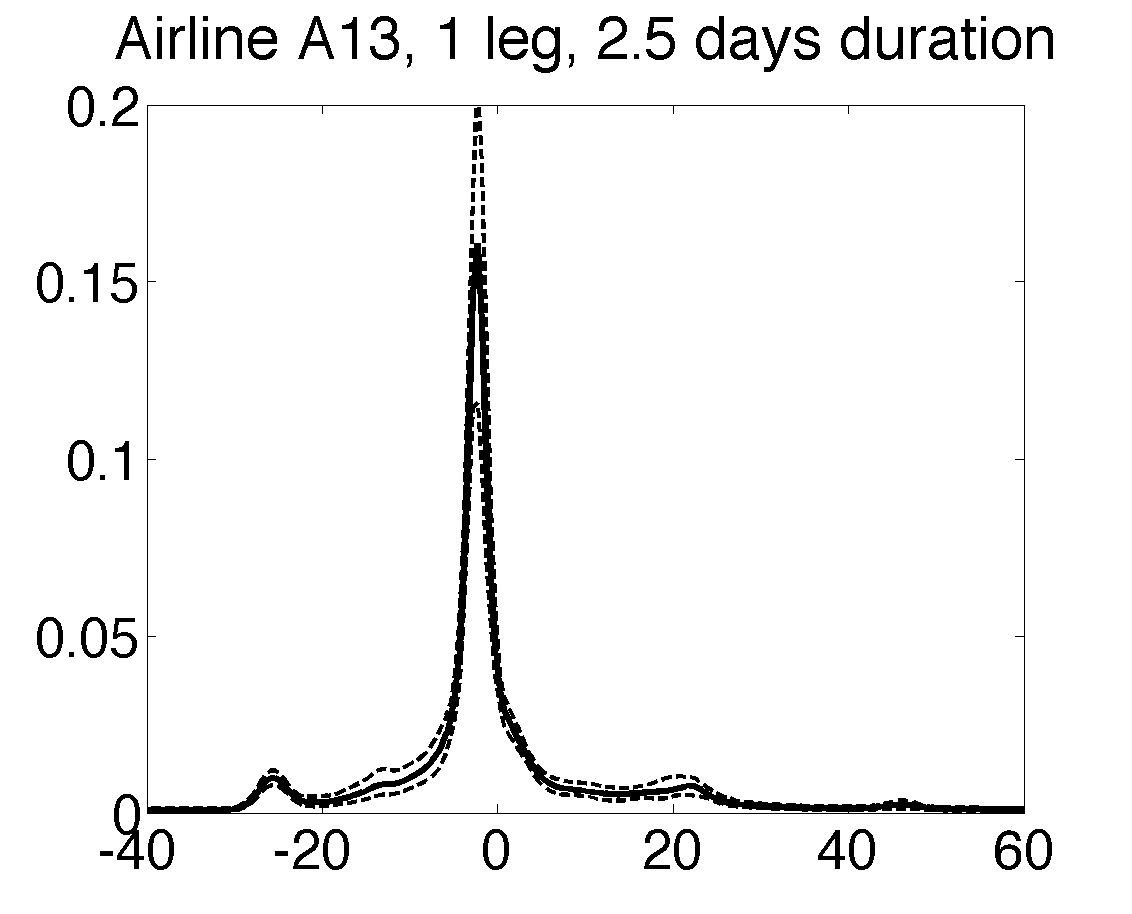}%
\end{minipage}
\par\end{centering}
\begin{centering}
\caption{PSBP predictive (solid curve) conditional response density $\hat{f}\left(y\mid x\right)$
with 95\% credible interval (dotted curve) of the \textit{normal }(bottom)
and \textit{speedy }(top) services from three airlines A6 (left),
A8 (middle) and A13 (right) on the route from Frankfurt (Germany)
to Atlanta (United States). For details of the calculation method,
please refer to ${\cal x}$5.2. }
\par\end{centering}
\centering{}\label{fig: predictive cpdf arwm}
\end{figure}
 \ref{fig: predictive cpdf arwm} are 6 choices as shown by the figure
titles, on the route from Frankfurt to Atlanta. The choices are randomly
picked from the data. We use the following three loss functions:
\begin{align*}
l_{1}\left(risk\right)=C_{1}\cdot risk &  & l_{2}\left(risk\right)=C_{2}\cdot{\bf 1}\left\{ risk>18\right\}  &  & l_{3}(risk)=C_{3}\cdot risk^{2}
\end{align*}
$l_{1}$ naturally arises when a risk neutral shipper is adverse to
delays while fond of early arrivals; $l_{2}$ is more proper when
a shipper is sensitive to extreme delays exceeding certain threshold
(18 hours in our example); $l_{3}$ is used when a shipper is risk
adverse and dislikes any deviations from the plan, neither negative
nor positive. Under these loss functions, the expected losses have
simple analytical forms 
\begin{align*}
Loss_{1}=C_{1}\cdot\mathsf{E}_{f} &  & Loss_{2}=C_{2}\cdot\left(1-F\left(18\right)\right) &  & Loss_{3}=C_{3}\cdot\left(\mathsf{Var}_{f}+\mathsf{E}_{f}^{2}\right)
\end{align*}
where $f$ is short for $f\left(risk\mid d,s\right)$ and $F$ is
the corresponding cumulative density function. Figure 
\begin{figure}[tbph]
\begin{centering}
\begin{minipage}[t]{0.33\columnwidth}%
\includegraphics[width=1\columnwidth]{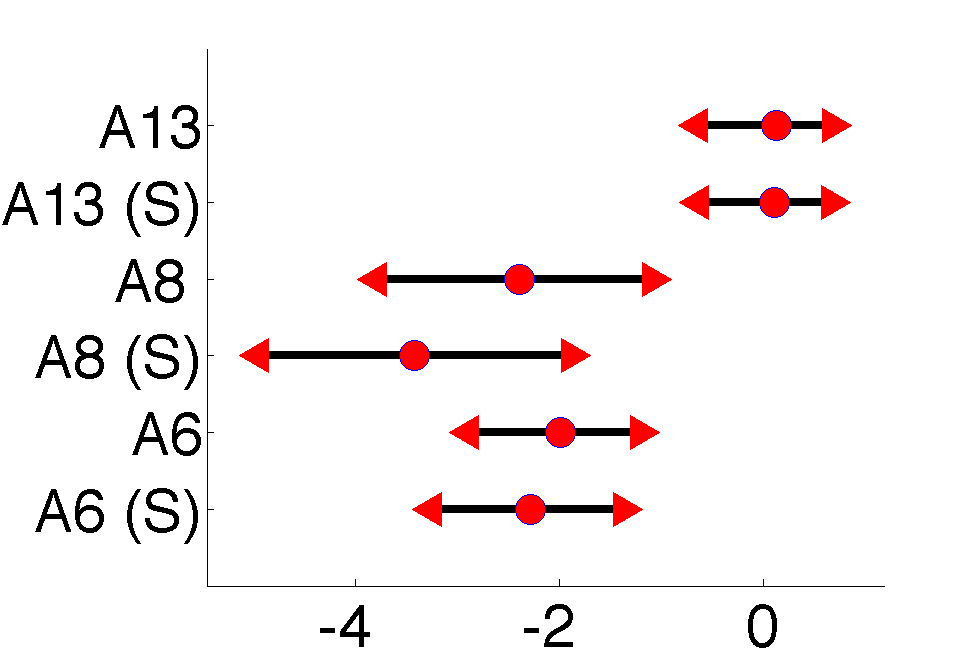}%
\end{minipage}%
\begin{minipage}[t]{0.33\columnwidth}%
\includegraphics[width=1\columnwidth]{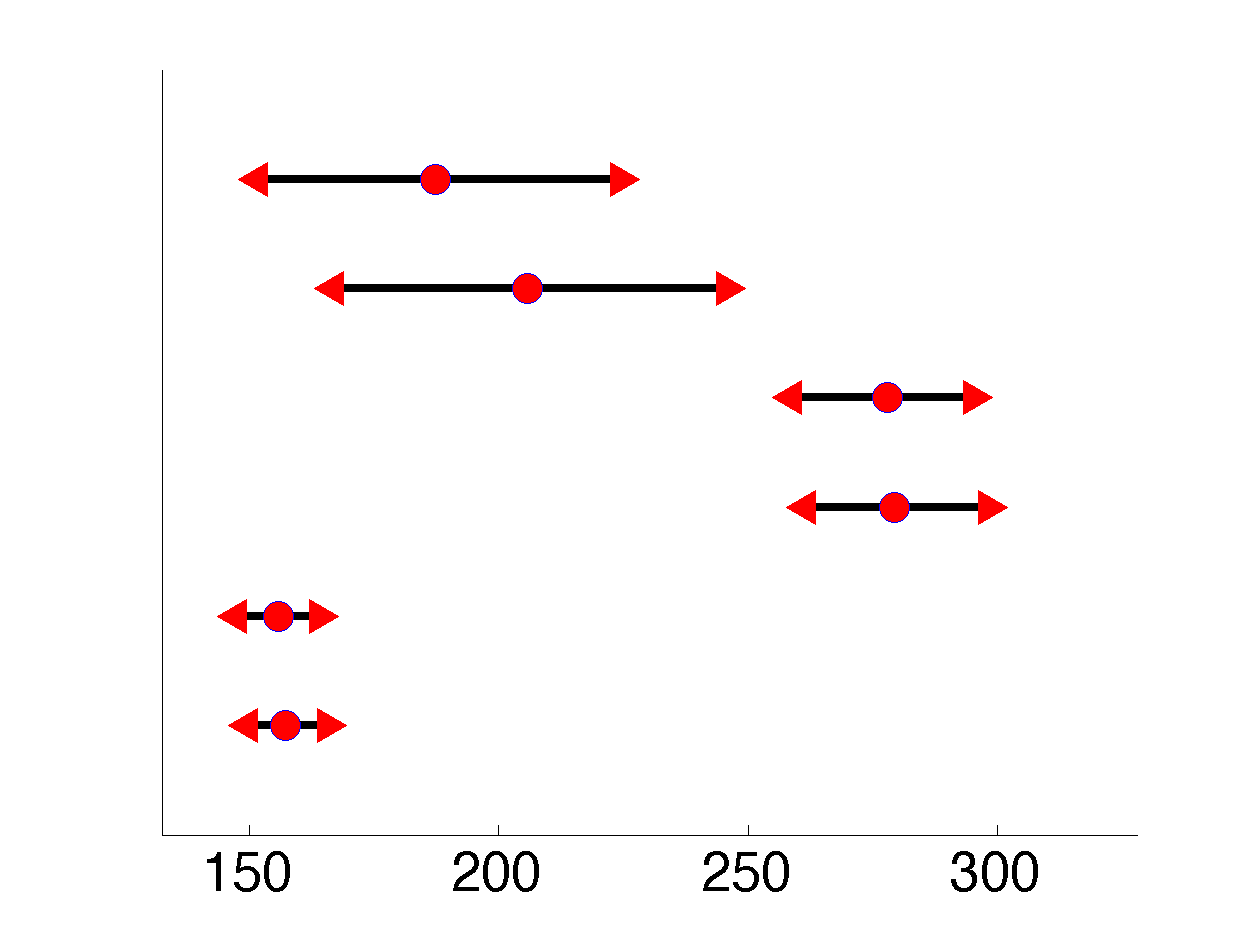}%
\end{minipage}%
\begin{minipage}[t]{0.33\columnwidth}%
\includegraphics[width=1\columnwidth]{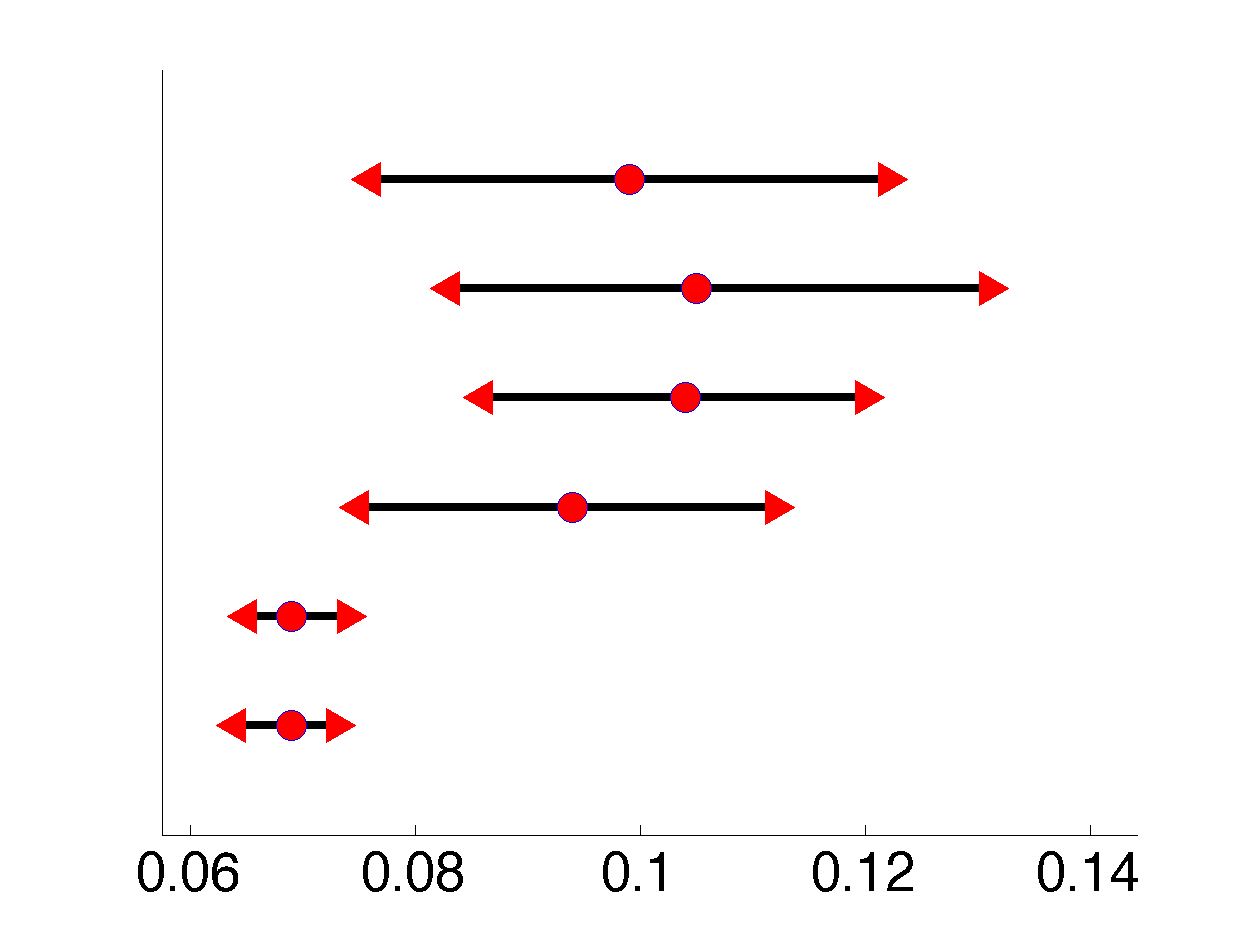}%
\end{minipage}
\par\end{centering}
\caption{Expected loss (with 95\% credible interval) of the \textit{normal}
and \textit{speedy} services from three airlines A6, A8 and A13 on
the route from Frankfurt (Germany) to Atlanta (United States). Left:
$Loss_{1}=C_{1}\cdot\mathsf{E}_{f}$; Middle: $Loss_{2}=C_{2}\cdot\left(1-F\left(18\right)\right)$;
Right: $Loss_{3}=C_{3}\cdot\left(\mathsf{Var}_{f}+\mathsf{E}_{f}^{2}\right)$}
\label{fig: risk comparison}
\end{figure}
 \ref{fig: risk comparison} presents the expected losses (with posterior
95\% probability intervals) calculated for the six choices under 3
risk functions with $C_{1}=C_{2}=C_{3}=1$, in which we use (S) to
indicate $speedy$ service. We observe (1) the rank of services in
terms of expected loss varies by loss functions; (2) choice of airlines
is playing a more dominant role than the choice between normal and
speedy services given an airline. 

With estimated expected loss of each choice, forwarders can offer
different price quotes to different types of shippers. In this example,
a forwarder can increase revenue by lowering A8's prices to attract
price-sensitive shippers and increasing A6's prices to attract quality-sensitive
shippers under loss function 2. 

\subsection{Supplier Ranking on Route or Higher Level}

Unlike a shipper, whose decision is made at the level of each shipment,
a forwarder plans its business at the route or higher level. To help
solve problems at high levels, the full predictive Cpdf should be
integrated. Specifically, let the full information set be $U=$\{$a$,
$r$, $m$, $leg$, $dur$, $dev_{start}$, $wgt$\}, for $U=U_{1}\cup U_{2}$
and $U_{1}\cap U_{2}=\phi$, then
\[
f\left(risk\mid U_{1}\right)=\int f\left(y\mid U_{1},U_{2}\right)f\left(U_{2}\right)dU_{2}
\]
where $U_{1}$ contains variables of central interest, and other variables
in $U_{2}$ are integrated out. For example, a practical problem faced
by a forwarder is whether to choose a carrier on a certain route and
how much capacity to reserve from it. For such decisions, an estimation
of the airline's service reliability is a critical input. In this
case airlines and routes are of interest, so we let $U_{1}=\left\{ a,r\right\} $
and $U_{2}=U-U_{1}$. By using Equation (\ref{eq: optimal choice})
with $c$ and $s$ replaced by $r$ and $a$, the forwarder can obtain
expected losses by each airline $a\in S\left(r\right)$, which, in
turn, can help make the right capacity reservation and pricing decisions. 

\subsection{Baseline Comparison }

Our result can also be used to generate baseline comparisons of various
factors. Baseline effect of a certain factor excludes the effects
of any other factors, thus allowing for a direct comparison between
factors of one type. 
\begin{figure}[tb]
\begin{centering}
\begin{minipage}[t]{0.33\columnwidth}%
\includegraphics[width=1\columnwidth]{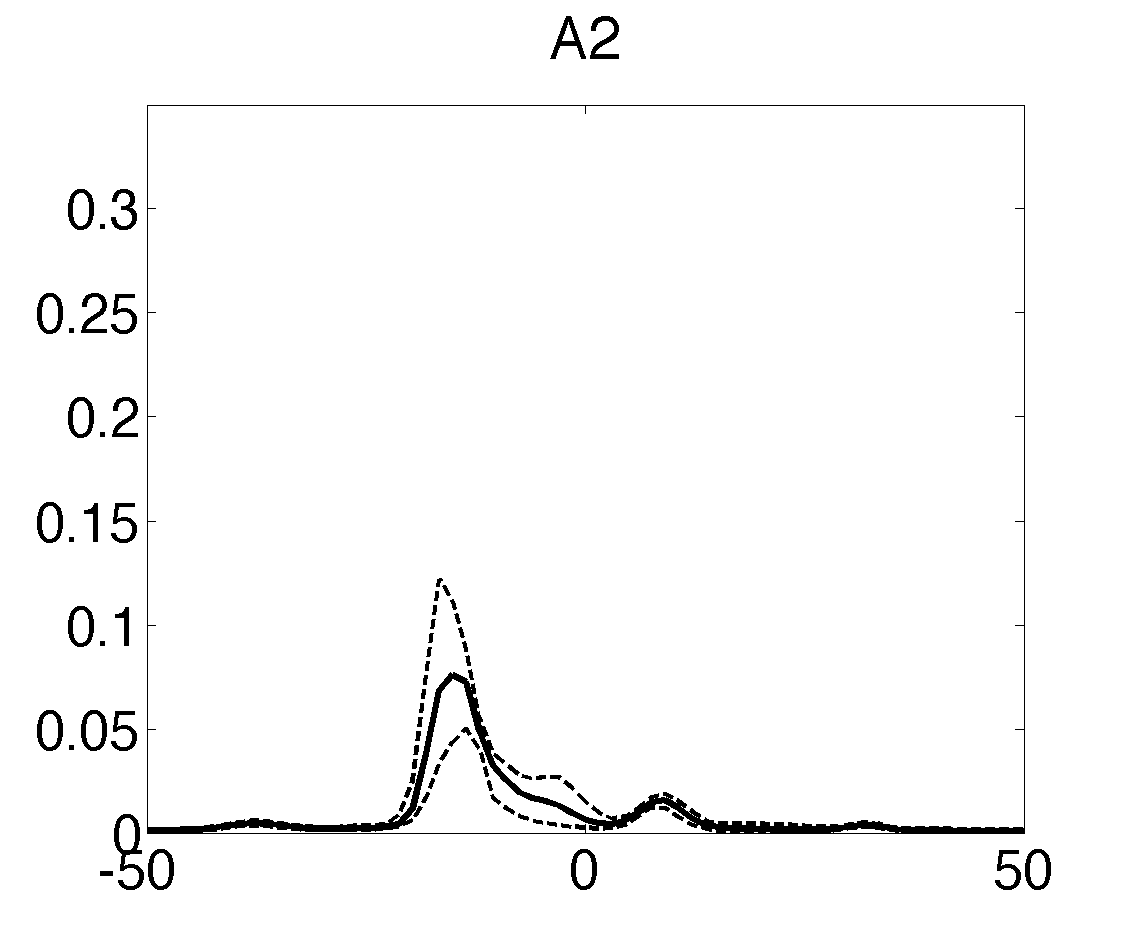}%
\end{minipage}%
\begin{minipage}[t]{0.33\columnwidth}%
\includegraphics[width=1\columnwidth]{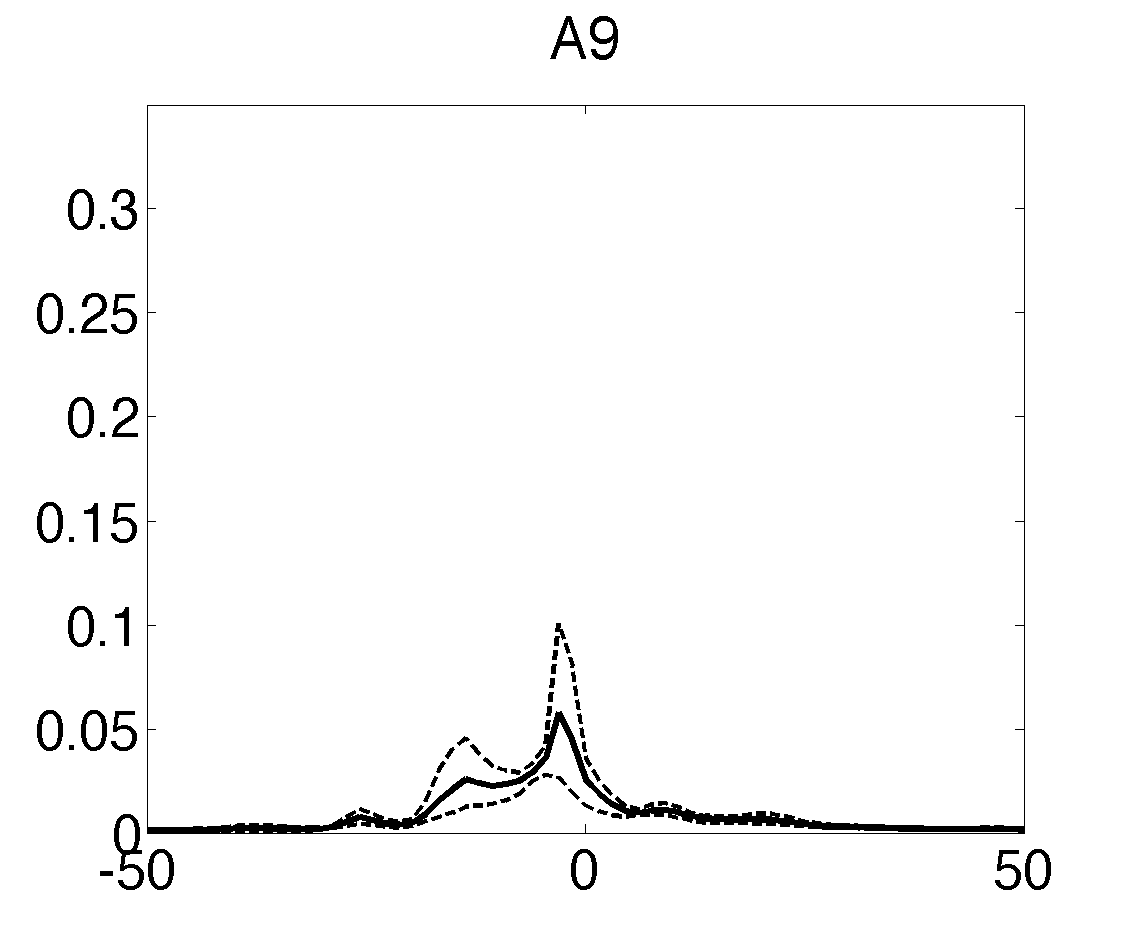}%
\end{minipage}%
\begin{minipage}[t]{0.33\columnwidth}%
\includegraphics[width=1\columnwidth]{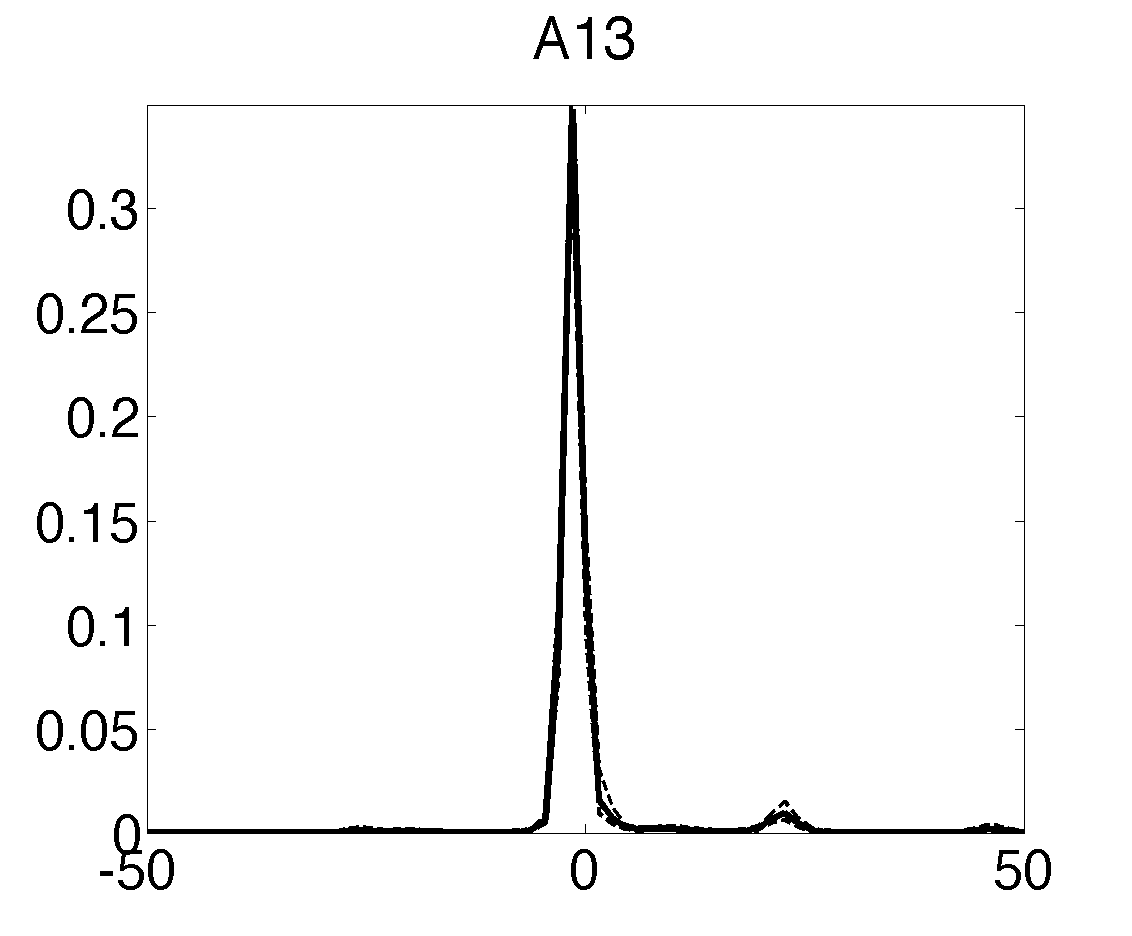}%
\end{minipage}
\par\end{centering}
\centering{}\caption{Reference performances of sample airlines A2 (left), A9 (middle) and
A13 (right) with predictive density mean (solid curve) and 95\% credible
interval (dotted curve)}
\label{fig: airline reference 1}
\end{figure}
One interesting example is to understand the baseline performance
of each airline, in which case a direct comparison is impossible due
to the fact that airlines serve different routes. To achieve this
baseline comparison, we use the average value for all other predictors,
except airline effects $\theta_{a}^{2}$, as their reference levels.
Then we plug these reference levels in the posterior samples of each
airline and then obtain the reference risk distribution for each airline
(See Figure \ref{fig: airline reference 1} for 3 samples from the
20 airlines. See Appendix ${\cal x}$C.2 for the remaining 17 baseline
distributions). From the plots we can directly compare airlines, which
differ from each other by the number, locations, and heights of peaks.
As such, our model allows baseline comparison based on distribution
knowledge. This offers a much richer comparison than those appearing
in the literature based on single average metrics. Meanwhile, the
richer tool allows us to obtain simple metric comparisons as special
cases.

For example, using a U.S. passenger flight data set, \citet{deshpande_impact_2012}
analyzed single-leg flight truncated block time, which is transport
risk plus planned duration minus initial deviation. Initial deviation
is defined as the positive delay of the previous flight by the same
craft if applicable and zero otherwise. The authors argue that if
the truncated block time is shorter than the scheduled block time,
the airline incurs an overage cost of $C_{o}$ per unit overage time.
Otherwise, the airline incurs an underage cost $C_{u}$ per unit shortage
time. The authors then estimate the overage to underage ratio, $\varphi=C_{0}/C_{u}$,
for each flight, and calculate the mean ratio of flights served by
a certain airline as the airline-wise overage to underage ratio, $\varphi_{a}$.
Using our international air cargo data, we can obtain an analogous
metric by replacing ``schedule block time'' and ``truncated block
time'' in their paper with $dur$ and $\left(dur+\mbox{arrival deviation}-[dev_{start}]^{+}\right)$.
One concern of estimating airline-wise ratio $\varphi_{a}$ by simply
calculating the average of flight-wise ratios is that the effects
from other factors, such as routes etc, cannot be excluded. Thus,
the calculated overage to underage ratio of each airline, $\varphi_{a}$,
cannot be used for direct comparison of airlines' intrinsic service
quality. Baseline distribution of airlines, on the other hand, is
a good solution to this problem. Specifically, the optimal $dur^{*}$
is defined by news-vendor solution that
\begin{eqnarray}
\mathsf{Prob}\left(dur^{*}+\mbox{arrival deviation}-\left[dev_{start}\right]^{+}\le dur^{*}\mid a\right) & = & \frac{1}{1+\varphi_{a}}\nonumber \\
\mathsf{Prob}\left(\mbox{arrival deviation}\le0\mid a\right) & = & \frac{1}{1+\varphi_{a}}\label{eq: overage to underage ratio}
\end{eqnarray}
where we use the fact that the reference level of $\left[dev_{start}\right]^{+}$,
calculated by the data average, is zero. Thus each airline's overage
to underage ratio is calculated by $\varphi_{a}=\frac{1}{F_{a}\left(0\right)}-1$
\begin{figure}[tbph]
\centering{}\includegraphics[width=1\columnwidth]{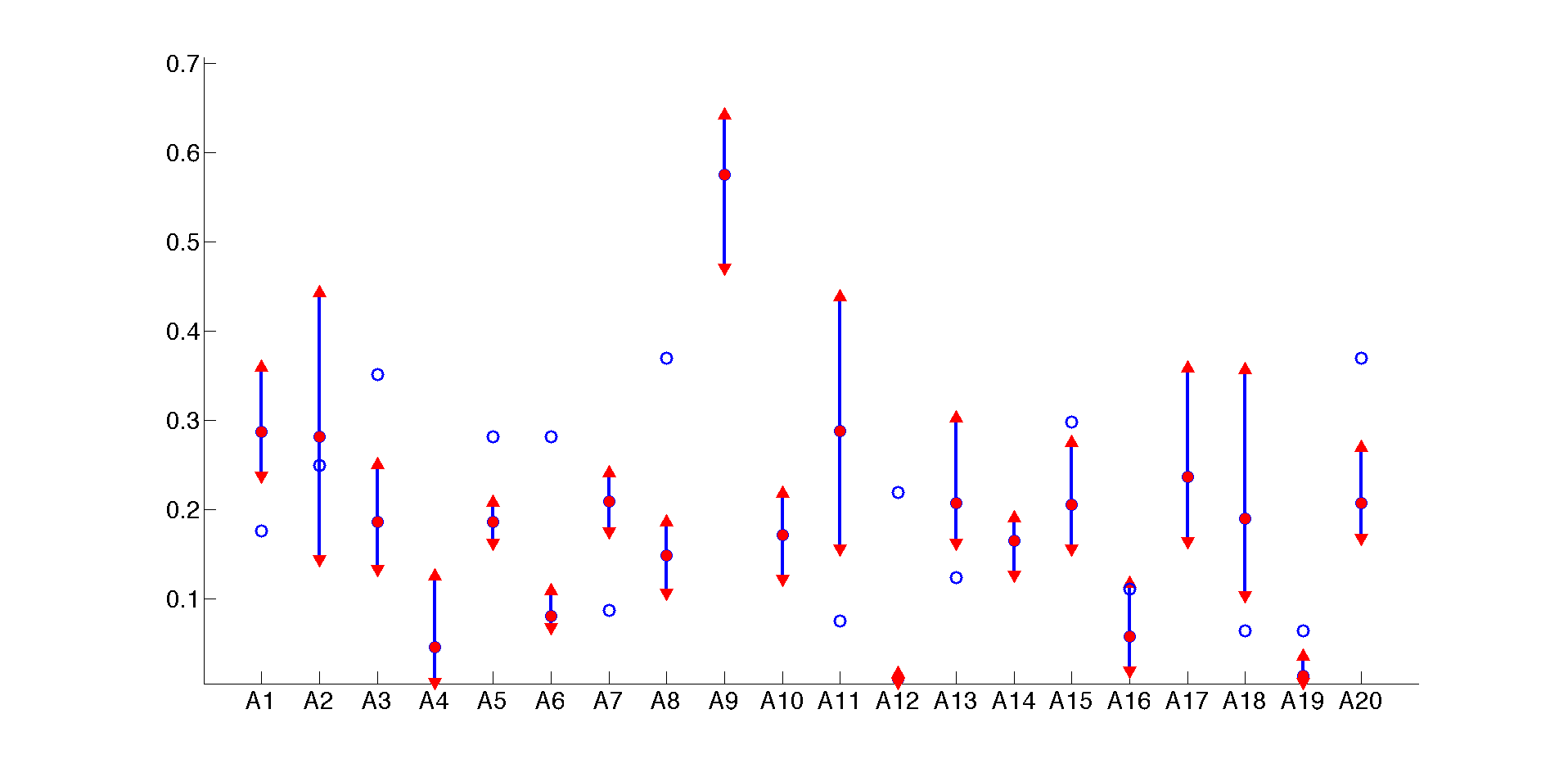}\caption{For each of the 20 airlines in the data: the red dot in the center
of the arrow is the mean overage to underage ratio calculated by PSBP;
the arrow represents the 95\% posterior probability interval; the
blue circle is the overage to underage ratio from the C2K Monthly
Statement issued by IATA (if there is no overage to underage ratio
reported in the C2K monthly statement, the blue circle is missing
for that airline).}
\label{fig: ratio airline}
\end{figure}
; see Figure \ref{fig: ratio airline} for the calculated overage
to underage ratios of 20 airlines with 95\% probability intervals. 

The overage to underage ratio $\varphi_{a}$ is related to airline's
on-time probability by Equation \ref{eq: overage to underage ratio}:
the higher the on-time rate the lower the ratio $\varphi_{a}$. We
compare our results to C2K Monthly Statement issued by IATA. In particular,
we choose monthly report issued in November 2012, the same period
of our data, and convert the reported airlines' on-time rates into
their overage/underage ratios (represented by the circles in Figure
\ref{fig: ratio airline}). The circles deviate from our estimations,
the solid dots, following no obvious rules. We believe this is because
IATA calculated the on-time rate by simply averaging on-time times
of an airline, which fails to exclude the impacts from factors other
than the airline, e.g., cargo weight, route, and thus results in unfair
comparison. The baseline distribution we calculated can also be used
to calculate many other metrics, such as variance, probability of
extreme disruptions etc, rather than the simple on-time rate reported
by IATA's monthly report. 

\section{Conclusions and Future Directions}

Using data from international air cargo logistics, we investigate
ways to assess and forecast transport risks, defined as the deviation
between actual arrival time and planned arrival time. To accommodate
the special multimodal feature of the data, we introduce a Bayesian
nonparametric mixture model, the Probit stick-breaking process (PSBP)
mixture model, for flexible estimation of conditional density function
of transport risk. Specifically, we build a linear structure, including
demand variables and decision variables, into kernel weights so that
the probability weights change with predictors. The model structure
is easily extended to account for other factors, such as long-term
effects, by allowing coefficients to change dynamically over time,
if data allows. Advantages of the PSBP include its generality, flexibility,
relatively simple sampling algorithm and theoretical support. 

Our results show that our method achieves much more accurate forecasts
than alternative models: naive linear model, generalized additive
model and flexible mixture model. Moreover, the linear model can lead
to misleading inferences. We also demonstrate how an accurate estimation
of transport risk Cpdf can help shippers to choose from multiple available
services, and help a forwarder to set targeting price, etc. In addition,
we show how to use the model to estimate baseline performance of a
predictor, such as an airline. We compare our findings with performance
reports issued by IATA and point out the shortcomings of IATA's simple
way of ranking airlines. We note that the usage of our method can
be much broader than the examples shown here. Indeed, any decisions
involving a distribution function require an estimated Cpdf. 

Our study serves as a stepping stone to deeper studies in the air
cargo transport industry, or more generally, the transportation industry,
which generates tons of data everyday yet lacks proper techniques
for data analysis. According to a 2011 McKinsey report \citep{manyika_big_2011},
in the transportation and warehousing sector, the main focus of our
paper, IT intensity is among the top 20\% and data availability is
among the top 40\% of all sectors, but the data-driven mind-set is
merely at the bottom 20\%. The authors' communication with leaders
in this industry, from whom we get the data supporting this research
project, confirms this situation, ``... we have plenty of data, or
we could say we have all the data possible, but we don't know how
to use the data...''. 

One of the interesting findings of our paper is that airlines have
critical impact on the shape of the transport risk distribution rather
than the mean focused on by linear models. For future research, we
hope to be able to obtain information regarding why and how airlines
are performing so differently on the same routes. By knowing the root
drivers of airline service performance, the service quality can be
improved for each airline rather than simply choose the best performer. 

\ACKNOWLEDGMENT{We are grateful for the generous support of industry sponsors Dr. Prof. Rod Franklin and Mr. Michael Webber. We also thank the area editor Anton Kleywegt, an anonymous associate editor and two referees for their valuable suggestions to make this paper better.}

\bibliographystyle{ormsv080}
\bibliography{KN_DSS}

\begin{APPENDICES}
\counterwithin{figure}{section} 
\counterwithin{table}{section}

\vfill{}
\clearpage{}

\section{Data }

\subsection{Data Cleaning}

After matching MUP with its baseline RMP, we obtain 155,780 shipments
(matching rate is higher than 95\%). After dropping (1) shipments
with extremely delayed milestones (usually caused by data input errors);
(2) shipments missing critical information (e.g., carrier); (3) shipments
missing weight or package information, 139,512 shipments are retained.
The 139,512 shipments are operated by 20 airlines on 11,282 routes
(we treat A to B and B to A as two distinct routes), and form 17,604
airline-route pairs. Since our analysis involves the airline-route
interaction term, in order to avoid the high noise caused by sparse
observations, we drop route-airline pairs containing less than 10
observations and routes containing less than 20 observations in the
observing period. After applying this filter, we have 86150 observations
left operated by 20 airlines on 1,333 routes. The filter is effective
in selecting large and profitable routes.

\subsubsection{Exception Records}

Exception codes are meant to facilitate (1) finding root causes of
delays and (2) identifying parties accountable for failures. Unfortunately,
as confirmed by the company as well as our data, exception codes are
not helpful in these regards. Less than 8\% of delays are assigned
exception codes, with only 10\% of delays of more than 1 day coded.
In addition, codes are ambiguous, with the most frequently appearing
code being ``COCNR'', denoting the carrier hasn't received the cargo.
Hence, we do not use exception data in our analysis.

\subsection{Data Illustration}

In Figure 
\begin{figure}[tbh]
\centering{}\includegraphics[scale=0.25]{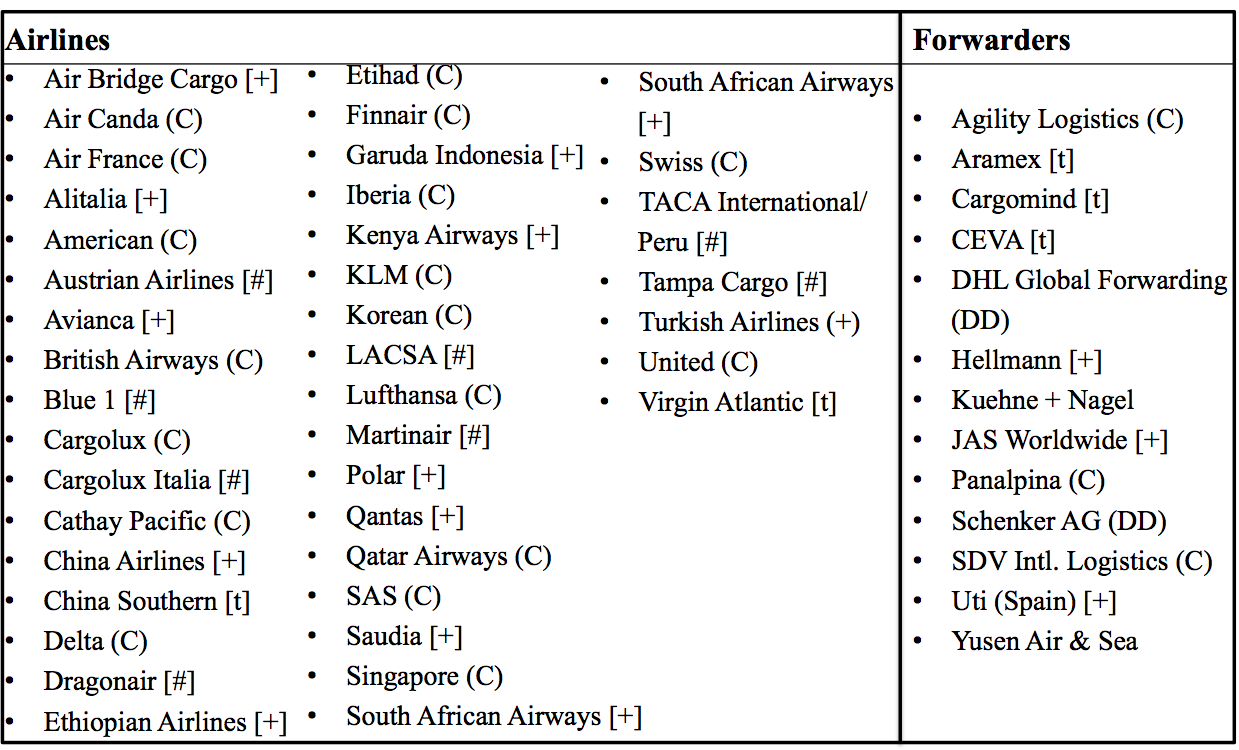} \caption{Cargo 2000 members}
\label{fig: c2k members} 
\end{figure}
 \ref{fig: c2k members} are the current members under C2K standards.
In Table A.1 
\begin{table}[tb]
\begin{centering}
{\small{}\caption{An example of a route map}
}
\par\end{centering}{\small \par}
\centering{}{\small{}}%
\begin{tabular}{cccccc}
\hline 
{\small{}Milestone } & {\small{}Time } & {\small{}Airport } & {\small{}Flight } & {\small{}Weight } & {\small{}Piece}\tabularnewline
\hline 
{\small{}RCS} & {\small{}06.12.2013 16:15:00 } & {\small{}NTE } & {\small{}\# } & {\small{}630 } & {\small{}2 }\tabularnewline
{\small{}DEP} & {\small{}06.12.2013 19:00:00 } & {\small{}NTE } & {\small{}AA 8854} & {\small{}630 } & {\small{}2 }\tabularnewline
{\small{}ARR} & {\small{}07.12.2013 08:52:00 } & {\small{}CDG } & {\small{}AA 8854 } & {\small{}630 } & {\small{}2 }\tabularnewline
{\small{}DEP} & {\small{}10.12.2013 09:21:00 } & {\small{}CDG } & {\small{}AA 0063 } & {\small{}630 } & {\small{}2 }\tabularnewline
{\small{}RCF } & {\small{}10.12.2013 21:26:00 } & {\small{}MIA } & {\small{}AA 0063 } & {\small{}630 } & {\small{}2 }\tabularnewline
{\small{}DEP} & {\small{}11.12.2013 14:58:00 } & {\small{}MIA } & {\small{}AA 0913 } & {\small{}630 } & {\small{}2 }\tabularnewline
{\small{}RCF} & {\small{}11.12.2013 21:46:00 } & {\small{}BOG } & {\small{}AA 0913 } & {\small{}630 } & {\small{}2 }\tabularnewline
{\small{}DLV} & {\small{}11.12.2013 22:40:00 } & {\small{}BOG } & {\small{}\# } & {\small{}630 } & {\small{}2 }\tabularnewline
\hline 
\end{tabular}\label{table: route map} 
\end{table}
is a typical route map for a shipment from Nantes (France) to Bogotá
(Columbia). In Figure 
\begin{figure}[tb]
\begin{centering}
\includegraphics[scale=0.24]{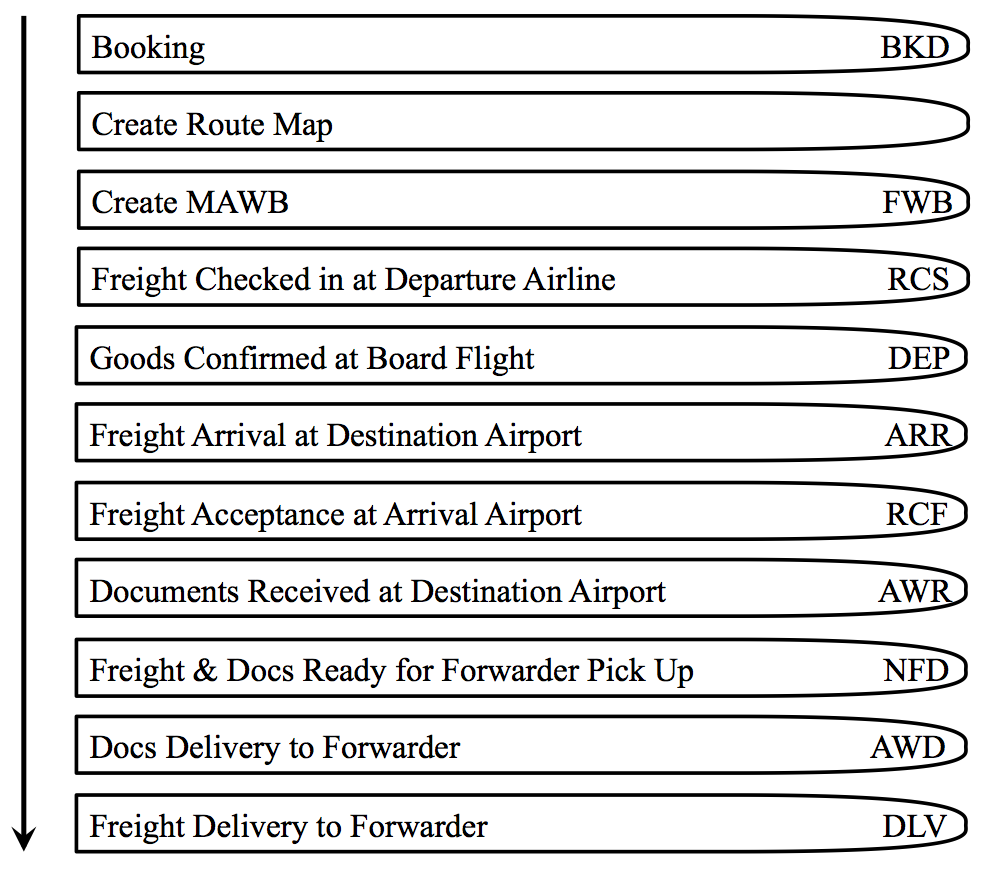}
\par\end{centering}
\centering{}\caption{Important milestones in a shipment with their short names}
\label{fig: milestone explanation}
\end{figure}
 \ref{fig: milestone explanation} are the milestone chain and explanation
of each milestone. In Table \ref{table: except} 
\begin{table}[h]
\begin{centering}
{\small{}\caption{A typical record of exception}
}
\par\end{centering}{\small \par}
\centering{}{\small{}}%
\begin{tabular}{ccccc}
\hline 
{\small{}Status } & {\small{}Exception } & {\small{}Time } & {\small{}Flight } & {\small{}Airport}\tabularnewline
\hline 
{\small{}DEP } & {\small{}COCSYMD } & {\small{}08.01.2013 05:05:00 } & {\small{}BA 0125 } & {\small{}LHR }\tabularnewline
\hline 
\end{tabular}\label{table: except} 
\end{table}
 is an typical record of an exception.

\subsection{Summary Statistics}

Figure \ref{fig: airline leg} shows the number of shipments for each
airline, and the percentage of shipments by the number of legs. In
Figure \ref{fig: from to} is the percentage of shipments between
the five continents (AF: Africa; AS: Asia; EU: Europe; NA: North America;
SA: South America). Figure \ref{fig: obs airline} shows the number
of airlines available for each shipment. Figure \ref{fig: obs airline leg}
depicts the choices between legs of each shipment. Table \ref{table: summary statistics}
provides summary statistics with predictors defined in Table \ref{table: predictors}.
\begin{table}[tbph]
\centering{}{\small{}\caption{Summary statistics}
\label{table: summary statistics}}%
\begin{tabular}{cc|>{\centering}p{2.3cm}>{\centering}p{1.6cm}>{\centering}p{2.1cm}>{\centering}p{2.3cm}>{\centering}p{1.7cm}>{\centering}p{1.7cm}}
\hline 
\multicolumn{3}{l}{\textbf{\small{}Dependent Variable}} &  &  &  &  & \tabularnewline
\hline 
 & \multicolumn{1}{c}{} & \multicolumn{1}{>{\centering}p{2.3cm}|}{} & {\small{}mean} & {\small{}std} &  &  & \tabularnewline
\cline{4-5} 
\multicolumn{3}{c|}{{\small{}transport risk (hour)}} & {\small{}-2.6} & {\small{}20.6} &  &  & \tabularnewline
\hline 
\multicolumn{4}{l}{\textbf{\small{}Predictors}} &  &  &  & \tabularnewline
\hline 
 & \multicolumn{4}{l}{\textit{\small{}Category Predictor}} &  &  & \tabularnewline
\cline{2-8} 
 &  & {\small{}airline} & {\small{}route} & {\small{}airline-route} & {\small{}month} & {\small{}airline-leg2} & {\small{}airline-leg3}\tabularnewline
\cline{3-8} 
 & {\small{}dimension} & {\small{}20} & {\small{}1336} & {\small{}588} & {\small{}7} & {\small{}20} & {\small{}16}\tabularnewline
\cline{2-8} 
 & \multicolumn{4}{l}{\textit{\small{}Continuous Predictor}} &  &  & \tabularnewline
\cline{2-8} 
 &  & {\small{}$dev_{start}$ (day)} & {\small{}$dur$ (day)} & {\small{}log$\left(wgt\right)$ (kg)} & {\small{}log$\left(pcs\right)$ (cbm)} &  & \tabularnewline
\cline{3-6} 
 & {\small{}mean} & {\small{}-0.327} & {\small{}1.75} & {\small{}4.91} & {\small{}1.29} &  & \tabularnewline
 & {\small{}std} & {\small{}0.648} & {\small{}1.30} & {\small{}2.4} & {\small{}1.43} &  & \tabularnewline
\hline 
\end{tabular}
\end{table}
 
\begin{figure}[tbph]
\begin{minipage}[t]{0.49\columnwidth}%
\includegraphics[width=1\columnwidth]{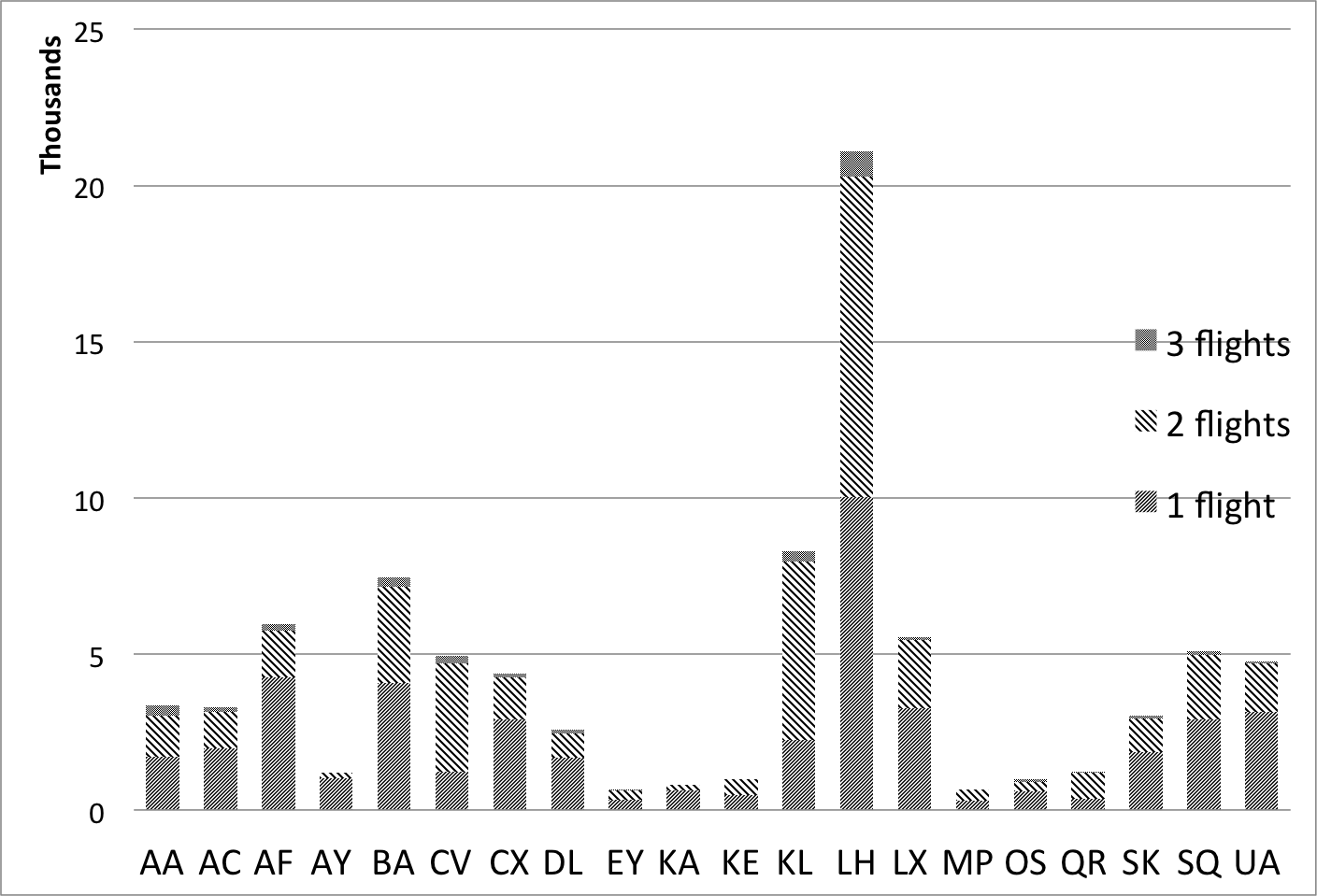}

\caption{Number of shipments by each airline with different number of legs
(1, 2 or 3)}
\label{fig: airline leg}%
\end{minipage}\hfill{}%
\begin{minipage}[t]{0.49\columnwidth}%
\includegraphics[width=1\columnwidth]{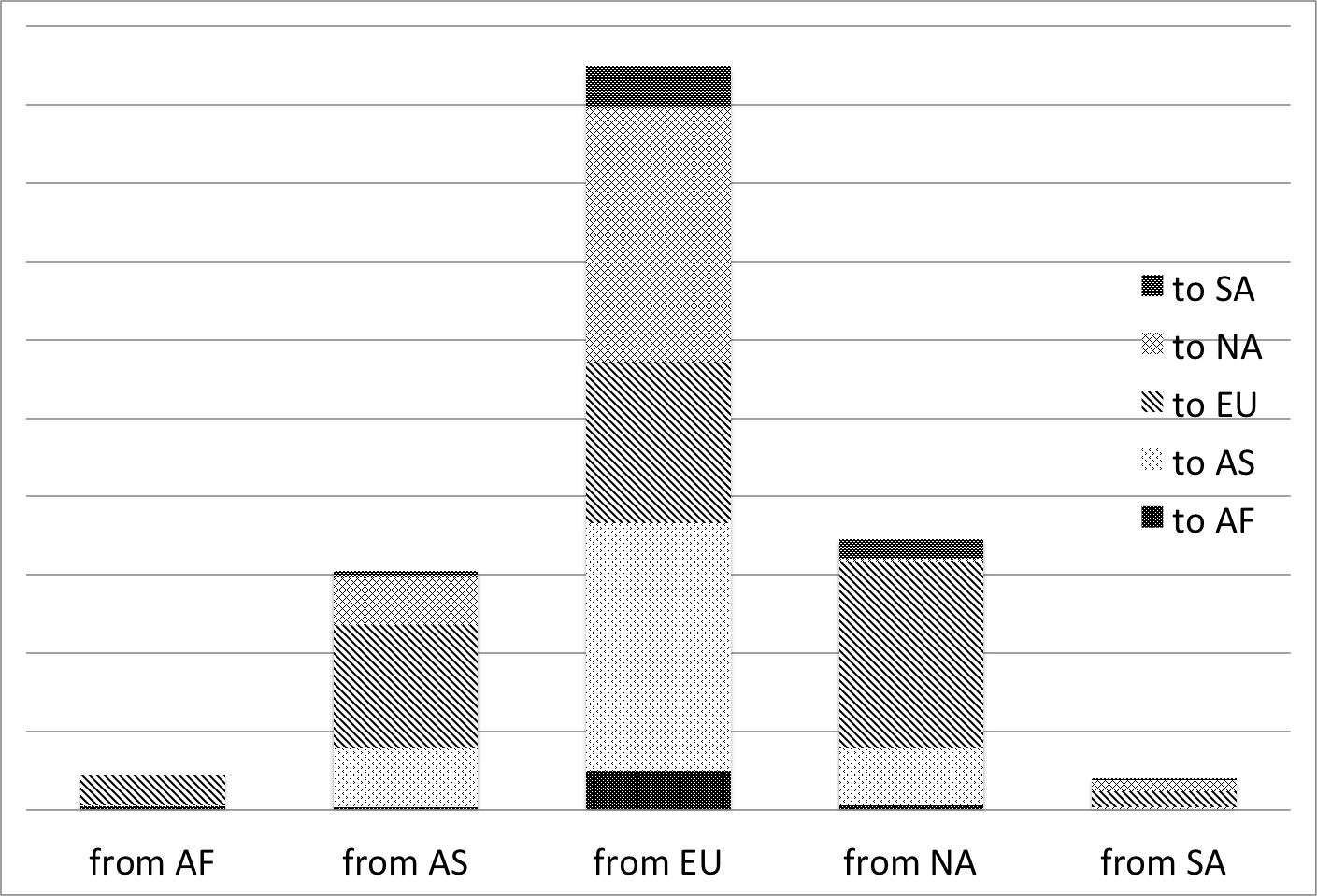}

\caption{Number of shipments between continents}
\label{fig: from to}%
\end{minipage}

\begin{minipage}[t]{0.49\columnwidth}%
\includegraphics[width=1\columnwidth]{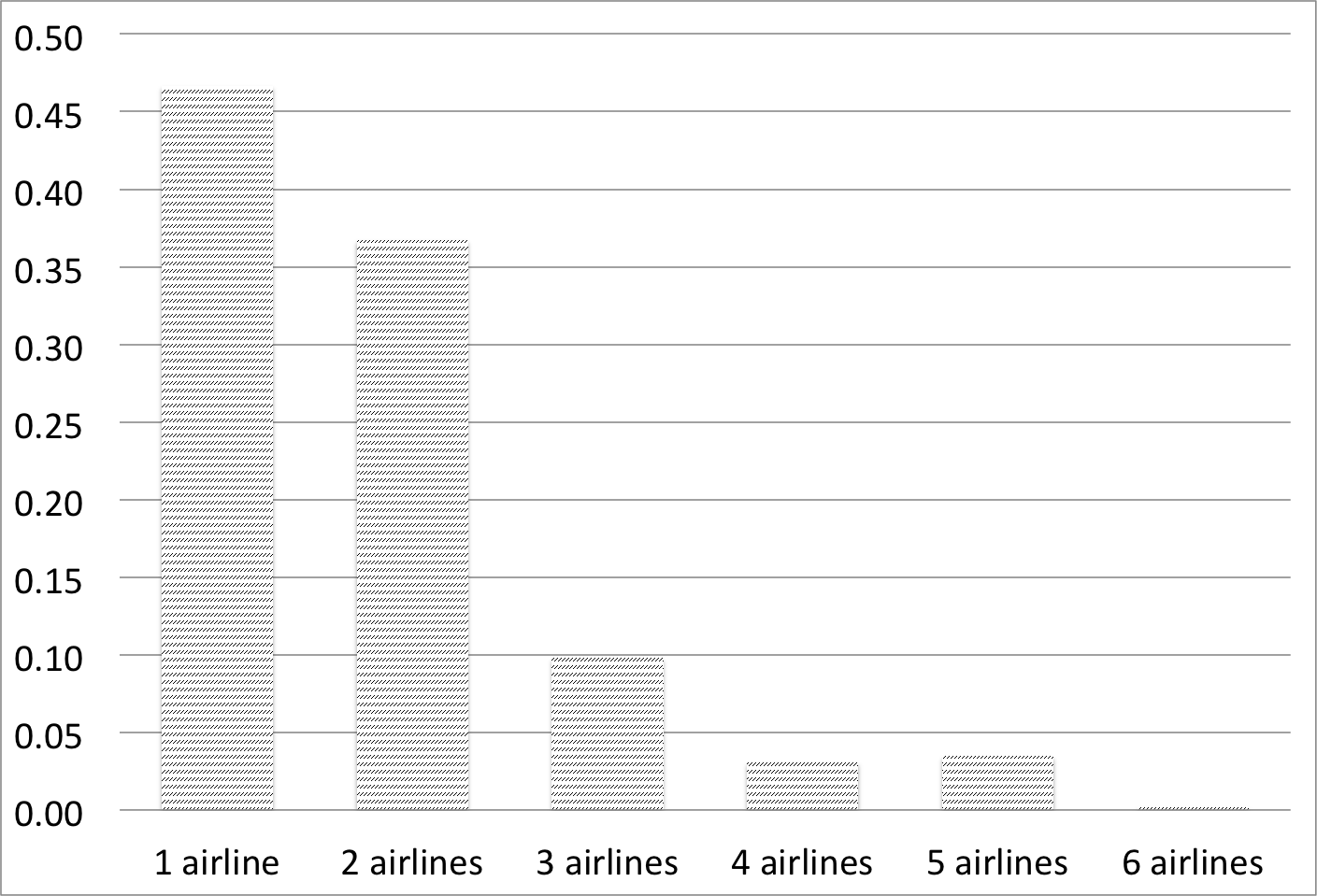}

\caption{The percentage of routes with different number of airline options.
The airline options vary from only 1 airline to as many as 6 airlines
serving the same route. }
\label{fig: obs airline}%
\end{minipage}\hfill{}%
\begin{minipage}[t]{0.49\columnwidth}%
\includegraphics[width=1\columnwidth]{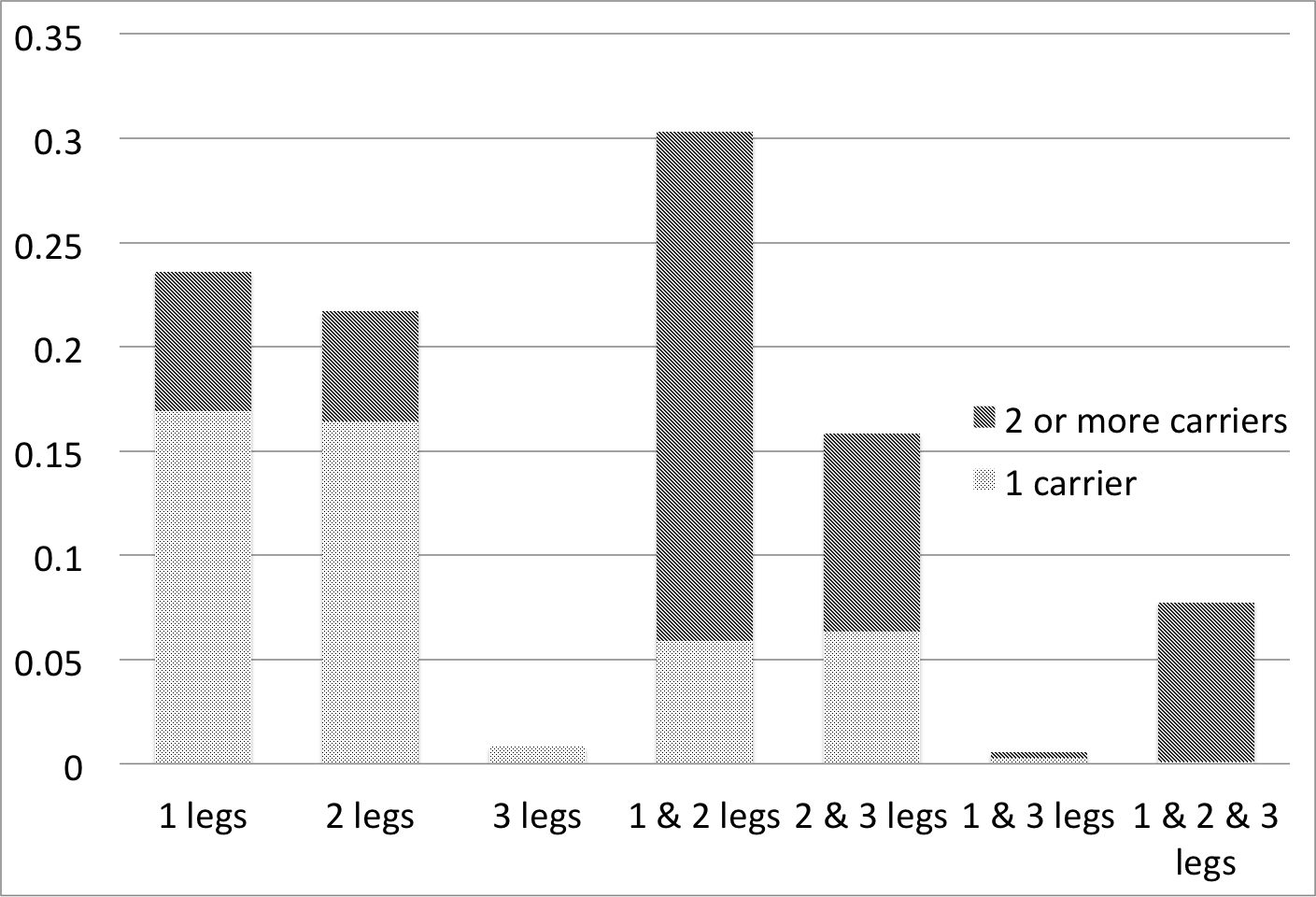}

\caption{The percentage of routes with different number of leg options (mainly
1 or 2 legs). The histogram is further classified into two categories:
whether the route is served by a single airline or multiple airlines. }
\label{fig: obs airline leg}%
\end{minipage}
\end{figure}

\vfill{}
\clearpage{}

\section{Supplementary Material of Computation and Model Checking }

\subsection{Gibbs Sampling}

\subsubsection{Gibbs Sampling for Kernel Parameters}

We use ``$\cdots$'' to indicate \textit{all the other parameters
and data}. The full conditional distribution of the component-specific
parameters, $\mu_{l}$ and $\phi_{l}$, is given by 

\noindent 
\[
p\left(\mu_{l},\phi_{l}\mid\cdots\right)\propto\mathsf{N}\mathsf{G}\left(\mu_{l},\phi_{l}\mid\zeta_{\mu},\xi_{\mu},a_{\phi},b_{\phi}\right)\prod_{({\bf x},j)\;s.t.\;s_{j}\left({\bf x}\right)=l}\mathsf{N}\left(y_{j}\mid\mu_{l},\phi_{l}\right)
\]
where $\propto$ represents ``proportional to'', $\mathsf{N}\mathsf{G}$
is the Normal-Gamma conjugate prior of $\mu_{l}$ and $\phi_{l}$.
Simplified by the conjugacy structure, the Gibbs sampling of kernel
mean $\mu_{l}$ is carried out by
\[
\mu_{l}\mid\cdots\sim\mathsf{N}\left(\left[\zeta_{\mu}+n_{l}\phi_{l}\right]^{-1}\left[\zeta_{\mu}\xi_{\mu}+h_{l}\phi_{l}\right],\xi_{\mu}+n_{l}\phi_{l}\right)
\]
where $n_{l}=\sum_{{\bf x}\in{\cal X}}\sum_{j=1}^{n({\bf x})}\mathbf{1}_{(s_{j}({\bf x})=l)}$
and $h_{l}=\sum_{{\bf x}\in{\cal X}}\sum_{j=1}^{n({\bf x})}y_{j}({\bf x})\mathbf{1}_{(s_{j}({\bf x})=l)}$.
Similarly, the Gibbs sampling of kernel precisions $\phi_{l}$ is
\[
\phi_{l}\mid\cdots\sim\mathsf{G}\left(a_{\phi}+\frac{n_{l}}{2};b_{\phi}+\frac{1}{2}\sum_{{\bf x}\in{\cal X}}\sum_{j=1}^{n({\bf x})}\left(y_{j}({\bf x})-\mu_{l}\right){}^{2}\mathbf{1}_{(s_{j}({\bf x})=l)}\right)
\]

\subsubsection{Gibbs Sampling for Weight Parameters: Latent Indicators}

Conditional on kernel parameters and the realized values of the weights
$\left\{ \omega_{l}\left({\bf x}\right),\forall{\bf x}\in{\cal X}\right\} _{l=1}^{L}$,
the distribution of the indicators is multinomial with probability
given by
\[
\mbox{Pr}(s_{j}({\bf x})=l|\cdots)\propto\omega_{l}({\bf x})\mathsf{N}\left(y_{j}({\bf x})|\mu_{l},\phi_{l}\right),
\]
So we can sample $s_{j}({\bf x})$ ($j=1,\cdots,n({\bf x}))$ from
a multinomial conditional distribution:
\[
Pr(s_{j}({\bf x})=l|\cdots)=\frac{\omega_{l}({\bf x})\mathsf{N}(y_{j}({\bf x})|\mu_{l},\phi_{l})}{\sum_{p=1}^{L}\omega_{p}({\bf x})\mathsf{N}(y_{j}({\bf x})|\mu_{l},\phi_{l})}
\]

\subsubsection{Gibbs Sampling for Weight Parameters: Latent Auxiliary Variable}

In order to sample the latent processes $\left\{ \gamma_{l}\left({\bf x}\right),\forall{\bf x}\in{\cal X}\right\} _{l=1}^{L}$
and the corresponding weights $\left\{ \omega_{l}\left({\bf x}\right),\forall{\bf x}\in{\cal X}\right\} _{l=1}^{L}$,
we \textbf{\textit{augment}} the data with a collection of conditionally
independent latent variables $z_{jl}({\bf x})\sim\mathsf{N}(\gamma_{l}({\bf x}),1)$
($j=1,\cdots,n({\bf x}))$. We aim to make the probability of observing
\{$z_{j1}\left({\bf x}\right)$, $\cdots$, $z_{jl}\left({\bf x}\right)$\}
equal to $\omega_{l}\left({\bf x}\right)$ in Equation (\ref{eq: stick-breaking}),
thus the event of observing \{$z_{j1}\left({\bf x}\right)$, $\cdots$,
$z_{jl}\left({\bf x}\right)$\} can represent the event of observing
$s_{j}({\bf x})=l$ as denoted at the beginning of ${\cal x}3.4$.
Specifically if $z_{jp}({\bf x})<0$ for all $p<l$ and $z_{jl}({\bf x})>0$,
we define $s_{j}({\bf x})=l$. For a finite $L$ case, we define $s_{i}({\bf x})=L$
if $z_{ip}({\bf x})<0$ for all $p\le L-1$. Then we have
\begin{eqnarray*}
\mbox{Pr}(s_{j}({\bf x})=l) & = & \mbox{Pr}\left(z_{jl}({\bf x})>0,z_{jp}({\bf x})<0\mbox{ for }p<l\right)\\
 & = & \Phi\left(\gamma_{l}({\bf x})\right)\prod_{p<l}\left\{ 1-\Phi\left(\gamma_{p}({\bf x})\right)\right\} 
\end{eqnarray*}
independently for $j=1,\cdots,n({\bf x})$. In this way, $\mbox{Pr}(s_{j}({\bf x})=l)$
equals to $\omega_{l}\left({\bf x}\right)$ as defined in Equation
(\ref{eq: stick-breaking}). This data augmentation scheme simplifies
computation as it allows us to implement the following Gibbs sampling
scheme
\[
z_{jl}({\bf x})\mid\cdots\sim\mathsf{N}\left(\gamma_{l}({\bf x}),1\right)\mathbf{1}_{\boldsymbol{\Omega}_{l}},\;\forall l\le\mbox{min}\{s_{j}({\bf x}),L-1\},
\]
with 
\[
\boldsymbol{\Omega}_{l}=\begin{cases}
\{z_{jl}({\bf x})<0\}, & \mbox{if }l<s_{j}({\bf x}),\\
\{z_{jl}({\bf x})\ge0\}, & \mbox{if }l=s_{j}({\bf x})<L
\end{cases}
\]
where $\mathsf{N}\left(\cdot\right){\bf 1}_{\Omega}$ denotes a normal
distribution truncated to the set $\Omega$. 

\subsubsection{Gibbs Sampling for Weight Parameters: Latent Processes}

The latent process $\left\{ \gamma_{l}\left({\bf x}\right),\forall{\bf x}\in{\cal X}\right\} $
is built on parameters $\Theta=$\{$\left\{ \theta_{l}^{1}\right\} $,
$\left\{ \theta_{a}^{2}\right\} $, $\left\{ \theta_{r}^{3}\right\} $,
$\left\{ \theta_{\left(a,r\right)}^{4}\right\} $, $\left\{ \theta_{m}^{5}\right\} $,
$\left\{ \theta_{leg}^{6}\right\} $, $\left\{ \theta_{\left(a,leg\right)}^{7}\right\} $,
$\boldsymbol{\theta}^{8}$, $\boldsymbol{\theta}^{9}$, $\boldsymbol{\theta}^{10}$,
$\boldsymbol{\theta}^{11}$, $\boldsymbol{\theta}^{12}$\} and hyper-parameters
$\Upsilon$ = \{$\epsilon^{i}$, $\forall i$ = =1, 2, $\cdots$,
7\}. The distribution of $\Theta$ and $\Upsilon$, conditional on
the augmented data, is given by

\noindent 
\[
p(\Theta,\Upsilon\mid\cdots)\propto\left[\prod_{{\bf x},j}p\left({\bf z}_{j}({\bf x})\mid\boldsymbol{\gamma}_{j}({\bf x})\right)\right]p(\Theta)p\left(\Upsilon\right)
\]
where $p\left(\Theta\right)$ is the prior distribution of $\Theta$
and $p\left(\Upsilon\right)$ is the prior distribution of $\Upsilon$,
and $j=1,\cdots,n({\bf x})$. The posterior sampling can be easily
implemented by taking advantage of the normal priors we choose. Due
to similarities of the Gibbs sampling schemes for $\Theta$ and $\Upsilon$,
here we only give updating schemes for two examples: one for coefficients
$\left\{ \theta_{l}^{1}\right\} _{l=1}^{L}\in\Theta$ and the other
one for hyper-parameter $\epsilon^{1}\in\Upsilon$. 
\begin{enumerate}
\item For $\theta_{l}^{1}$ ($l=1,2,\cdots,L$), the posterior Gibbs sampling
follows normal distribution given by 
\begin{eqnarray*}
\theta_{l}^{1}\mid\cdots & \propto & \mathsf{N}\left(\mu_{\theta_{l}^{1}},\phi_{\theta_{l}^{1}}\right)
\end{eqnarray*}
where $\mu_{\theta_{l}^{1}}=\left\{ \Phi^{-1}\left(\frac{1}{L-l+1}\right)+\sum_{{\bf x}\in{\cal X}}\sum_{j}\left[z_{jl}\left({\bf x}\right)-\Delta_{jl}\left({\bf x}\right)\right]{\bf 1}\left(s_{j}\left({\bf x}\right)\ge l\right)\right\} /\left(n_{l}+1\right)$,
$\phi_{\theta_{l}^{1}}=\left(n_{l}+\epsilon^{1}\right)/\left(n_{l}+1\right)$,
$n_{l}=\sum_{{\bf x}\in{\cal X}}\sum_{j}{\bf 1}\left(s_{j}\left({\bf x}\right)\ge l\right)$
and $\Delta_{jl}\left({\bf x}\right)=\left(\gamma_{j}\left({\bf x}\right)-\theta_{l}^{1}\right){\bf 1}\left(s_{j}\left({\bf x}\right)\ge l\right)$. 
\item For $\epsilon^{1}$ the posterior Gibbs sampling follows Gamma distribution
given by
\begin{eqnarray*}
\epsilon^{1}\mid\cdots & \propto & \mathsf{G}\left(c_{1}+\frac{L}{2},d_{1}+\frac{\sum_{l=1}^{L}\theta_{l}^{1}\cdot\theta_{l}^{1}}{2}\right)
\end{eqnarray*}
 
\end{enumerate}
In the case $L=\infty$, we can easily extend this algorithm to generate
a slice sampler, as discussed in \citet{papaspiliopoulos_note_2008}.
Alternatively, the results in \citet{rodriguez_nonparametric_2011}
suggest that a finite PSBP with a large number of components (30 \textendash{}
40, depending on the value of $\mu$) can be used instead (\citealt{ishwaran_dirichlet_2002}).
So we use $L=50$ as the number of components in this paper; this
provides a conservative upper bound as many of these components may
not be utilized.

In general, in the conditional method, the Markov chain Monte Carlo
algorithm has to explore multimodal posterior distributions. Therefore,
we need to add label-switching moves, which assist the algorithm in
jumping across modes. This is particularly important for large data
sets, where the modes are separated by areas of negligible probability.
We use the framework developed in \citet{papaspiliopoulos_retrospective_2008}
to design our label switching moves. These label switching moves greatly
improved the convergence of the chain. 

\subsubsection{Label Switching Moves}

The label switching moves with infinite mixture models are listed
as follows:
\begin{enumerate}
\item From $1,2,\dots,L$ choose two elements $l_{1}$ and $l_{2}$ uniformly
at random and change their labels with probability 
\[
\min\left(1,\Pi_{{\bf x}\in{\cal X}}\left(\frac{\omega_{l_{1}}\left({\bf x}\right)}{\omega_{l_{2}}\left({\bf x}\right)}\right)^{n_{l_{2}}\left({\bf x}\right)-n_{l_{1}}\left({\bf x}\right)}\right)
\]
where $n_{l}\left({\bf x}\right)=\sum_{j}s_{j}\left({\bf x}\right)=l$
($j=1,\cdots,n\left({\bf x}\right)$)
\item Sample a label $l$ uniformly from $1,2,\dots,L-1$ and propose to
swap the labels $l$, $l+1$ and corresponding stick-breaking weights
$\gamma_{l}$, $\gamma_{l+1}$ with probability 
\[
\min\left(1,F\times\Pi_{{\bf x}\in{\cal X}}\frac{\left(1-\Phi\left(\gamma_{l+1}\left({\bf x}\right)\right)\right)^{n_{l}\left({\bf x}\right)}}{\left(1-\Phi\left(\gamma_{l}\left({\bf x}\right)\right)\right)^{n_{l+1}\left({\bf x}\right)}}\right)
\]
where 
\[
F=\frac{\mathsf{N}\left(\theta_{l}^{1}\mid\Phi^{-1}\left(\frac{1}{L-l}\right),1\right)\cdot\mathsf{N}\left(\theta_{l+1}^{1}\mid\Phi^{-1}\left(\frac{1}{L-l+1}\right),1\right)}{\mathsf{N}\left(\theta_{l}^{1}\mid\Phi^{-1}\left(\frac{1}{L-l+1}\right),1\right)\cdot\mathsf{N}\left(\theta_{l+1}^{1}\mid\Phi^{-1}\left(\frac{1}{L-l}\right),1\right)}
\]
is the change of prior probability since the prior of $\theta^{1}$
is not symmetric. 
\end{enumerate}
Label switching moves for finite mixture models are listed as follows:
\begin{enumerate}
\item Sample a label $l$ uniformly from $1,2,\dots,L-1$ and propose to
swap the labels $l$, $l+1$ and corresponding stick-breaking weights
$\gamma_{l}$, $\gamma_{l+1}$ with probability 
\[
\min\left(1,F\times\Pi_{{\bf x}\in{\cal X}}\frac{\left(1-\Phi\left(\gamma_{l+1}\left({\bf x}\right)\right)\right)^{n_{l}\left({\bf x}\right)}}{\left(1-\Phi\left(\gamma_{l}\left({\bf x}\right)\right)\right)^{n_{l+1}\left({\bf x}\right)}}\right),\;\mbox{if }l\le L-2
\]
where 
\[
F=\frac{f\left(\alpha_{l}\mid\Phi^{-1}\left(\frac{1}{L-l}\right),1\right)f\left(\alpha_{l+1}\mid\Phi^{-1}\left(\frac{1}{L-l+1}\right),1\right)}{f\left(\alpha_{l}\mid\Phi^{-1}\left(\frac{1}{L-l+1}\right),1\right)f\left(\alpha_{l+1}\mid\Phi^{-1}\left(\frac{1}{L-l}\right),1\right)}
\]
is the change of prior probability and $f(\cdot\mid\mu,\phi)$ is
the probability density function of $\mathsf{N}\left(\cdot\mid\mu,\phi\right)$.
If $l=L-1$, the Metropolis-Hasting probability is: 
\[
\min\left(1,\Pi_{{\bf x}\in{\cal X}}\left[\frac{\Phi\left(\gamma_{l}\left({\bf x}\right)\right)}{1-\Phi\left(\gamma_{l}\left({\bf x}\right)\right)}\right]^{n_{l+1}\left({\bf x}\right)-n_{l}\left({\bf x}\right)}\right),\;\mbox{if }l=L-1
\]
\end{enumerate}

\subsection{Prior Elicitation}

First, we consider eliciting hyper-parameters $\left\{ \zeta_{\mu_{l}}\right\} _{l=1}^{L}$
and $\left\{ \xi_{\mu_{l}}\right\} _{l=1}^{L}$, corresponding to
the location of the Normal components, and $a_{\phi}$ and $b_{\phi}$,
corresponding to their precisions. These hyper-parameters need to
be chosen to ensure that the mixture spans the expected range of observed
values with high probability. In our case, we have all prior means
$\left\{ \zeta_{\mu_{l}}\right\} _{l=1}^{L}$ equal to the global
mean (or global median) of all observations, -2.64, and set all $\left\{ 1/\xi_{\mu_{l}}\right\} _{l=1}^{L}$
equal to half the range of the observed data (a rough estimate of
dispersion), 189.6. Sensitivity was assessed by halving and doubling
the values of $\xi_{\mu_{l}}$. Under a similar argument, $a_{\phi}$
and $b_{\phi}$ should be chosen so that $E\left(1/\phi_{l}\right)=b_{\phi}/\left(a_{\phi}-1\right)$
is also around half the range of the observations, so we choose $a_{\phi}=1.25,\;b_{\phi}=47.5$.
In every scenario we have employed proper priors, as weakly informative
proper priors lead to improved performance and improper priors can
lead to paradoxical behavior in mixture models, similar to the well
known Bartlett-Lindley paradox in Bayesian model selection. 

Next, we consider the prior structure on the weights $\omega_{l}\left({\bf x}\right)$.
As discussed above, the use of a continuation ratio Probit model along
with normal priors for the transformed weights is convenient, as it
greatly simplifies implementation of the model. In particular, the
transformed mixture weights $\{\gamma_{l}\left({\bf x}\right)\}$
can be sampled by the algorithm shown in ${\cal x}$3.2.3 above from
conditionally normal distributions. Hyper-parameter choice is also
simplified. A common assumption of basic mixture models for $i.i.d.$
data is that all components have the same probability a priori. In
the current context in which mixture weights are predictor dependent,
a similar constraint can be imposed on the baseline conditional distribution
by setting $E(\theta_{l}^{1})=\Phi^{-1}\left(1/\left(L-l+1\right)\right)$.
Since we build a hierarchy above heterogeneity parameters to allow
information borrowing, the variance of $\theta^{i}$ ($i=1,2,\cdots,7$),
is controlled by the distribution of hyper parameters $\epsilon^{i}$.
In order to make sure the continuation ratio $\Phi(\gamma_{l}\left({\bf x}\right))$
is between 0.001 and 0.998 with 0.99 probability, we would expect
$\text{Var}\left(\theta^{i}\right)\approx1$. Smaller values for $V(\theta^{i})$
lead to strong restrictions on the set of weights, discouraging small
ones (especially for the first few components in the mixture). On
the other hand, larger variances can adversely affect model selection.
For the hyper parameter $\epsilon^{i}$ of $\theta^{i}$, in order
to make sure $\text{Var}\left(\theta^{i}\right)\approx1$, we let
$c_{i}=6,\;d_{i}=5$ so that $\text{E}\left(1/\epsilon^{i}\right)=1$.
This yields a prior sample size of 6, which gives some stability while
very small restrictions. 

\subsection{Implementation}

The data were analyzed using the models described in ${\cal x}$3.1.
Fifty mixture components were judged sufficient to flexibly characterize
changes in the density across predictors, while limiting the risk
of over-fitting. Inferences were robust in our sensitivity analysis
for $L$ ranging between 40 and 60, but the quality of the fit, as
assessed through the plots described in ${\cal x}$3 and ${\cal x}$5,
was compromised for $L<40$. 

The Gibbs samplers were run for 100,000 iterations following a 70,000
iteration burn-in period. Code was implemented in Matlab, and the
longest running time was 118h on a 2.96-GHz Intel Xeon E5-2690 computer
with 32 cores. This run time could be dramatically reduced by improving
code efficiency and relying on recent developments in scalable Bayesian
computation, but we preferred to use standard Gibbs sampling instead
of new and less well established computational methods. Examination
of diagnostic plots showed adequate mixing and no evidence of lack
of convergence. 

\subsection{Cross Validation for Variable Selection}

To balance computation time and accuracy, we use 3-fold cross validation
based on predictive log likelihood. Specifically, we partition the
original data into three equal sized subsamples, with two of the subsamples
used as training data for parameter estimation. Then, the estimated
parameters are used to calculate the log likelihood, as shown in Equation
(\ref{eq: likelihood}), of the left out subsample, also known as
the validation data. This process is repeated for 3 times so that
every subsample is used as validation data once. The reason why we
choose to calculate the log likelihood of the validation data that
log likelihood is a strictly proper scoring rule for density forecasts
as in our study, as explained in \citet{gneiting_strictly_2007}.
We compare the cross validation value of many models and list 10 models
in Table \ref{table: CV}.
\begin{table}[tbh]
\centering{}{\small{}\caption{Cross validation for model comparison}
\label{table: CV}}%
\begin{tabular}{l|>{\raggedright}p{3.2cm}>{\raggedright}p{1.3cm}|l|>{\raggedright}p{4cm}>{\raggedright}p{1.3cm}}
\hline 
 & {\small{}Model} & {\small{}-$LL$} &  & {\small{}Model} & {\small{}-$LL$}\tabularnewline
\hline 
{\small{}1} & {\small{}$\Xi$} & {\small{}324235} & {\small{}6} & {\small{}$\Xi-\theta_{\left(a,leg\right)}^{7}-\theta_{\left(a,r\right)}^{4}$} & {\small{}326687}\tabularnewline
{\small{}2} & {\small{}$\Xi-\theta_{\left(a,leg\right)}^{7}$} & {\small{}318101} & {\small{}7} & {\small{}$\Xi-\theta_{\left(a,leg\right)}^{7}-\theta_{leg}^{6}-\theta^{11}$} & {\small{}319515}\tabularnewline
{\small{}3} & {\small{}$\Xi-\theta_{\left(a,leg\right)}^{7}-\theta_{leg}^{6}$} & {\small{}320239} & {\small{}8} & {\small{}$\Xi-\theta_{\left(a,leg\right)}^{7}-\theta_{m}^{5}-\theta^{11}$} & {\small{}318684}\tabularnewline
{\small{}4} & {\small{}$\Xi-\theta_{\left(a,leg\right)}^{7}-\boldsymbol{\theta}^{11}$} & {\small{}317894} & {\small{}9} & {\small{}$\Xi-\theta_{\left(a,leg\right)}^{7}-\theta_{leg}^{6}-\theta_{m}^{5}$} & {\small{}327174}\tabularnewline
{\small{}5} & {\small{}$\Xi-\theta_{\left(a,leg\right)}^{7}-\theta_{m}^{5}$} & {\small{}318894} & {\small{}10} & {\small{}$\Xi-\theta_{\left(a,leg\right)}^{7}-\theta_{\left(a,r\right)}^{4}-\theta_{a}^{2}$} & {\small{}328721}\tabularnewline
\hline 
\end{tabular}
\end{table}

For each model, the log-likelihood of its three subsamples is calculated
by averaging over 10,000 posterior samples with the first 10,000 posterior
samples dropped as burn-in. In Table \ref{table: CV} we use $\Xi$
to indicate the full model, as shown in Equation (\ref{eq: full model}),
and use ``$-$'' to indicate dropping certain predictors. We use
$LL$ to indicate sum of the three subsample log-likelihood for each
model. Since we are comparing $-LL$ in Table \ref{table: CV}, smaller
values suggest stronger predictive capability. So we choose model
(4) in the Table \ref{table: CV}, which is equivalent to Equation
(\ref{eq: used model - appendix}).
\begin{eqnarray}
\gamma_{l}({\bf x}) & = & \theta_{l}^{1}+\theta_{a}^{2}+\theta_{r}^{3}+\theta_{\left(a,r\right)}^{4}+\theta_{m}^{5}+\theta_{leg}^{6}+f_{1}\left(dev_{start}\mid\boldsymbol{\theta}^{8}\right)+f_{2}\left(dur\mid\boldsymbol{\theta}^{9}\right)+f_{3}\left(\mbox{log}\left(wgt\right)\mid\boldsymbol{\theta}^{10}\right)\label{eq: used model - appendix}
\end{eqnarray}

Since there are 11 kinds of predictors in the full model (i.e., Equation
(\ref{eq: full model})), which are $2^{11}=2048$ different possible
variable combinations, it is impossible for us to compare every model
in a relatively short research time frame. We choose to use a backward
fitting process. Specifically, we start with the full model, then
every time we drop one predictor and observe how the $-LL$ change.
If the $-LL$ decreases, then we keep this predictor removed, otherwise,
we add this predictor back. For example, from model 1 to 2 in Table
\ref{table: CV}, while the interaction term of airline and legs is
dropped the $-LL$ drops, so we remove the interaction of airline
and leg. From model 2 to 3, we further drop predictor ``legs'',
however $-LL$ increases, as a result, we add predictor ``legs''
back. We tested many more models than those listed in Table \ref{table: CV}
with the best model shown in Equation \eqref{eq: used model - appendix}.
The 10 models listed in Table \ref{table: CV} are shown for illustration. 

As we have mentioned in $\mathsection$3.2, $f_{1}$, $f_{2}$ , $f_{3}$
in Equation \eqref{eq: used model - appendix} ($f_{4}$ in Equation
\eqref{eq: full model} is similar) are spline functions expressed
as a linear combination of B-splines, or basis spline, of degree 4.
The usefulness of B-spline lies in the fact that any spline function
of order $n$ on a given set of knots can be expressed as a unique
linear combination of B-splines, hence the name basis spline. The
knots of $f_{1}$, $f_{2}$, $f_{3}$ and $f_{4}$ have been listed
in $\mathsection$3.2, the knots are chosen based on both the meaning
in reality (e.g., for better interpretation) and to make sure the
data amount in each range of support is roughly comparable. The final
expression of the basis spline of a higher order (3 or higher) can
be very complicated, so here I only list the recursion formula: 
\begin{eqnarray*}
B_{i,1}(x) & = & \begin{cases}
1 & \mbox{if }t_{i}\le x<t_{i+1}\\
0 & \mbox{otherwise}
\end{cases}\\
B_{i,k}(x) & = & \frac{x-t_{i}}{t_{i+k-1}-t_{i}}B_{i,k-1}(x)+\frac{t_{i+k}-x}{t_{i+k}-t_{i+1}}B_{i+1,k-1}(x)
\end{eqnarray*}
where vector ${\bf t}$ is the knots and $k$ is the order of B-spline.
We have tried different orders of B-splines (i.e., $3\sim6$) with
slightly changed knots, however, the predictive power is almost the
same. So we decide to use a B-spline of order 4. 

\subsection{Model Checking and Comparison}

In this section, $\mathsection$B.5.1 and $\mathsection$B.5.2 are
on model checking while $\mathsection$B.5.3 is about model comparison. 

\subsubsection{Posterior Predictive Checks}

We first use posterior predictive checks (PPC), devised in \citet{rubin_bayesianly_1984}
and expanded in \citet{gelman_posterior_1996}. PPC provide a popular
approach for goodness-of-fit assessments of Bayesian models. In implementing
PPC, one first generates multiple data sets of the same structure
as the observed dataset from the posterior predictive distribution.
This is done by generating parameters from the posterior distribution,
plugging these parameters into the likeihood, and then sampling new
data from this likelihood. If the model does not fit the data well,
data generated from the posterior predictive distribution will deviate
systematically from the observed data. As an illustration, we implemented
a PPC check of a naive linear model (referred to as LM), a method
widely used in previous research on flight delays. The specific form
of LM is as follows: 
\begin{eqnarray}
y & \sim & N(\mu,\sigma^{2})\nonumber \\
\mu & = & \theta^{1}+\theta_{a}^{2}+\theta_{r}^{3}+\theta_{\left(a,r\right)}^{4}+\theta_{m}^{5}+\theta_{leg}^{6}+\theta^{8}\cdot dev_{start}+\theta^{9}\cdot dur+\theta^{10}\cdot\mbox{log}\left(wgt\right)\label{eq: OLS}
\end{eqnarray}
where $y$ represents the dependent variable transport risk. All the
other predictors $\boldsymbol{\theta}$ are the same as explained
in $\mathsection3.2$ except the original vector $\theta^{1}$ now
doesn't have the subscript $l$ for each cluster and is just a scalar
intercept. 

\begin{figure}[tbh]
\begin{centering}
\begin{minipage}[t]{0.33\columnwidth}%
\includegraphics[width=1\columnwidth]{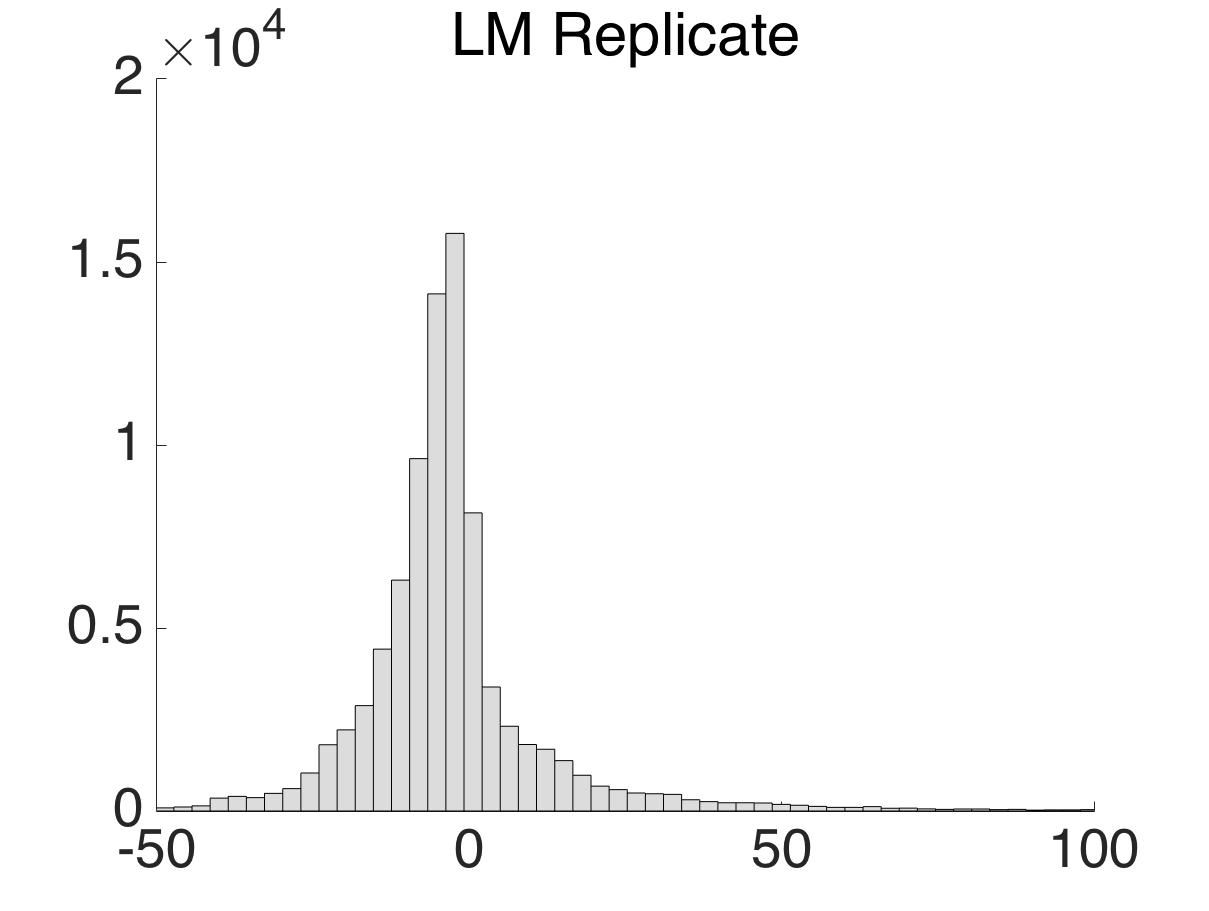}%
\end{minipage}%
\begin{minipage}[t]{0.33\columnwidth}%
\includegraphics[width=1\columnwidth]{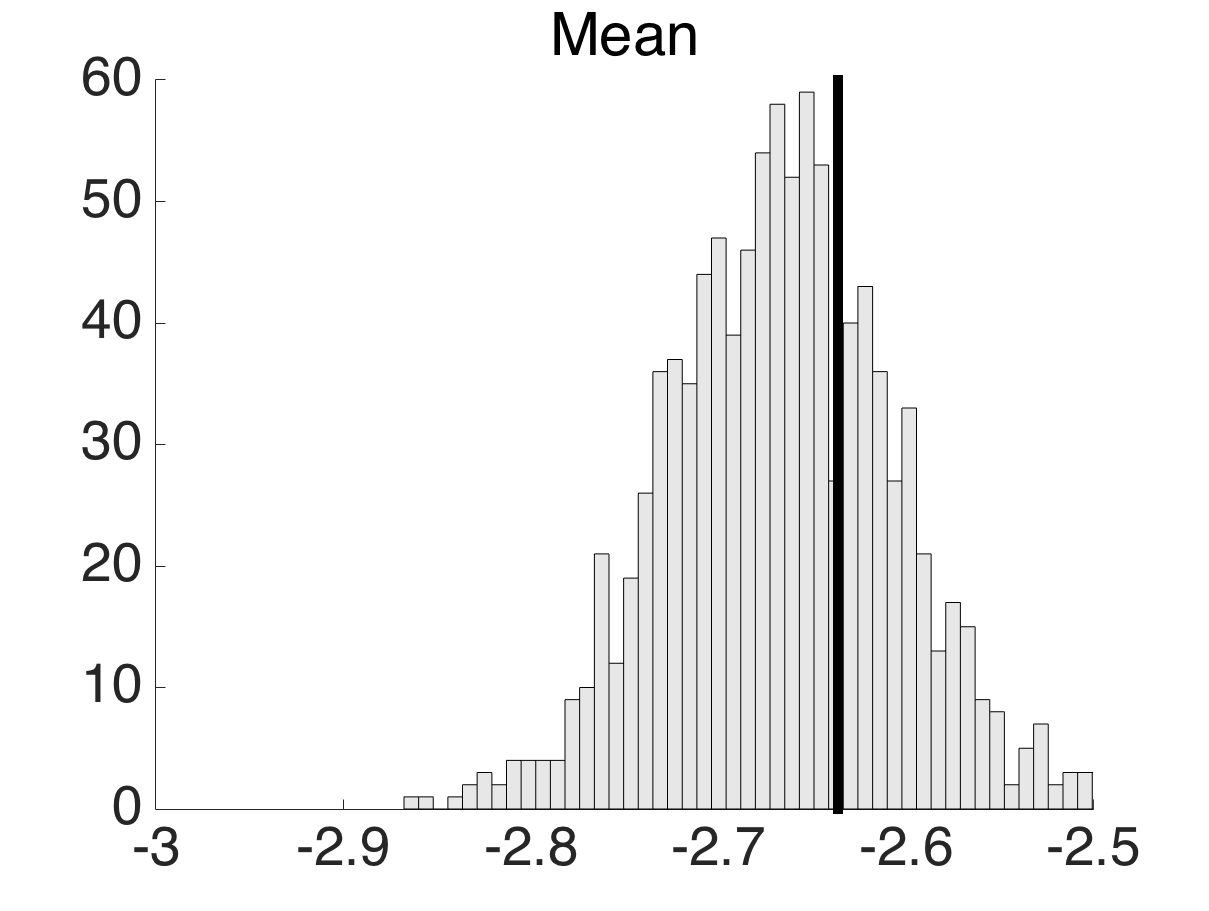}%
\end{minipage}%
\begin{minipage}[t]{0.33\columnwidth}%
\includegraphics[width=1\columnwidth]{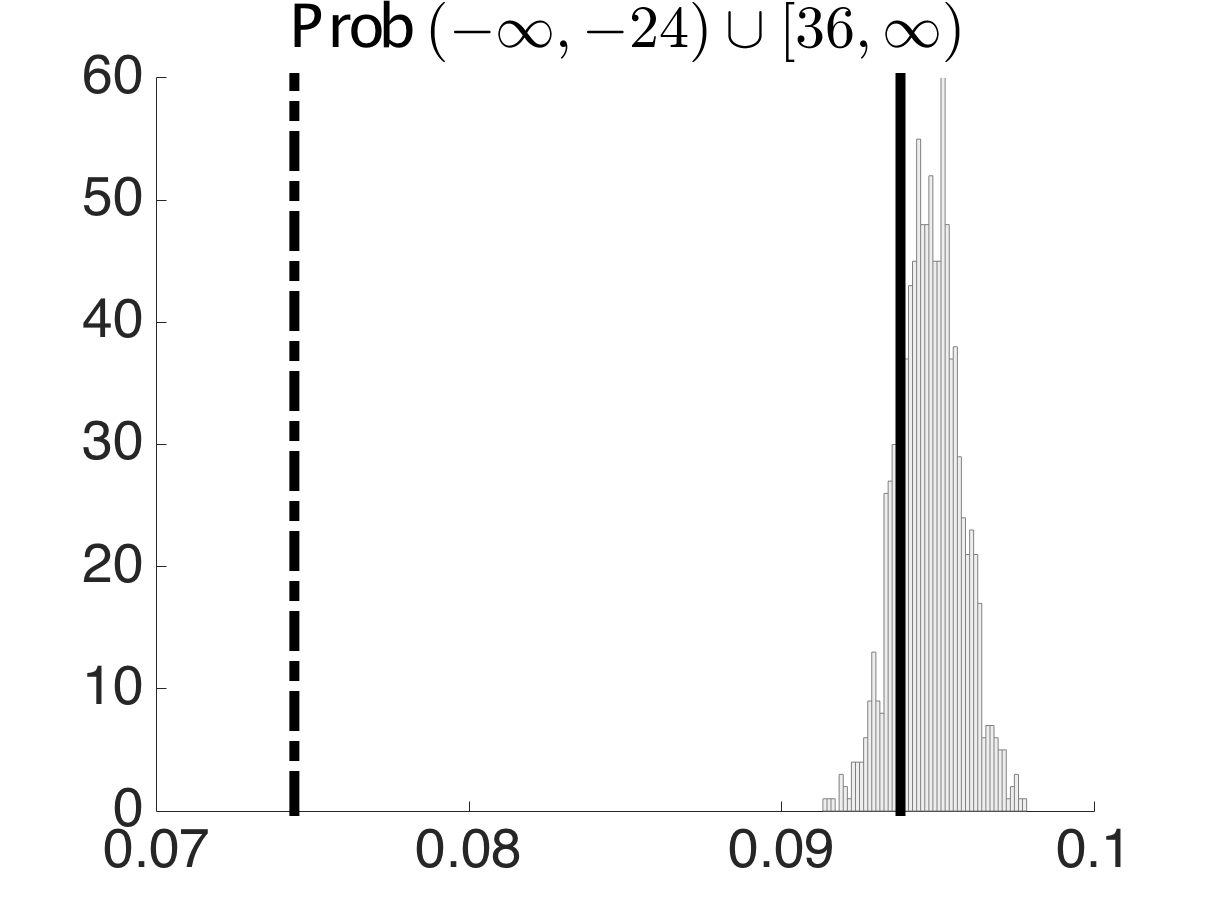}%
\end{minipage}
\par\end{centering}
\begin{centering}
\begin{minipage}[t]{0.33\columnwidth}%
\includegraphics[width=1\columnwidth]{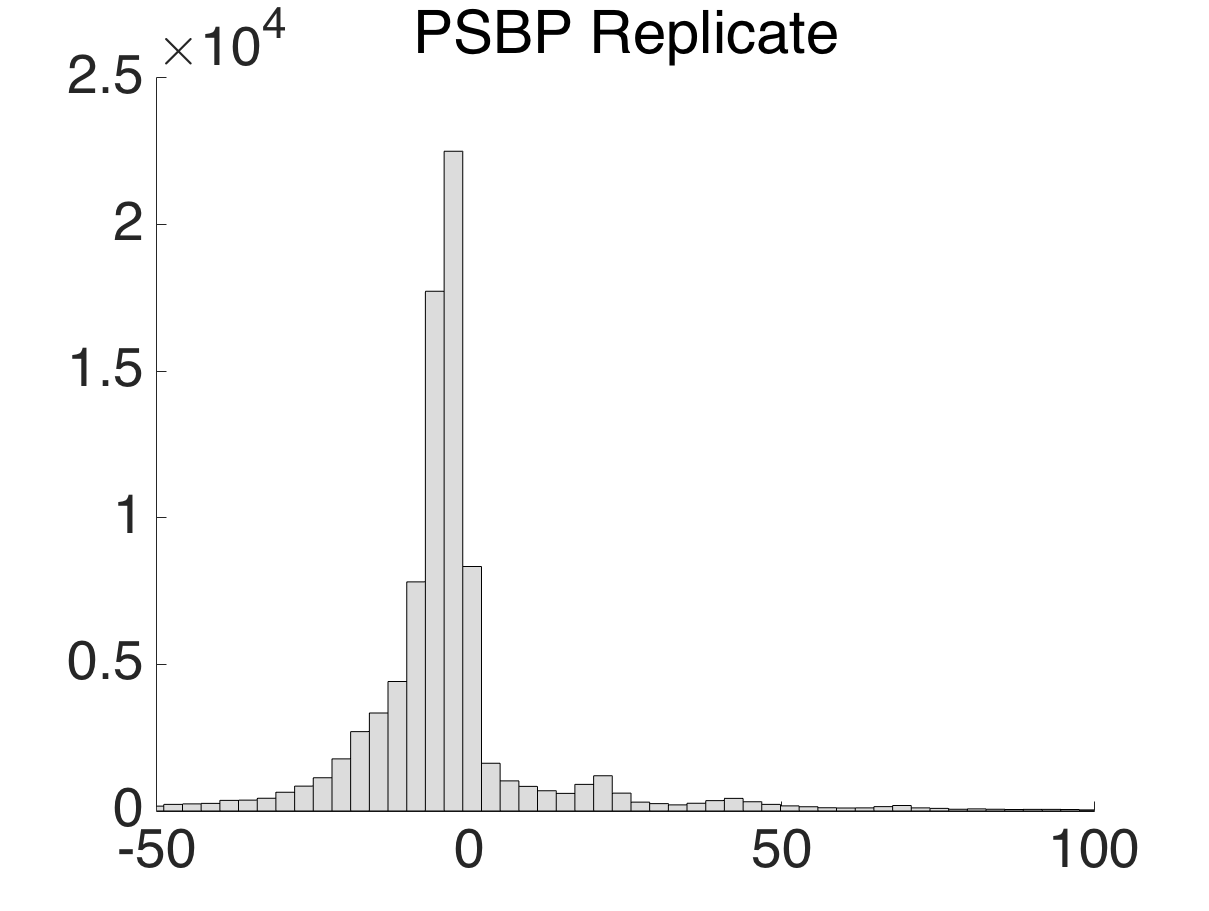}%
\end{minipage}%
\begin{minipage}[t]{0.33\columnwidth}%
\includegraphics[width=1\columnwidth]{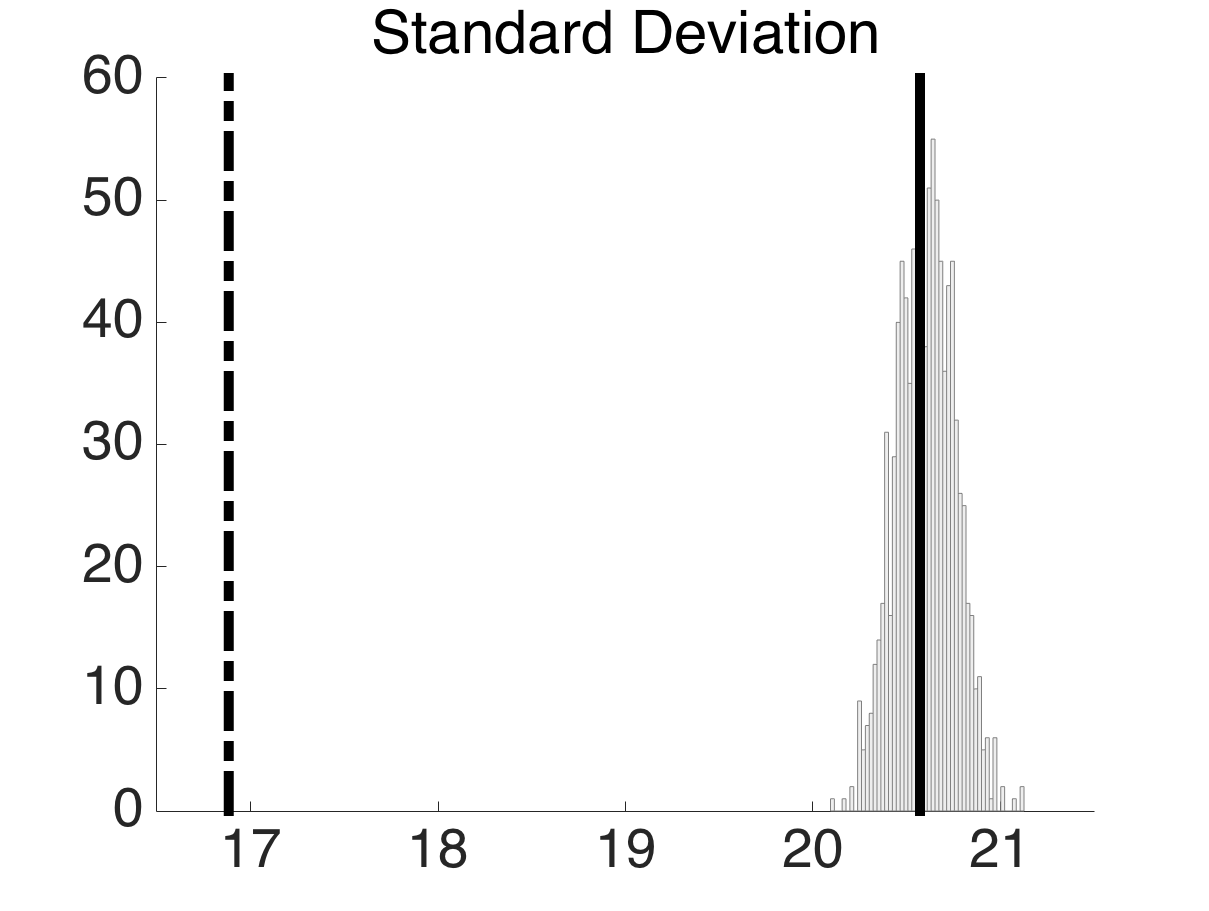}%
\end{minipage}%
\begin{minipage}[t]{0.33\columnwidth}%
\includegraphics[width=1\columnwidth]{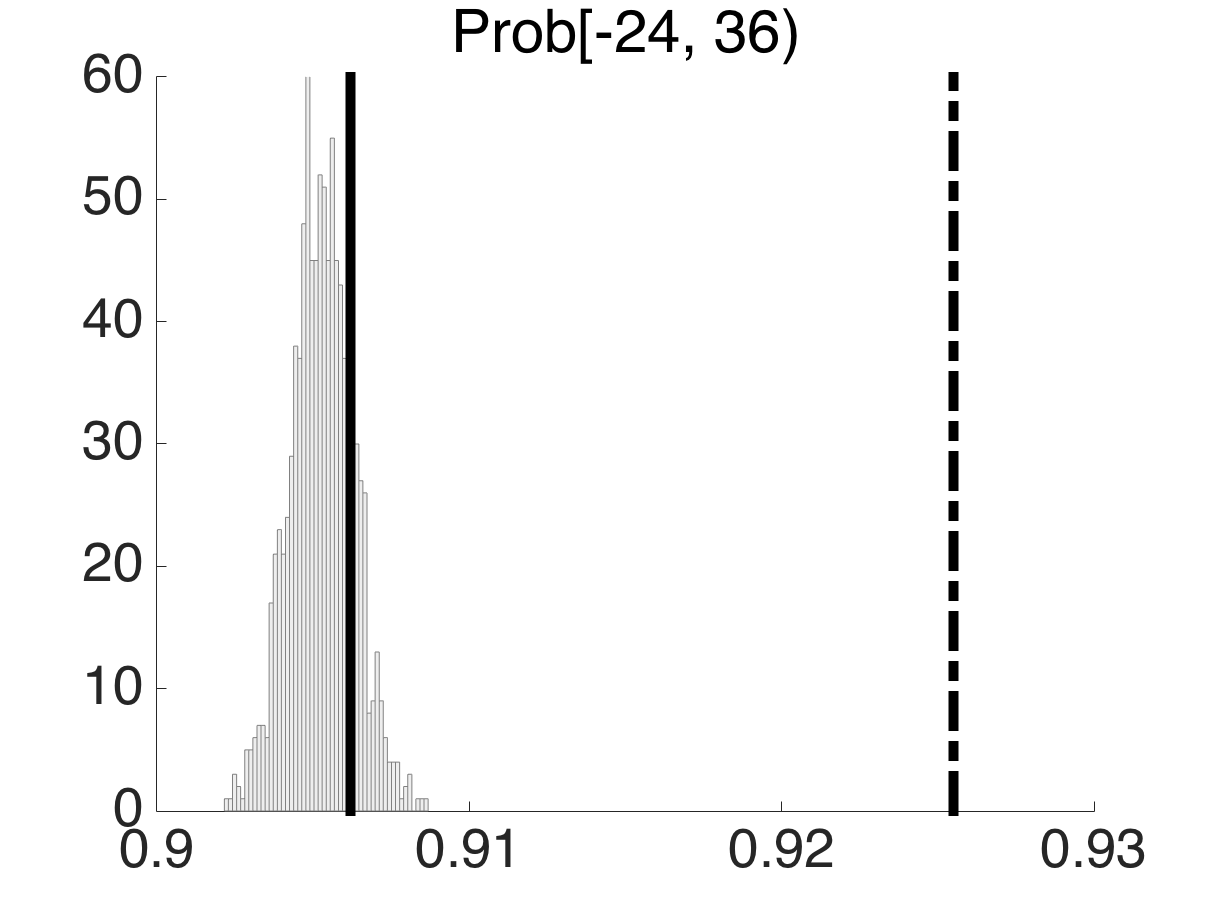}%
\end{minipage}
\par\end{centering}
\centering{}\caption{Posterior predictive model checking. In ``LM Replicate'' and ``PSBP
Replicate'', the histogram is the predicted response, transport risk.
In the other four plots, the title is the target value for posterior
predictive checking; the histogram is PSBP predictive values; the
solid line is the value calculated from the real data; the dash-dotted
line is the predictive value by LM. }
\label{fig: PPC}
\end{figure}
To follow the steps of PPC, we replicate two data sets: one predicted
by LM and one predicted by our model. We use these two replicated
data to calculate summary statistics including mean, standard deviation,
$\mathsf{Prob}\left(\hat{y}<-24\mbox{ or }\hat{y}\ge36\right)$ (i.e.,
the probability of transport disruption) and $\mathsf{Prob}\left(-24\le\hat{y}<36\right)$
(i.e., the probability of recurrent transport risks), then compare
the values to the statistics calculated using the true data. If the
model fits well, we expect the empirical values of these statistics
to not deviate significantly from the simulated distribution under
the assumed model. From Figure \ref{fig: PPC} we found that LM predicts
the mean accurately (the dashed line and solid line overlays), however
largely underestimates extreme situations like more than 24 hours
earliness or more than 36 hours delays, while overestimating recurrent
risk such as deviations between -24 to 36 hours. In addition, standard
deviation is substantially underestimated. On the other hand, the
posterior predictive statistics by PSBP closely surrounding the empirical
values. Moreover, if we just look at the empirical distribution of
the two replica data as compared to the observed data empirical distribution
(the first column in Figure \ref{fig: PPC}), we can see that the
replicated data by LM resembles the shape of the real data poorly.
Whereas, the replicated data by our model resembles the real data
very well. This clearly shows the model inadequacy of linear model
in our situation and confirms our PSBP model to fit the data well. 

\subsubsection{Visual Inspection}

We further check the model at a more granular level \textendash{}
airline-route level. In Figure \ref{fig: sample routes}, the histogram
is drawn from real data, the solid line is the predictive conditional
density by PSBP (the posterior 95$\%$ probability intervals are too
narrow to be visible in this figure), while the dashed line is predicted
by LM. PSBP captures the location and weights of peaks accurately
while linear model predicts badly. 

\subsubsection{Model Comparison}

We further formally compare our model with alternative models including
LM, generalized additive model (GAM) and flexible mixture model (Flexmix)
using predictive residuals. GAM, as shown in Equation \eqref{eq: GAM}
\begin{eqnarray}
y & \sim & N(\mu,\sigma^{2})\nonumber \\
\mu & = & \theta^{1}+\theta_{a}^{2}+\theta_{r}^{3}+\theta_{\left(a,r\right)}^{4}+\theta_{m}^{5}+\theta_{leg}^{6}+s(dev_{start}\mid\theta^{8})+s\left(dur\mid\boldsymbol{\theta}^{9}\right)+s\left(\mbox{log}\left(wgt\right)\mid\boldsymbol{\theta}^{10}\right)\label{eq: GAM}
\end{eqnarray}
generalizes LM by incorporating nonlinear forms of the continuous
predictors. Here $s(\cdot\mid\boldsymbol{\theta})$ represents the
smooth function (also the nonparametric or nonlinear form) and $\boldsymbol{\theta}$
are the parameters. We estimate the model using the ``bam'' function
in R package ``mgcv'' \citep{wood_fast_2011}, in which the estimation
of GAMs is conducted via a penalized likelihood frequentist approach.
The flexible mixture model we consider is shown as follows (we use
the same number of clusters as in PSBP mixture model):
\begin{eqnarray}
y & \sim & \sum_{l}\omega_{l}N(\mu_{l},\sigma_{l})\nonumber \\
\mu_{l} & = & \theta_{l}^{1}+\theta_{a}^{2}+\theta_{r}^{3}+\theta_{\left(a,r\right)}^{4}+\theta_{m}^{5}+\theta_{leg}^{6}+s(dev_{start}\mid\theta^{8})+s\left(dur\mid\boldsymbol{\theta}^{9}\right)+s\left(\mbox{log}\left(wgt\right)\mid\boldsymbol{\theta}^{10}\right)\label{eq: Flexmix}
\end{eqnarray}
This expression is similar to Equation \eqref{eq: likelihood}. However,
the regression part is now in the Gaussian mean rather than the mixture
weights. We estimate the model using ``flexmix'' function in R package
``flexmix'' \citep{gruen_flexmix_2008}, in which maximum likelihood
estimation is conducted via an EM algorithm. 
\begin{table}[b]
\noindent \centering{}\caption{Residual checks for model comparison}
{\small{}\label{tab: residual checks}}%
\begin{tabular}{c|>{\centering}p{3cm}>{\centering}p{3cm}>{\centering}p{3cm}>{\centering}p{3cm}}
\hline 
 & {\small{}LM} & {\small{}GAM} & {\small{}Flexmix} & {\small{}PBSP Mixture}\tabularnewline
\hline 
{\small{}I/O{*} RMSE} & {\small{}18.6/19.1} & {\small{}17.8/18.4} & {\small{}14.9/15.7} & {\small{}14.1/15.0}\tabularnewline
{\small{}I/O MAE} & {\small{}9.66/10.00} & {\small{}9.01/9.31} & {\small{}7.73/8.02} & {\small{}7.17/7.56}\tabularnewline
{\small{}I/O Log-likelihood} & {\small{}373090/387755} & {\small{}370547/385183} & {\small{}314477/325575} & {\small{}307124/317894}\tabularnewline
\hline 
\multicolumn{5}{l}{{\footnotesize{}{*} I/O: in-sample/out-of-sample}}\tabularnewline
\end{tabular}
\end{table}
Table \ref{tab: residual checks} shows the root mean squared error
(RMSE), mean absolute error (MAE) and log likelihood of the in-sample
and out-of-sample prediction of the four models. Here the out-of-sample
prediction is the average RMSE/MAE/log-likelihood value from the three
3-fold cross-validation methods explained in Appendix $\mathsection$B.4. 

From the table, the PSBP Mixture model clear beats the alternatives
in all metrics in both in-sample and out-of-sample tests. The under-performance
of LM and GAM are easy to understand, as likely arising from the lack
of ability to allow the delay distribution to shift flexibly with
predictors. Flexmix is a more realistic competitor in this respect,
but has some clear disadvantages relative to our Bayesian PSBP mixture
approach. Both models are richly parameterized and even though the
sample size is large overall, there can be sparsity in local regions
of the predictor space. Hence, maximum likelihood estimation may have
inflated errors relative to a penalization approach. The weakly informative
priors we advocate protect against overfitting and reduce mean square
error in estimating coefficients. Another advantage of PSBP is the
form of the model in which the kernels are fixed and the weights vary
with predictors. As explained in $\mathsection$3.1, the locations
of peaks of the multi-modal distribution of the dependent variable,
transport risk, are almost constant. However, the heights of the peaks
change greatly with predictors (e.g., route, airline, demand variables).
Our model can more parsimoniously account for such changes. 

In terms of the computational time, we have explained those of our
model in Appendix $\mathsection$B.3. LM, GAM and Flexmix are implemented
in R using packages listed above on a server of 128G memory. The computational
time of LM and GAM are fairly short, 3 minutes and 15 minutes respectively.
On the other hand, the estimation of Flexmix takes 135h to reach final
convergence. As the sample size and number of predictors increase,
the rate of convergence and time per iteration can both increase substantially
for standard algorithms for fitting mixture models; both frequentist
and Bayesian. We note that our code has not been optimized and we
have not attempted to use recently developed algorithms for scaling
up Bayesian computation to bigger problem sizes. We also note that
our PSBP mixture model can be implemented using a frequentist optimization
approach; although maximum likelihood estimation via the EM algorithm
would be one possibility, the number of predictors and parameters
in the model makes it important to include penalties. We do not consider
such possibilities further in this article.

\vfill{}
\clearpage{}

\section{Supplementary Material of Results}

\subsection{Model Parameter Estimation}

Table \ref{table: posterior parameters} shows the posterior mean
and 95\% probability interval of (selected) model parameters. 
\begin{table}[tbph]
\centering{}{\small{}\caption{Posterior summaries of model parameters}
}%
\begin{tabular}{>{\raggedright}p{0.5cm}>{\centering}p{2.2cm}>{\centering}p{2.2cm}>{\centering}p{2.2cm}>{\centering}p{2.2cm}>{\centering}p{2.2cm}>{\centering}p{2.2cm}}
\hline 
\multicolumn{7}{l}{{\small{}Kernel Parameters}}\tabularnewline
 & \multicolumn{2}{l}{{\small{}$\mu_{l}$ $\left(l=1,2,\cdots,50\right)$}} & \multicolumn{4}{l}{{\small{}min$\left(\mu_{l}\right)=-79.6,\qquad$ max$\left(\mu_{l}\right)=76.01$}}\tabularnewline
 & \multicolumn{2}{l}{{\small{}$\nicefrac{1}{\sqrt{\phi_{l}}}$ $\left(l=1,2,\cdots,50\right)$}} & \multicolumn{4}{l}{{\small{}min$\left(\nicefrac{1}{\sqrt{\phi_{l}}}\right)=0.72,\qquad$
max$\left(\nicefrac{1}{\sqrt{\phi_{l}}}\right)=84.4$}}\tabularnewline
\hline 
\multicolumn{7}{l}{{\small{}Parameters in Weight $\gamma$}}\tabularnewline
\hline 
 & \multicolumn{2}{l}{{\small{}Category Predictors}} &  &  &  & \tabularnewline
 & \multicolumn{2}{l}{{\small{}$\theta_{l}^{1}$ $\left(l=1,2,\cdots,49\right)$}} & \multicolumn{4}{l}{{\small{}min$\left(\theta_{l}^{1}\right)=-10.9,\qquad$max$\left(\theta_{l}^{1}\right)=6.74$}}\tabularnewline
\cline{2-7} 
 & \multicolumn{2}{l}{{\small{}$\theta_{a}^{2}$ $\left(a=1,2,\cdots,20\right)$}} &  &  &  & \tabularnewline
 & {\small{}$\theta_{1}^{2}$ A1} & {\small{}$\theta_{2}^{2}$ A2} & {\small{}$\theta_{3}^{2}$ A3} & {\small{}$\theta_{4}^{2}$ A4} & {\small{}$\theta_{5}^{2}$ A5} & \tabularnewline
 & {\small{}0}{\small \par}

{\small{}(0, 0)} & {\small{}0.03}{\small \par}

{\small{}(-0.40, 0.61)} & {\small{}-5.27}{\small \par}

{\small{}(-5.86, -4.83)} & {\small{}5.15}{\small \par}

{\small{}(4.31, 6.11)} & {\small{}3.09}{\small \par}

{\small{}(2.89, 3.26)} & \tabularnewline
 & {\small{}$\theta_{6}^{2}$ A6 } & {\small{}$\theta_{7}^{2}$ A7 } & {\small{}$\theta_{8}^{2}$ A8 } & {\small{}$\theta_{9}^{2}$ A9} & {\small{}$\theta_{10}^{2}$ A10} & \tabularnewline
 & {\small{}1.16}{\small \par}

{\small{}(0.84, 1.53)} & {\small{}8.53}{\small \par}

{\small{}(8.19, 8.91)} & {\small{}2.54}{\small \par}

{\small{}(2.01, 2.98)} & {\small{}-0.82}{\small \par}

{\small{}(-1.22, -0.40)} & {\small{}2.90}{\small \par}

{\small{}(2.23, 3.64)} & \tabularnewline
 & {\small{}$\theta_{11}^{2}$ A11} & {\small{}$\theta_{12}^{2}$ A12} & {\small{}$\theta_{13}^{2}$ A13} & {\small{}$\theta_{14}^{2}$ A14} & {\small{}$\theta_{15}^{2}$ A15} & \tabularnewline
 & {\small{}-3.35}{\small \par}

{\small{}(-4.02, -2.74)} & {\small{}5.74}{\small \par}

{\small{}(5.44, 5.97)} & {\small{}-2.96}{\small \par}

{\small{}(-3.19, -2.67)} & {\small{}2.74}{\small \par}

{\small{}(2.27, 2.98)} & {\small{}-2.82}{\small \par}

{\small{}(-3.26, -2.36)} & \tabularnewline
 & {\small{}$\theta_{16}^{2}$ A16} & {\small{}$\theta_{17}^{2}$ A17} & {\small{}$\theta_{18}^{2}$ A18} & {\small{}$\theta_{19}^{2}$ A19} & {\small{}$\theta_{20}^{2}$ A20} & \tabularnewline
 & {\small{}4.95}{\small \par}

{\small{}(4.35, 5.50)} & {\small{}-3.16}{\small \par}

{\small{}(-3.50, -2.76)} & {\small{}-5.36}{\small \par}

{\small{}(-6.59, -4.41)} & {\small{}6.23}{\small \par}

{\small{}(5.79, 6.67)} & {\small{}-2.34}{\small \par}

{\small{}(-2.58, -2.12)} & \tabularnewline
\cline{2-7} 
 & \multicolumn{2}{l}{{\small{}$\theta_{leg}^{5}$ ($leg=2,3$)}} &  &  &  & \tabularnewline
 & {\small{}$\theta_{2}^{5}$} & {\small{}$\theta_{3}^{5}$} &  &  &  & \tabularnewline
 & {\small{}-0.29}{\small \par}

{\small{}(-0.38, -0.21)} & {\small{}-0.34}{\small \par}

{\small{}(-0.47, -0.21)} &  &  &  & \tabularnewline
\cline{2-7} 
 & \multicolumn{2}{l}{{\small{}Hyper-parameters}} &  &  &  & \tabularnewline
 & {\small{}$1/\sqrt{\epsilon^{1}}$} & {\small{}$1/\sqrt{\epsilon^{2}}$} & {\small{}$1/\sqrt{\epsilon^{3}}$} & {\small{}$1/\sqrt{\epsilon^{4}}$} & {\small{}$1/\sqrt{\epsilon^{5}}$} & {\small{}$1/\sqrt{\epsilon^{6}}$}\tabularnewline
 & {\small{}4.86}{\small \par}

{\small{}(3.98, 5.93)} & {\small{}3.39}{\small \par}

{\small{}(2.62, 4.44)} & {\small{}6.26}{\small \par}

{\small{}(5.86, 6.63)} & {\small{}7.02}{\small \par}

{\small{}(6.46, 7.60)} & {\small{}0.64}{\small \par}

{\small{}(0.46, 0.90)} & {\small{}0.74}{\small \par}

{\small{}(0.51, 1.10)}\tabularnewline
\hline 
\end{tabular}{\small{}\label{table: posterior parameters}}
\end{table}

\subsection{Application: Baseline Distributions}

In Figure \ref{fig: airline reference cont.} and Figure \ref{fig: airline reference performance }
are the baseline risk distributions of the remaining 17 airlines.
\begin{figure}[tb]
\begin{centering}
\begin{minipage}[t]{0.33\columnwidth}%
\includegraphics[width=1\columnwidth]{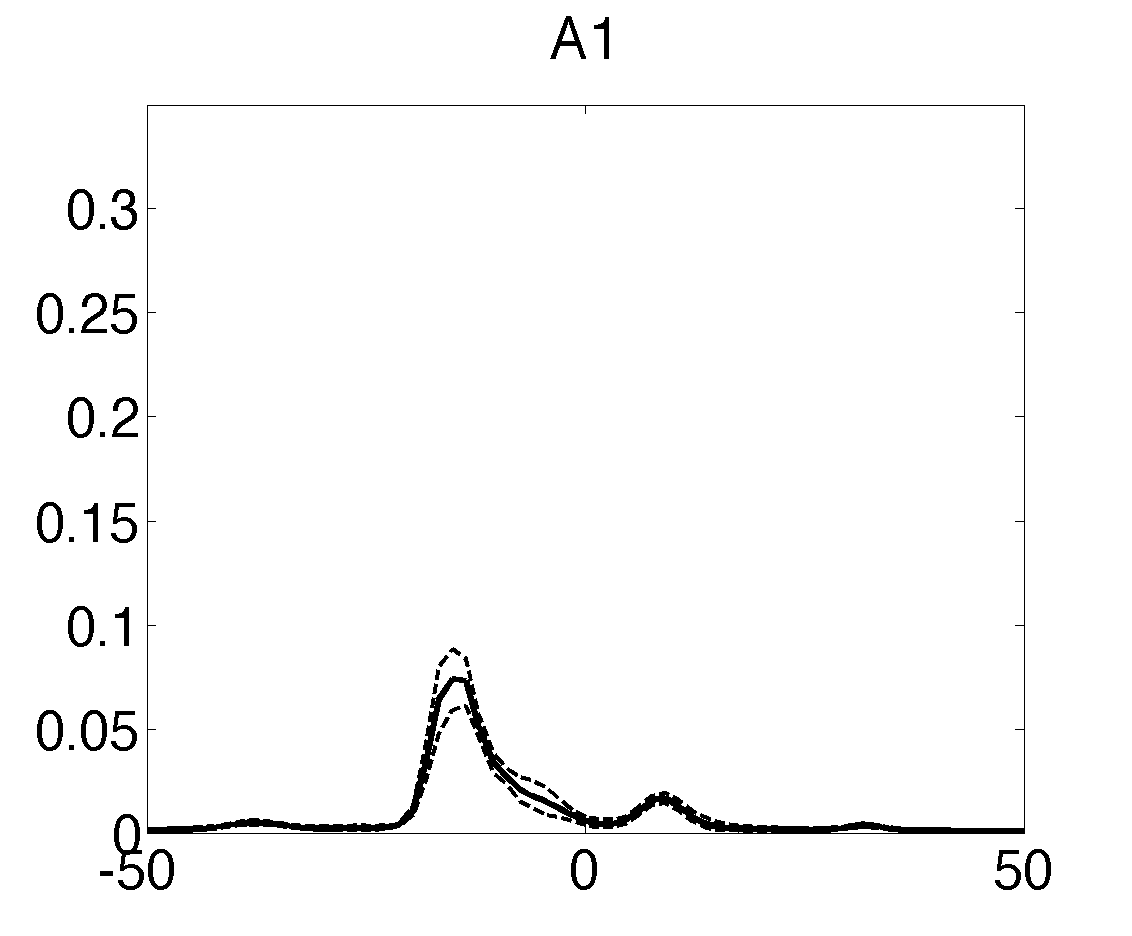}%
\end{minipage}%
\begin{minipage}[t]{0.33\columnwidth}%
\includegraphics[width=1\columnwidth]{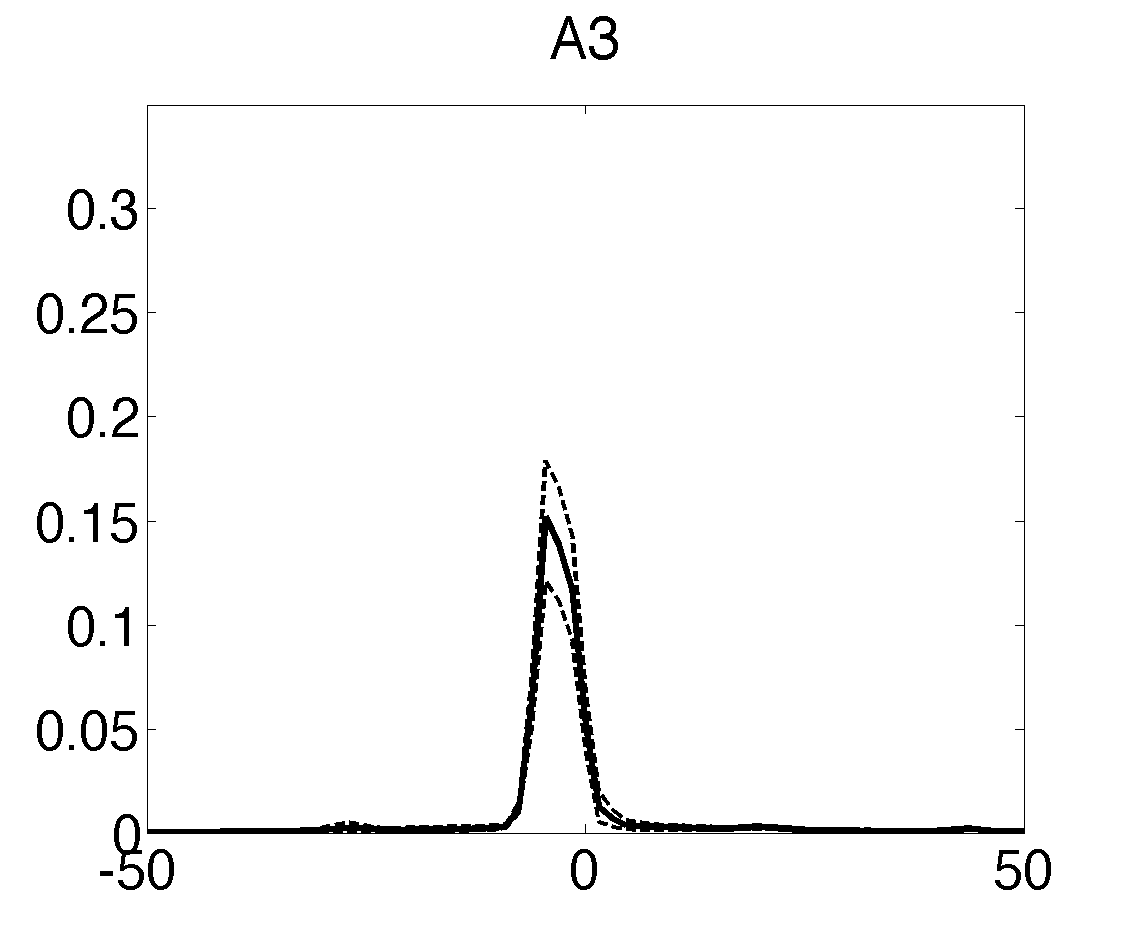}%
\end{minipage}%
\begin{minipage}[t]{0.33\columnwidth}%
\includegraphics[width=1\columnwidth]{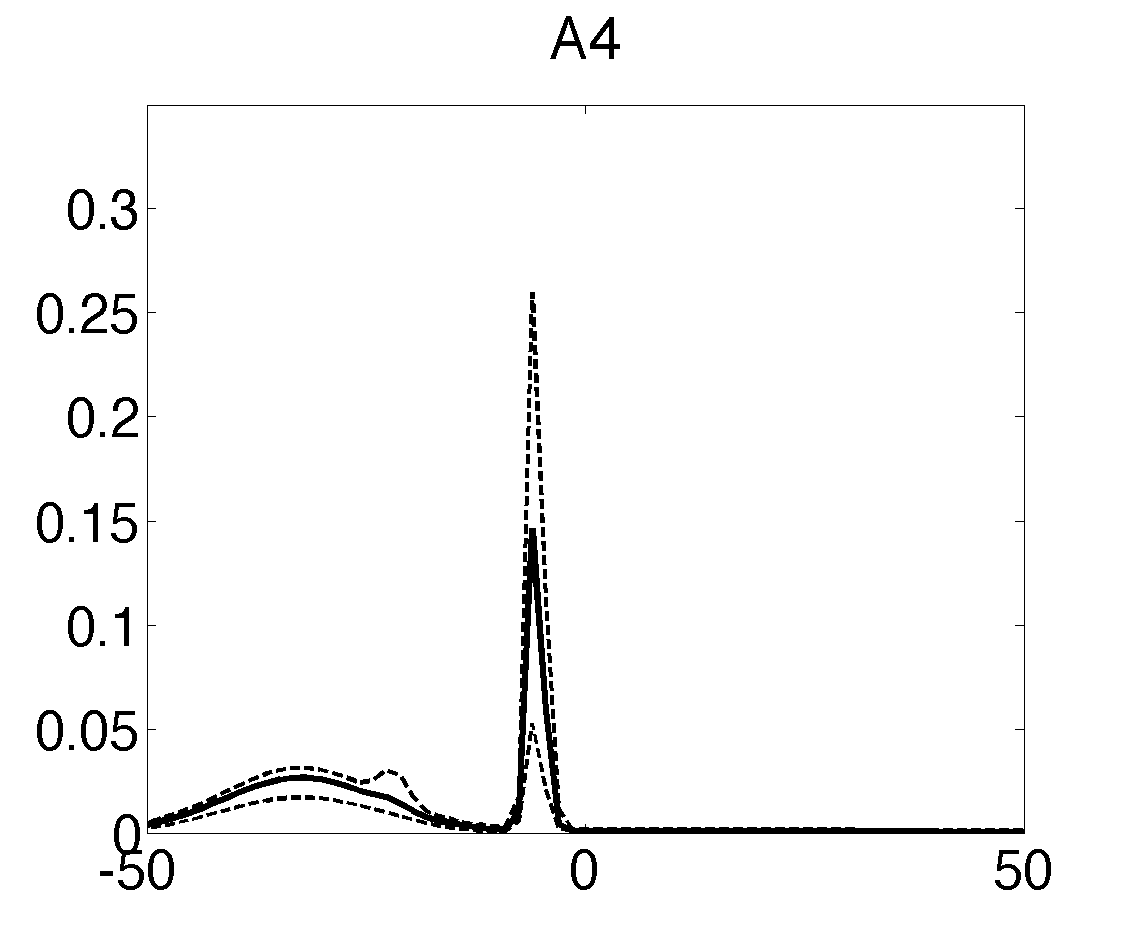}%
\end{minipage}
\par\end{centering}
\begin{centering}
\begin{minipage}[t]{0.33\columnwidth}%
\includegraphics[width=1\columnwidth]{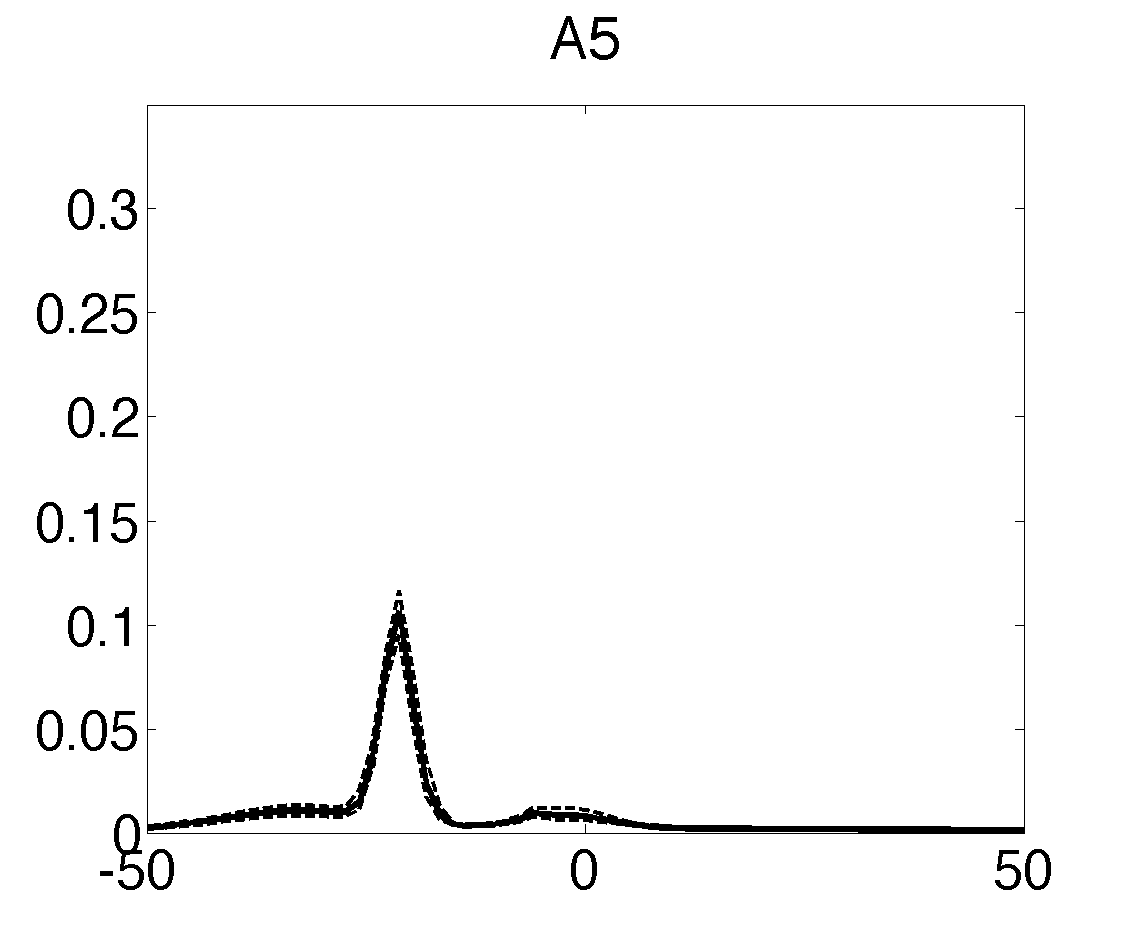}%
\end{minipage}%
\begin{minipage}[t]{0.33\columnwidth}%
\includegraphics[width=1\columnwidth]{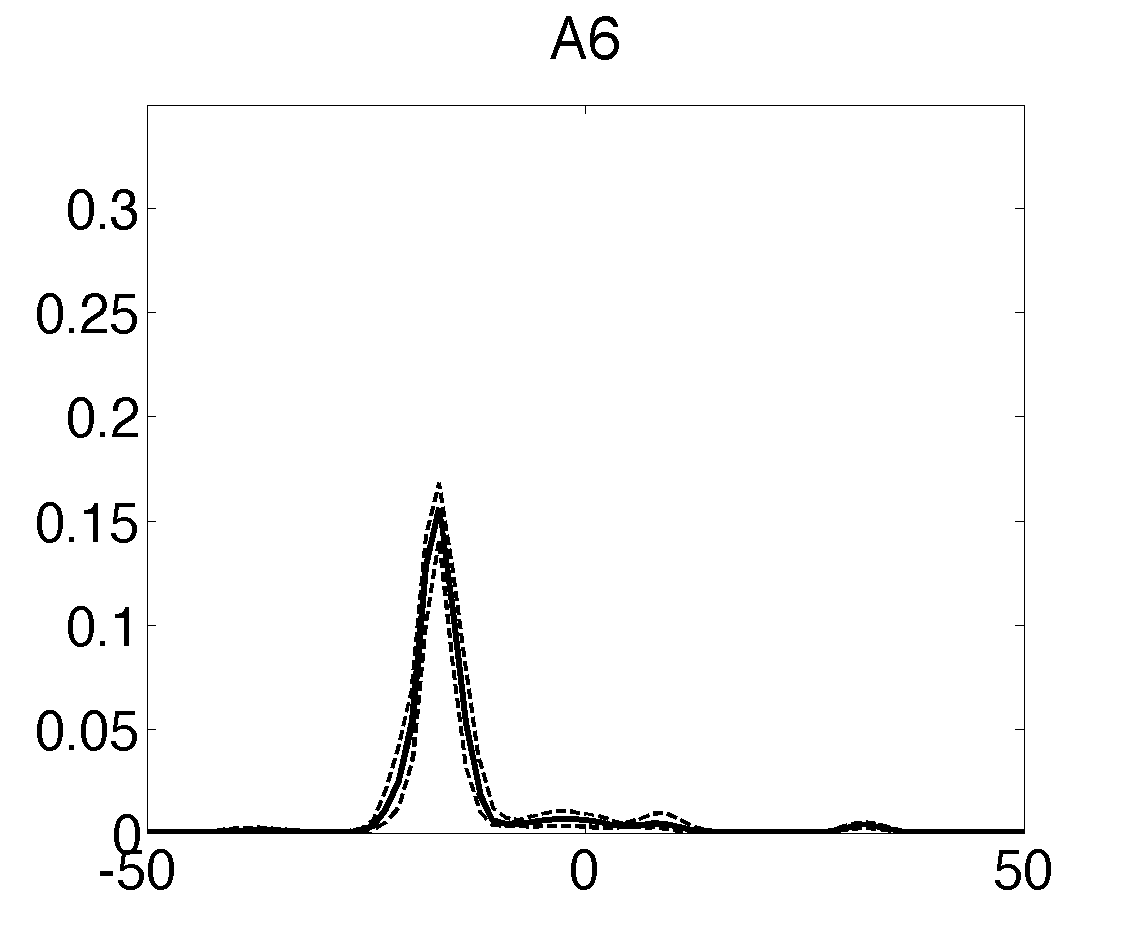}%
\end{minipage}%
\begin{minipage}[t]{0.33\columnwidth}%
\includegraphics[width=1\columnwidth]{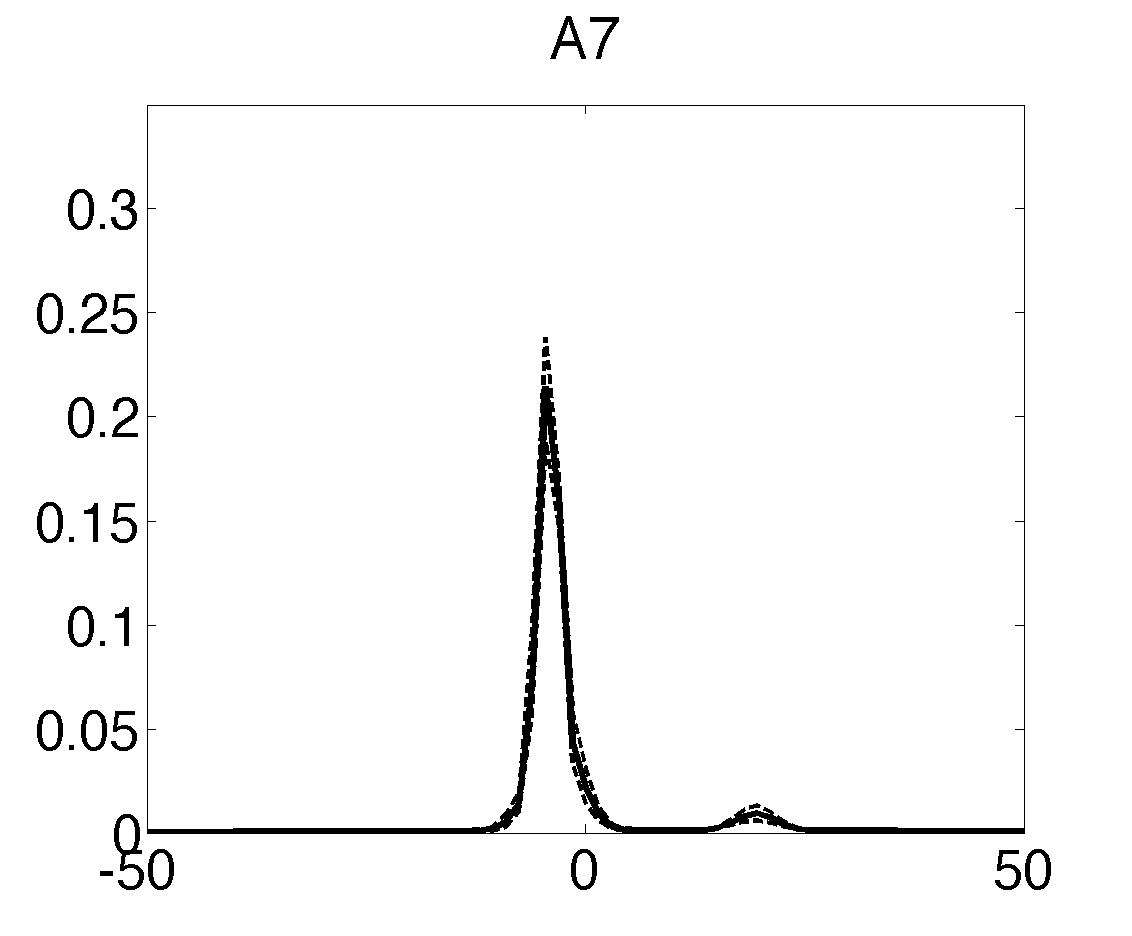}%
\end{minipage}
\par\end{centering}
\begin{centering}
\begin{minipage}[t]{0.33\columnwidth}%
\includegraphics[width=1\columnwidth]{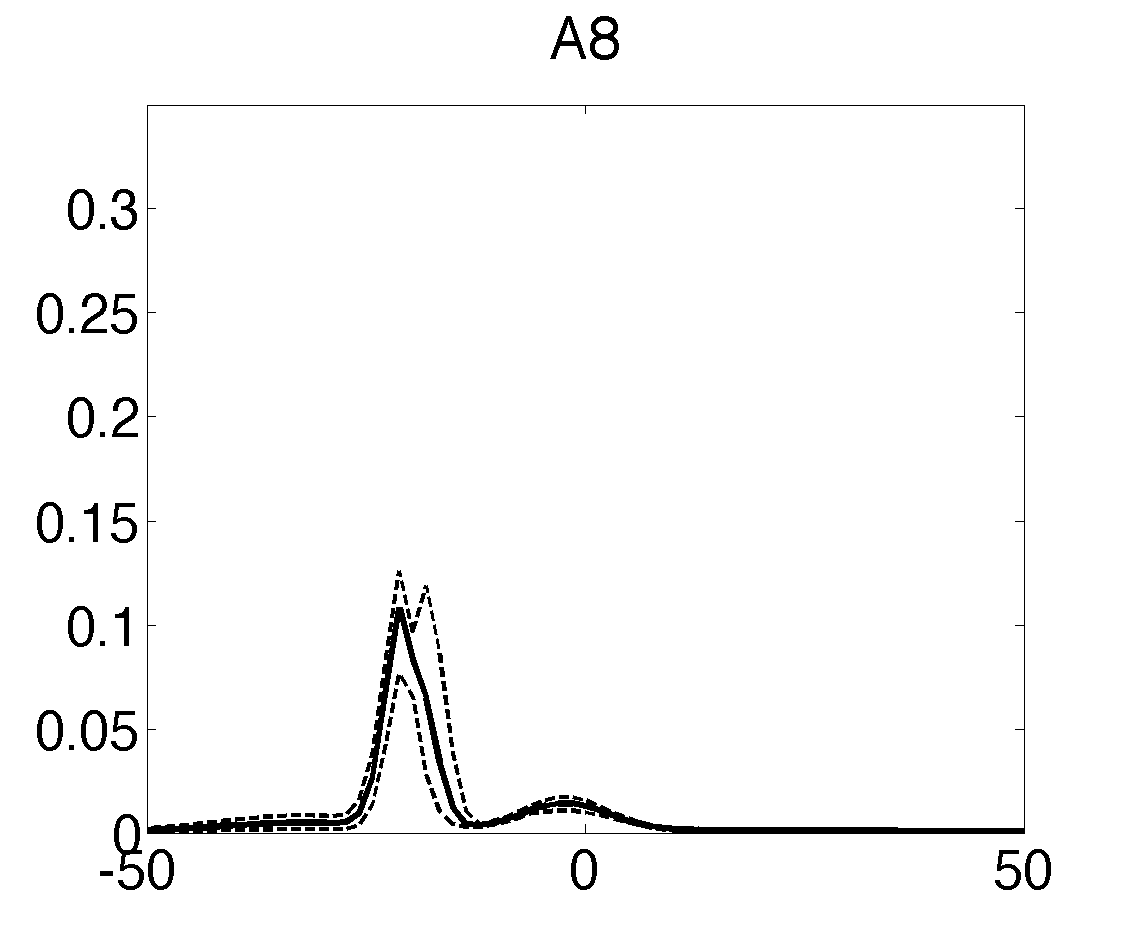}%
\end{minipage}%
\begin{minipage}[t]{0.33\columnwidth}%
\includegraphics[width=1\columnwidth]{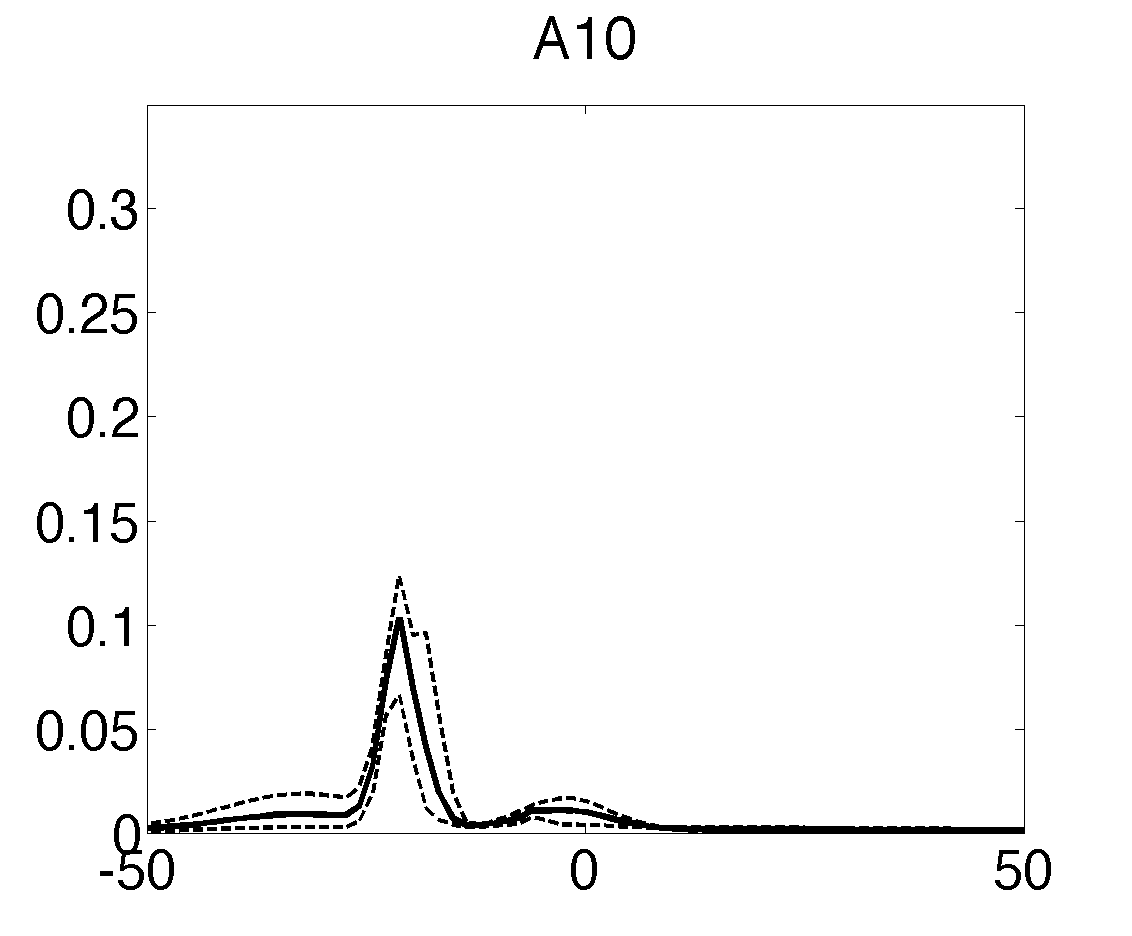}%
\end{minipage}%
\begin{minipage}[t]{0.33\columnwidth}%
\includegraphics[width=1\columnwidth]{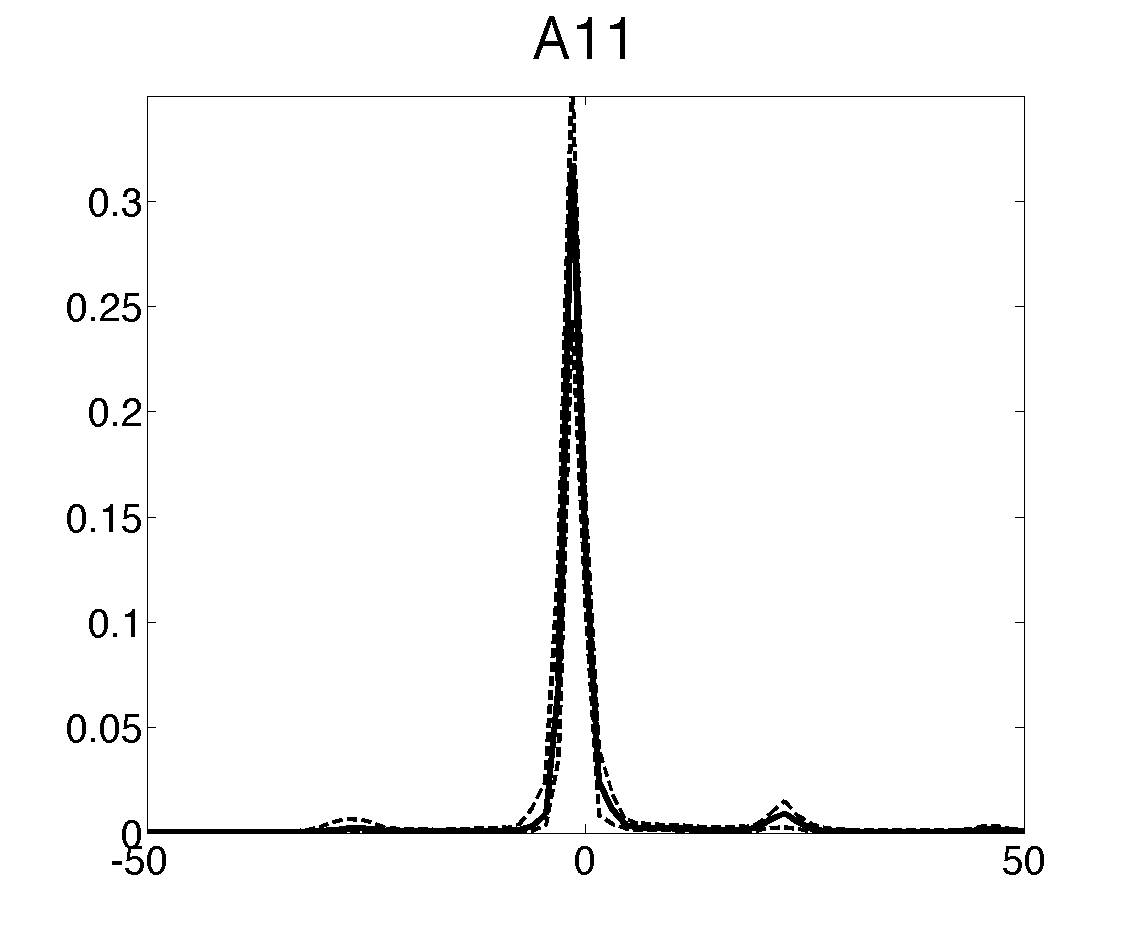}%
\end{minipage}
\par\end{centering}
\begin{centering}
\begin{minipage}[t]{0.33\columnwidth}%
\includegraphics[width=1\columnwidth]{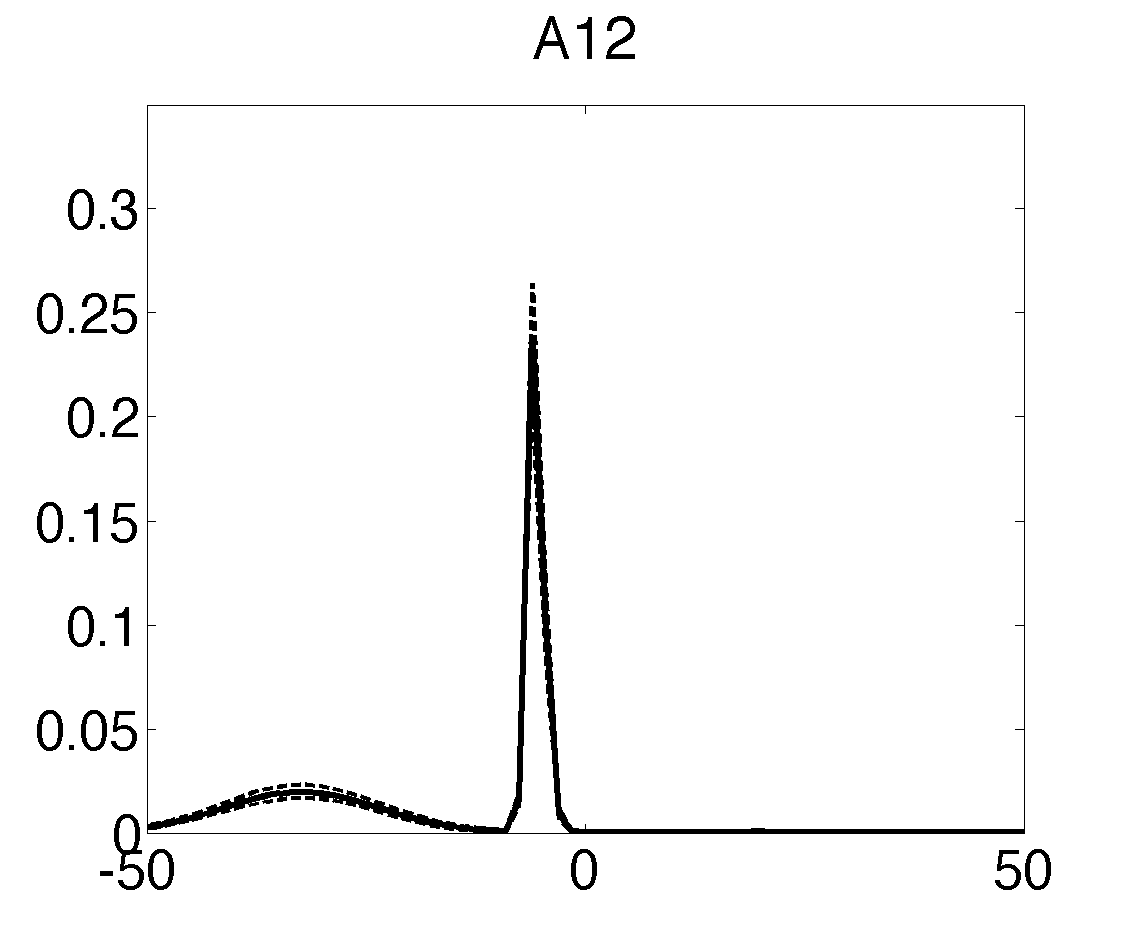}%
\end{minipage}%
\begin{minipage}[t]{0.33\columnwidth}%
\includegraphics[width=1\columnwidth]{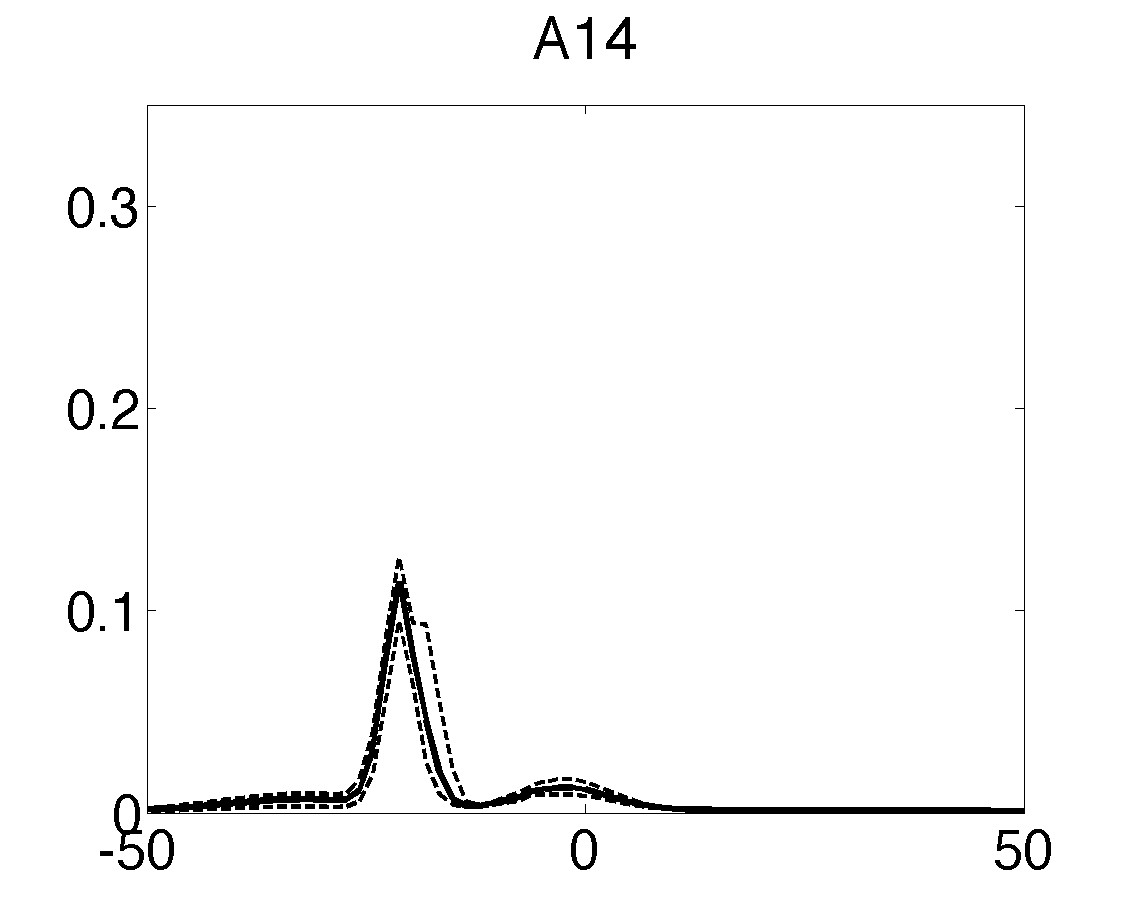}%
\end{minipage}%
\begin{minipage}[t]{0.33\columnwidth}%
\includegraphics[width=1\columnwidth]{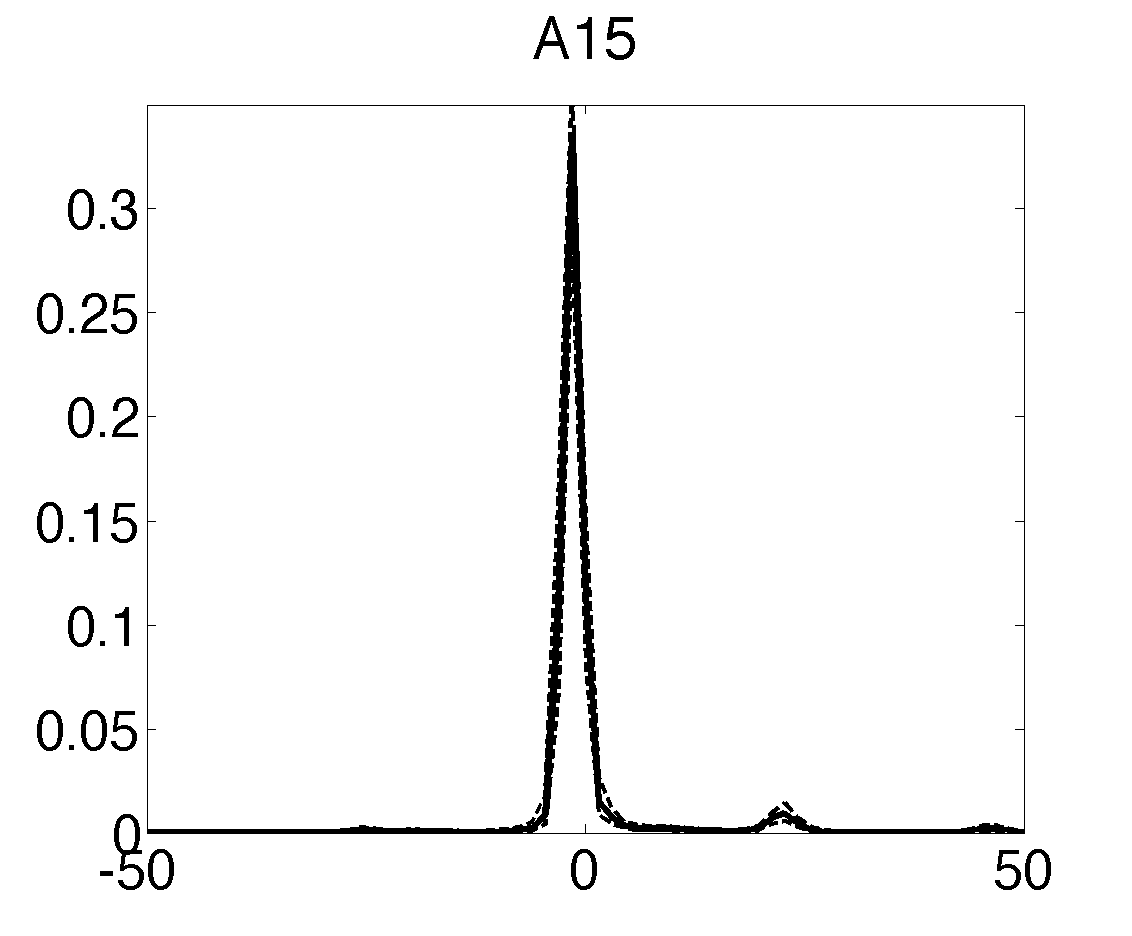}%
\end{minipage}
\par\end{centering}
\centering{}\caption{Reference performances of sample airlines with predictive density
mean (solid) and 95\% credible interval (dotted) (cont.)}
\label{fig: airline reference cont.}
\end{figure}
\begin{figure}[tbph]
\begin{centering}
\begin{minipage}[t]{0.33\columnwidth}%
\includegraphics[width=1\columnwidth]{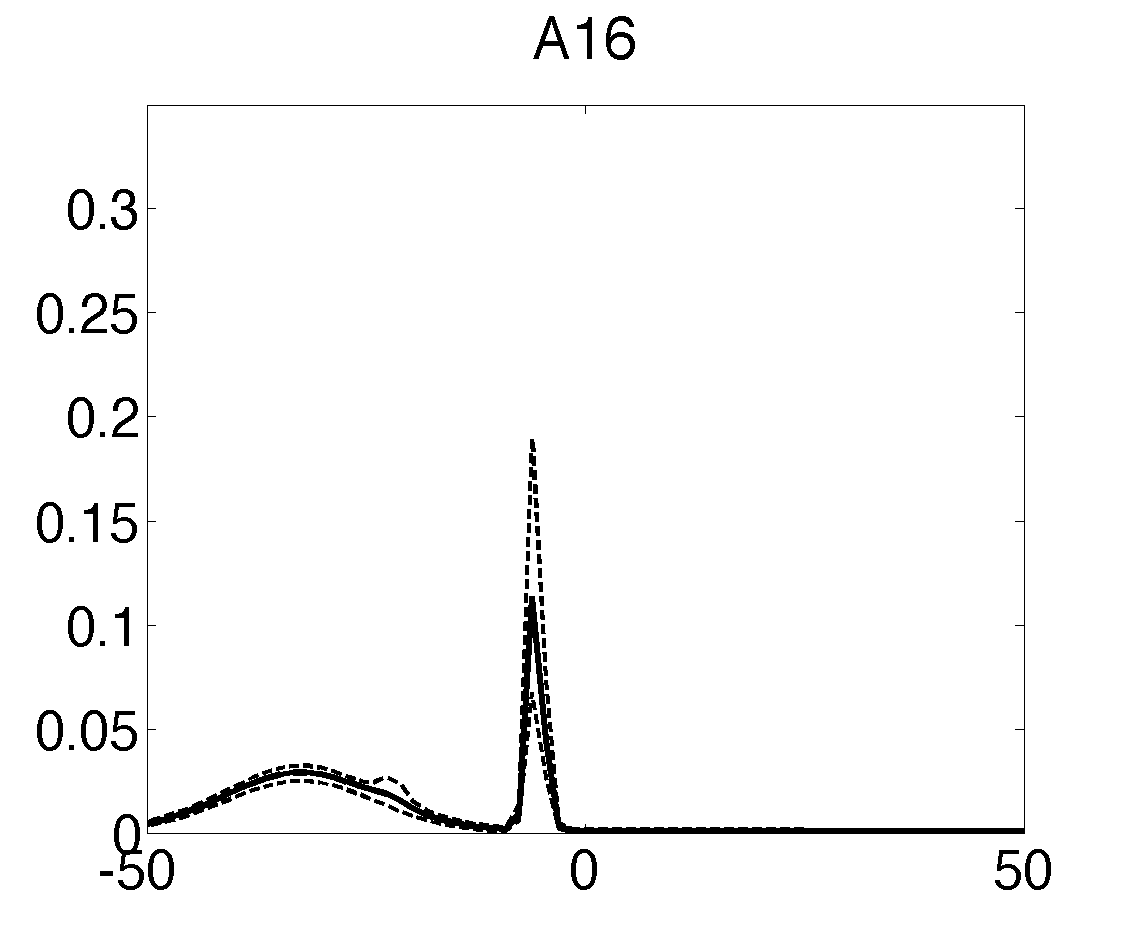}%
\end{minipage}%
\begin{minipage}[t]{0.33\columnwidth}%
\includegraphics[width=1\columnwidth]{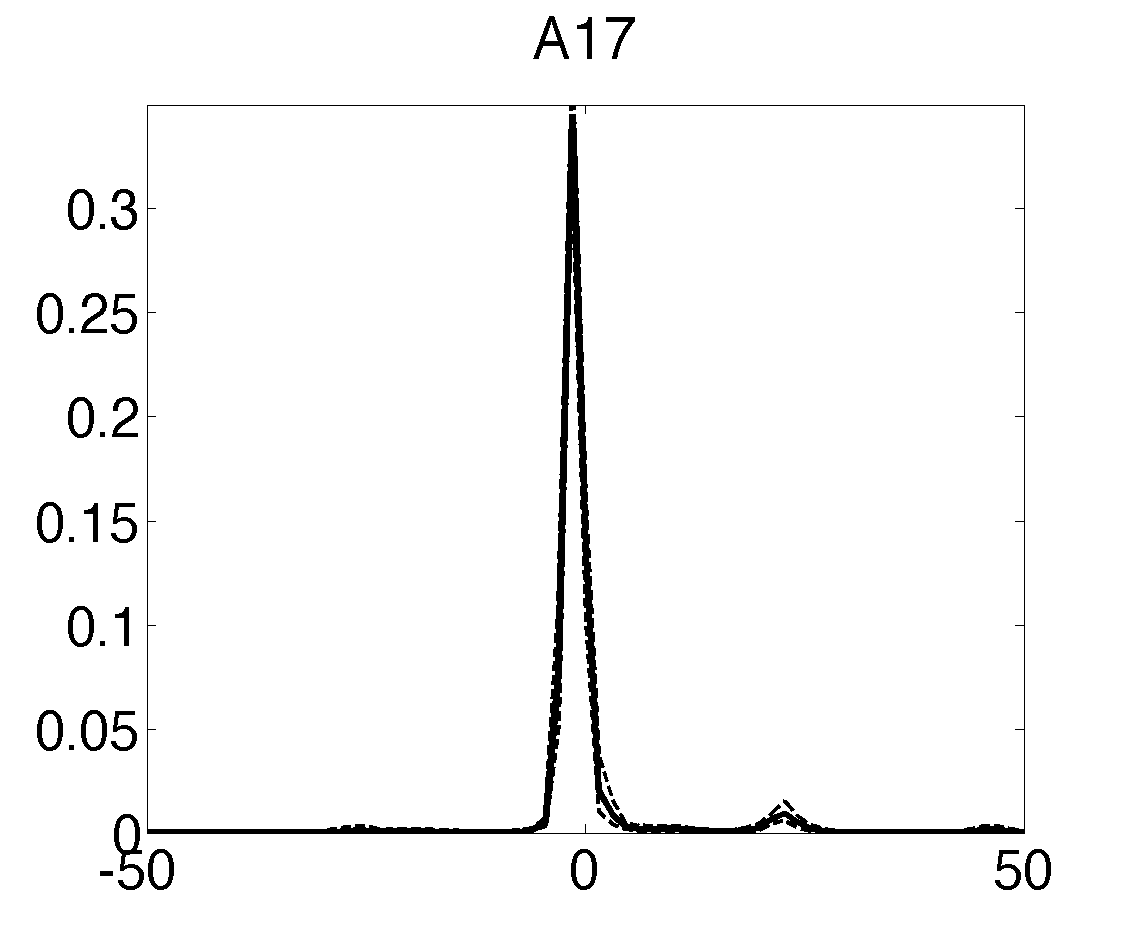}%
\end{minipage}%
\begin{minipage}[t]{0.33\columnwidth}%
\includegraphics[width=1\columnwidth]{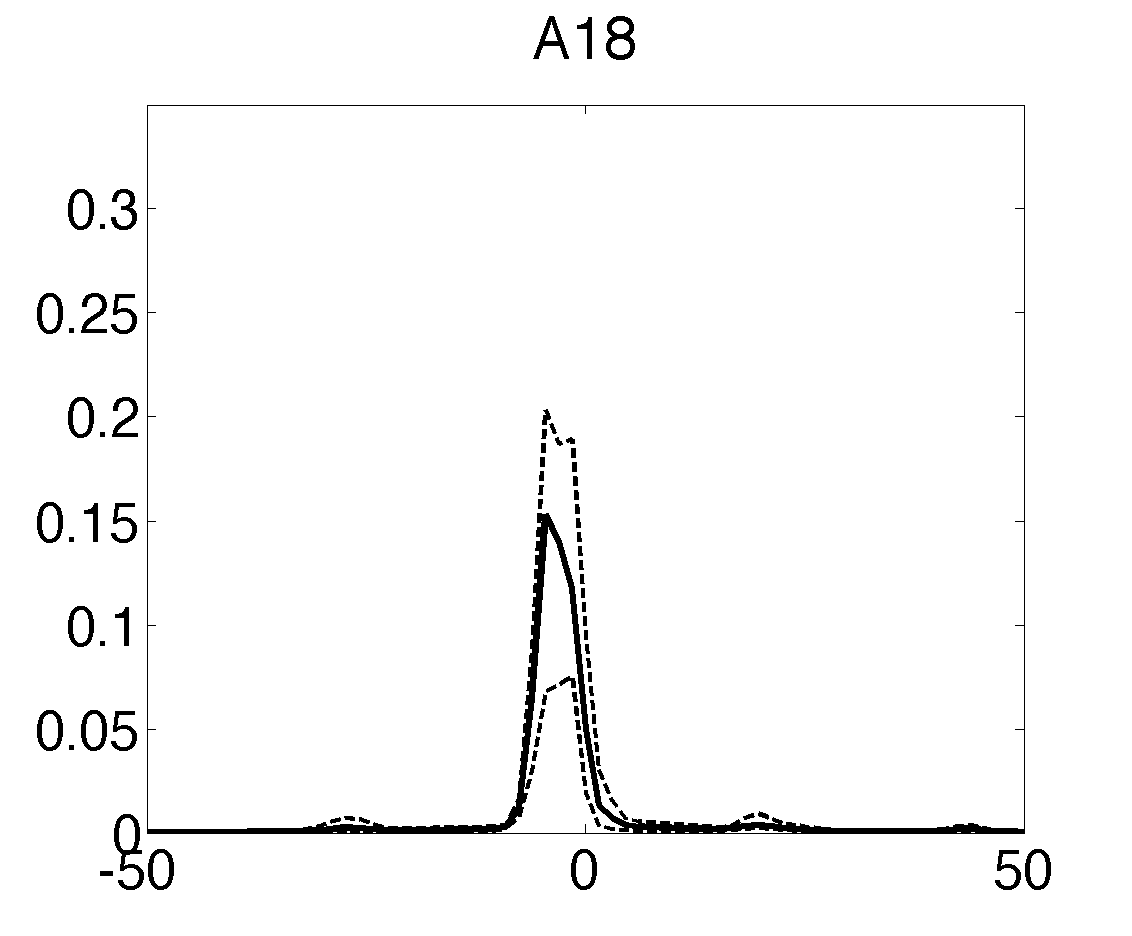}%
\end{minipage}
\par\end{centering}
\begin{centering}
\begin{minipage}[t]{0.33\columnwidth}%
\includegraphics[width=1\columnwidth]{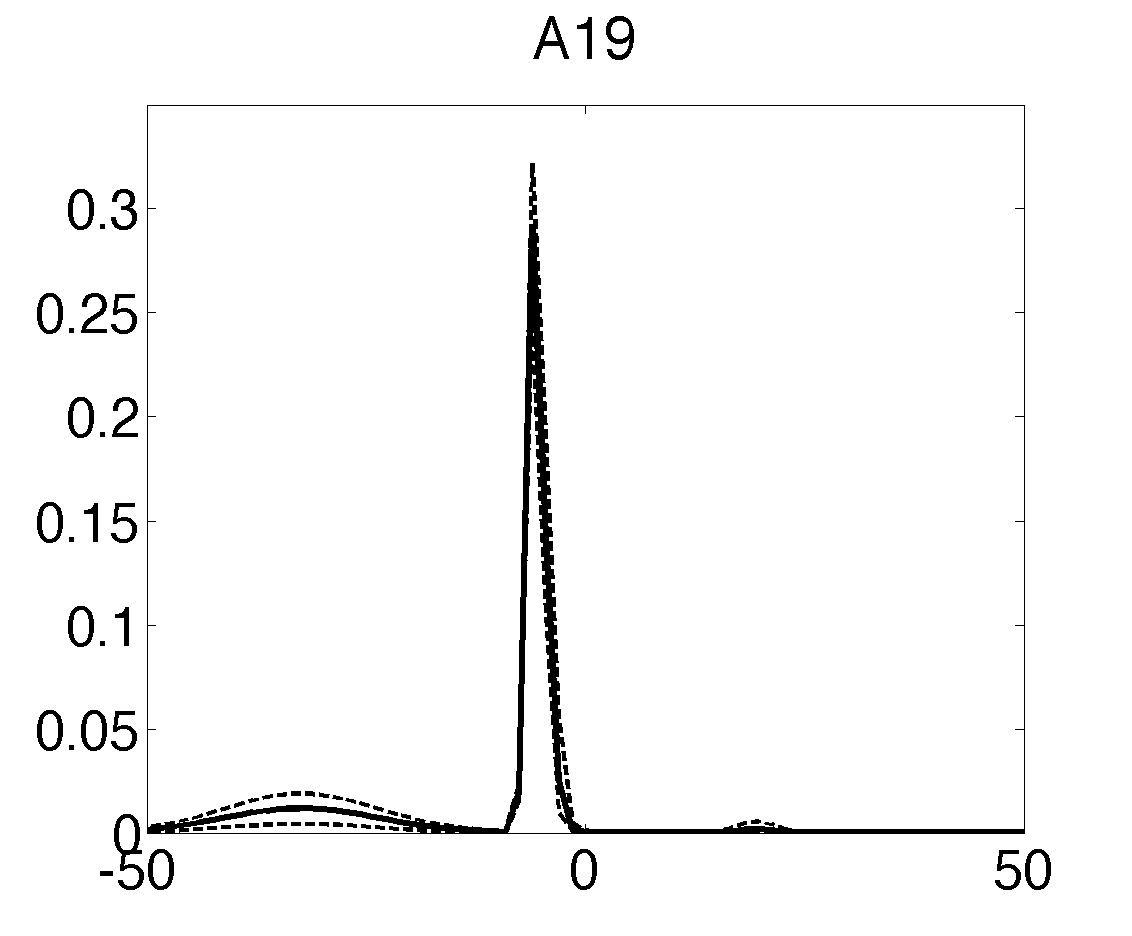}%
\end{minipage}%
\begin{minipage}[t]{0.33\columnwidth}%
\includegraphics[width=1\columnwidth]{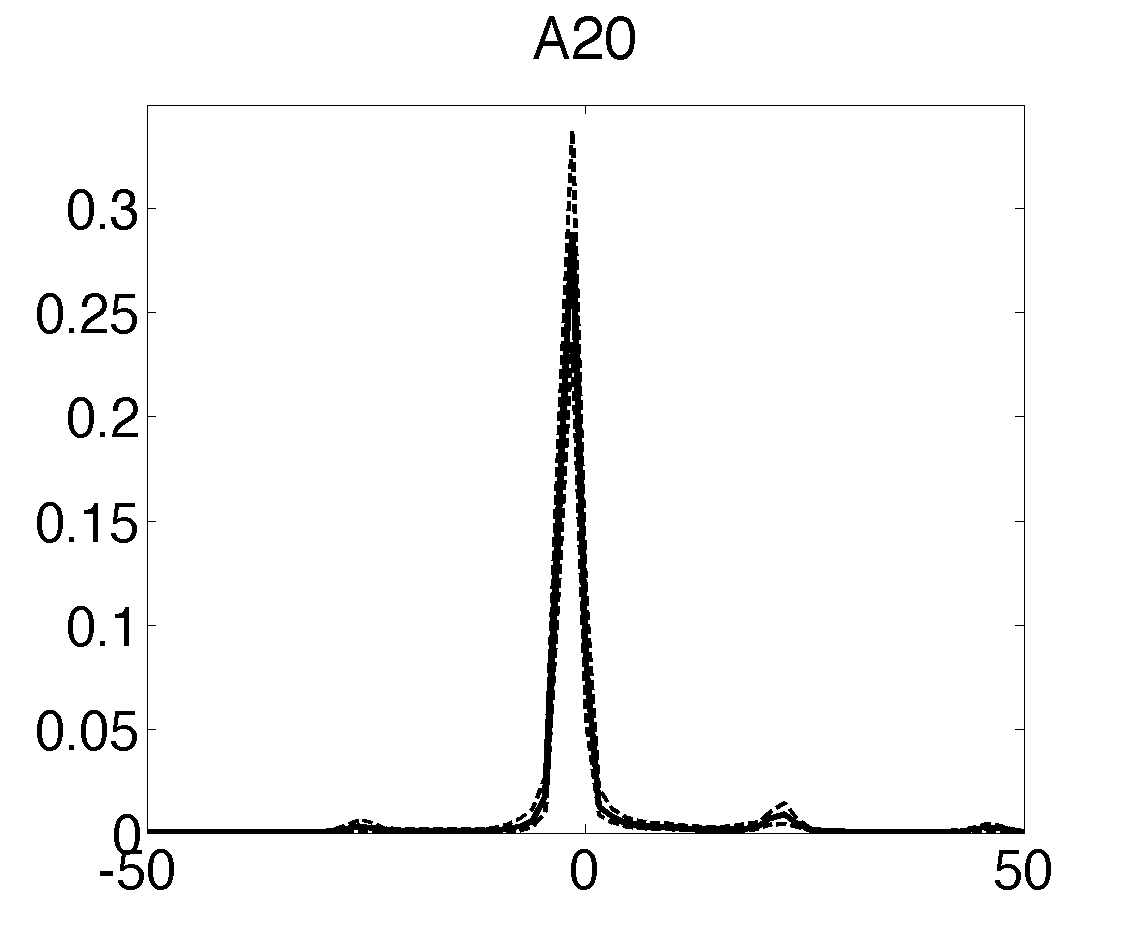}%
\end{minipage}
\par\end{centering}
\caption{Reference performances of sample airlines with predictive density
mean (solid) and 95\% credible interval (dotted)}
\label{fig: airline reference performance }

\end{figure}

\end{APPENDICES} 
\end{document}